\documentclass[opre,nonblindrev]{informs3_no_remark} 

\OneAndAHalfSpacedXI 

\usepackage{natbib}
 \bibpunct[, ]{(}{)}{,}{a}{}{,}%
 \def\bibfont{\small}%
 \usepackage{tikz,calc}
 \usetikzlibrary{decorations.markings}
 \usepackage{soul}
 \setul{}{0.75pt}
 \setuloverlap{1.5pt}
 \usepackage{endnotes}

\usepackage{subfigure}
\usepackage{mwe}
\usepackage{lscape,booktabs,longtable}
\usepackage{amsmath}
\usepackage[pdftex,colorlinks=true,urlcolor=blue,citecolor=blue,pdfstartview=FitH]{hyperref}
\usepackage{comment}
\usepackage{enumerate}
\usepackage[ruled,vlined]{algorithm2e}
\usepackage[graphicx]{realboxes}

\newcommand{\edit}{}
\newcommand{\editTwo}{}

\usepackage[normalem]{ulem}

\newcommand{\E}[1]{{\mathrm{E}\left[#1\right]}}

\newcommand{\PP}[1]{\mathrm{P}\left( #1 \right)}

\newcommand{\IU}{{\pmb{\cap}{}}}
\newcommand{\U}{{\pmb{\cup}{}}}

\newenvironment{myquote}%
  {\vspace{0.05in}\list{}{\leftmargin=0.3in\rightmargin=0.3in}\item[]}%
  {\endlist \vspace{0.05in}}

\usepackage[colorinlistoftodos]{todonotes}
\TheoremsNumberedThrough     
\ECRepeatTheorems

\EquationsNumberedThrough    

\begin{document}


\RUNAUTHOR{Daw and Yom-Tov}

\RUNTITLE{Asymmetries of Service}

\TITLE{
\textsf{Asymmetries of Service:}\\ \textsf{Interdependence and Synchronicity}
}
\ARTICLEAUTHORS{%
\AUTHOR{Andrew Daw\footnote{University of Southern California Marshall School of Business, Data Sciences and Operations (\texttt{andrew.daw@usc.edu})} and Galit B.\ Yom-Tov\footnote{Technion---Israel Institute of Technology, Faculty of Data and Decision Sciences}}

} 



\ABSTRACT{On many dimensions, services can be seen to exist along spectra measuring the degree of interaction between customer and agent. 
For instance, every interaction features some number of contributions by each of \edit{those} two sides, creating a spectrum of \emph{interdependence}. 
Additionally, each interaction is further characterized by the \edit{relative} pacing of these contributions, implying a spectrum of \emph{synchronicity}. 
Where a service falls on such spectra can be a consequence of its design, but it can also be a function of its state. \edit{For instance, as} broadly evidenced empirically, an agent with several concurrent interactions will be slowed in each individual interaction, altering the service's synchronicity. Here, we study a Hawkes cluster \edit{model of the service interaction}, which we show captures \edit{the interdependence and synchronicity spectra and their resulting customer-agent (a)symmetries}. We find insightful connections to behavioral operations, such as proving the occurrence of non-monotonic performance (e.g., inverted-U throughput) from concurrency-driven asynchrony. Hence, we can prescribe the agent's optimal concurrency level. Furthermore, we show how the service design dictates the efficacy of these operational improvements, proving that the concurrency-optimized throughput is itself non-monotonic as a function of the interdependence.
\edit{Of possible} independent interest methodologically, we establish an interpretable \edit{temporal} decomposition for Hawkes clusters.}



\maketitle



%


\section{Introduction}\label{intro}




By comparison to other classic areas of operations research such as manufacturing or production, service is characterized by an interaction between two (or more) people: together, the customer and agent co-produce the service \citep[c.f.][]{fuchs1968service}. Specifically, the success of a service interaction relies on contributions from the agent \emph{and} the customer\edit{, and these two sides of the interaction} rely on one another to complete the service. 
The customer depends on the superior technical knowledge, \edit{authority, or} resources available to the agent, and the agent \edit{draws upon} the customer's ability to clearly articulate their problem and provide relevant information throughout.  As a classic example of service being co-produced by the customer and agent, consider tax preparation services \citep[e.g.,][]{chase1978where,roels2014optimal}. Here, the agent --- an accountant or other tax professional --- has subject matter expertise in filing taxes and qualifying for possible deductions, whereas the customer --- a taxpayer --- has insider knowledge of their income information and financial circumstances. Together, they collaborate to punctually and accurately submit the tax return and, ideally, maximize the taxpayer's refund. Hence, through their interaction, the customer and agent collectively achieve the original service goal across a series of contributions from the two sides.

However, not all service interactions are \editTwo{alike}. For example, in some services, the agent may be expected to do nearly all of the work in the interaction, whereas, in others, the customer may make the vast majority of the contributions; in others still, the customer and agent may contribute roughly the same amount. Often, the exact distribution of work is  dictated by the company as part \edit{of} the design of their service offering, and a given industry typically features many choices across this spectrum. In meal-preparation services, for example, the variety ranges from providing a full, ready-to-eat meal (e.g.,~\href{https://www.factor75.com/}{Factor} or \href{https://www.everytable.com}{Everytable}) to supplying the ingredients and recipe and then letting the customer be the chef (e.g.,~\href{https://www.hellofresh.com}{HelloFresh} or \href{https://www.blueapron.com/}{Blue Apron}). Similar differentiation can be seen in many other arenas, such as education (where the variety of student-teacher arrangements include ``sage on a stage,'' flipped classrooms, private tutoring, massively open online courses, etc.) or fitness (where workouts can coached individually, in live classes, or through apps).
Additionally, some firms even offer multiple designs of the same service to constitute multiple service products. Returning to the hallmark tax preparation example, the platform \href{https://turbotax.intuit.com/}{TurboTax} now hosts three types of tax assistance: the classic TurboTax software (which has been labelled ``do your own taxes''), Live Assisted (``experts help as you do your taxes''), and Live Full Service (``we do your taxes for you'').\endnote{Labels are as seen on the \href{https://turbotax.intuit.com/}{TurboTax} website in January 2024.} Naturally, these three create a spectrum for the allocation of tasks within the interaction, from primarily self-service by the customer, to collaborative service between the customer and the tax expert, to largely agent-driven service.  


In a similar but separate manner, the customer and agent can also differ from one another in the pace of their contributions. That is, in some services, the customer may complete their tasks much more quickly or more slowly than the agent does, whereas, in others, the two sides may be relatively in sync. By comparison to designed division of labor, the difference of speed may be dynamic and dependent on the state of the system. For example, myriad service contexts have shown that the agent's overall workload, broadly defined, will impact the pace at which they contribute to the interaction \citep[e.g.,][or see also Section~\ref{litReview}]{Delasay2017LoadEffect}. Again invoking the tax preparation example, a tax expert on an online platform like TurboTax realistically is interacting with several customers concurrently.  As the agent is assigned more concurrent customers, their pace of response to any individual customer will likely slow due to the heightened division of their attention and the increased demand of their focus. Hence, as the agent's concurrency grows, their pace will slow relative to the customer's, which does not (directly) depend on the total customer load. This creates a state-dependent spectrum of disparate speeds, where the agent may contribute more quickly than the customer when operating under a small concurrency, roughly on pace with the customer under a moderate concurrency, and more slowly than the customer under a large concurrency. 

\subsection{\edit{Research Question and Context}}\label{questionSec}

This paper is principally concerned with these spectra measuring the cardinality and pace of the contributions made by the customer and agent across the duration of a given service interaction. Keeping to a general setting, our goal is to understand how these service dimensions are influenced \edit{both by the design (i.e.,~offline model parameters) of the service and by the state of the system (e.g.,~the agent's concurrent load)}, where the latter may offer opportunity to improve managerial decision making in light of such behavior. Building from the modeling and empirical analysis in \citet{daw2021co}, we approach this pursuit through a history-driven stochastic model of concurrent service interactions, and we draw inspiration from classic service theory, service design, and behavioral operations management in our analysis.

First, the scope of our stochastic model is guided by principles from the field of behavioral operations. Following the reasoning that ``a  robust understanding of individual‐level behaviors will have the most impact when it can meaningfully connect to the system‐level behavior and metrics that operations managers tend to care about, such as average wait times, throughput, or utilization,'' we are principally concerned with modeling and analyzing at the level of an individual customer-agent service interaction \citep[][pg. 325]{allon2018behavioral}. That is, by comparison to much of the service stochastic modeling literature, we model each individual service duration as a stochastic process itself.
Following recent evidence on the way that service unfolds dynamically and reciprocally, the interaction model we study here is a Hawkes self-exciting cluster model.
We aim to employ  this interaction-level model to yield fundamental, analytical understanding of system-level behavior.
Much like the quote from \citet{allon2018behavioral} suggests, we are interested in how the throughput of the service system will be shaped by the characteristics of the interaction, namely, by the spectra measuring the respective contributions from the customer and the agent.

Turning to service theory and design, let us recognize that these spectra, while largely absent from the queueing and stochastic models literature, are not contributions of this work, but rather concepts sourced from theory on services, organizations, and relationships. Classic service literature has held that the level of customer ``contact'' during service is a critical determinant of the system performance. For instance, \citet{kellogg1995constructing} (and references therein) distilled contact into three main elements: coupling, {interdependence}, and {information richness}. Coupling, which we will deem \emph{synchronicity} to avoid confusion with the common technique of analysis for stochastic models, refers to whether the customer and agent  share the same time frames (synchrony) or predominantly operate on separate time scales (asynchrony). Or, in the words of \citet[pg.\ 1736]{kellogg1995constructing}, asynchrony is when ``parties affect each other suddenly, occasionally, negligibly, and eventually,'' whereas synchrony is when ``parties affect each other continuously, constantly, significantly, and immediately.'' In the language of our preceding examples, this is captured in the relative (average) paces of contributions from the two sides. Then, \emph{interdependence} denotes whether the two parties collaborate and rely upon one another significantly (co-production), or, alternatively, the service goal is primarily achieved by one party alone (self-production). Connecting again to our prior discussion, this is (a distribution over) the possible allocations of contributions or tasks between the two sides. Finally, \citet{kellogg1995constructing} described \emph{information richness} as the nuance and variability underlying the interaction, which creates randomness throughout the service. 
Naturally, this uncertainty and variability can be readily captured in a stochastic model.

The confluence of randomness with the notions of synchronicity and interdependence leads to our modeling framework. As alluded to in the motivating examples, we will think of \edit{interdependence} as being determined by the nature or design of the service, meaning that we will identify interdependence as a fixed parameter of the interaction model. On the other hand, we will connect \edit{synchronicity} to the state of the service system, thus allowing the broader operation to impact the dynamics of individual interactions through \edit{the relative pacing of the customer and agent}. When viewing an interaction as isolated in a vacuum, our model will represent synchronicity as a parameter much like the way in which it represents interdependence. Then, in order to meaningfully connect the individual-level to the system, in the course of our analysis we will move to view this parameter as a function of the agent's concurrent customer load. Specifically, we will let the value of the concurrency act as a functional modulation on the agent's pace of contributions, delineating the direct impact of the load onto the agent's side of the service interaction. Hence, we offer a novel stochastic modeling perspective on how concurrency impacts service performance.


Yet, let us recognize that this discussion begs the question: ``is there not already a model of concurrent services?'' Indeed, one of queueing theory's most well-studied models and service disciplines, the processor sharing queue, is by definition devoted to the notion of sharing service capacity \citep[see, e.g.,][or references therein]{yashkov2007processor}. Originating with \citet{kleinrock1967time}, the processor sharing queueing model idealizes round-robin processing of jobs. \editTwo{For} $k$ concurrent jobs in the service system and a total service capacity of $C$ jobs per time unit, each job ``receives continuous processing at a rate $C/k$'' \citep[][pg. 243]{kleinrock1967time}. Hence, as long as there is any positive number of jobs present, the evenly divided capacity implies that the throughput of the system is $C$ at any level of concurrency.

Relative to the opening discussion of this paper, this language of ``jobs'' in the previous paragraph is intentional: the history of processor sharing is in machines and computation. Quite like how cloud computing is viewed presently, the motivating scenario described in \citet{kleinrock1967time} is a ``time-shared'' computing system that seems like a personal processor to any individual user, yet actually divides a single machine's capability across all jobs from all present users. 
In concurrent service systems, a growing stream of queueing models for multitasked services updates the processor sharing discipline to be both limited and state-dependent \citep[see, e.g.,][for both recent advances and a review, or see also Section~\ref{litReview}]{storm2023diffusion}. 
We use ``limited'' in the sense such as in \citet{zhang2009law}, meaning there is a maximal possible concurrency for each agent, and any number of simultaneously present customers above this threshold must wait. Similarly, by ``state-dependent,'' we refer to a structure such as in \citet{tezcan2014routing}, meaning that an agent's throughput, or total service completion rate across all their concurrent customers, may change as the number of customers changes. In a multi-server system, this construction applies in a distributed fashion, where each individual agent is modeled \editTwo{with} this limited, state-dependent processor sharing protocol. 


These generalizations away from the \citet{kleinrock1967time} notion of processor sharing reflect a departure from computation, conducted by machines, to services, delivered by people. The combination of these limited and state-dependent constructions allows the models to assume a \emph{multitasking effect}, meaning that further dividing the agent's attention with the assignment of an additional concurrent customer may come with \edit{decreasing} marginal returns in efficiency. Taken to an extreme, this implies that, at some point, the system productivity may actually be worse off as the agent's concurrency grows. For example, in the model of chat-based contact center services in \citet[pg. 946]{tezcan2014routing}, it is assumed ``that [the agent's throughput]  should be increasing in [the concurrency] until a certain level,'' after which it should be decreasing. Hence, by contrast to the constant --- or, more generally, monotonic --- throughput implied by classical processor sharing, these state-dependent conditions suppose that the throughput is non-monotonic as a function of the agent's concurrency. Moreover, by the accompanying assumption of limited-ness, it is implicitly taken that an optimal concurrency can be identified, and, as phrased by \citet[pg. \edit{2685}]{storm2023diffusion}, the maximal concurrency can be managerially curtailed ``up to the point that multitasking is no longer efficient.''



To see that this supposed non-monotonic performance is indeed well-motivated, one can simply return to the behavioral operations management literature.  Indeed, over the last decade or so, ample empirical evidence has emerged to show that, when comparing real-world service observations with the implications of classical processor sharing, there is a mismatch between data and theory. For example, in settings as diverse as hospitals \citep[e.g,.][]{Kc2013DoesDepartment,berry2016past}, contact centers \citep[e.g.,][]{luo2013staffing,Goes2017WhenMultitasking}, and court systems \citep[e.g.,][]{coviello2014time,bray2015multitasking}, the agent- and/or system-level throughput has been seen to be non-monotonic as a function of concurrency. 
More specifically, and more clearly in contrast with processor sharing, these data sets reveal an \emph{inverted-U shape} of throughput. Like in the assumptions which generalize processor sharing for models of concurrent service systems \citep[such as quoted from][]{tezcan2014routing}, this means that the agent's (or, generally, the system's) throughput rises and then falls as the concurrency increases.

\subsection{Organization of Paper and Summary of Contributions}

In total, we take the documented data-theory discrepancy as an impetus to revisit the bedrock notion of concurrency in services. As motivated by these observed violations of the canonical stochastic model of multitasked service and as measured along the dimensions of interdependence and synchronicity,  our aim in this work is to approach concurrency through a novel perspective gained by modeling at the level of the customer-agent interaction. 
After reviewing related fields and relevant literature in Section~\ref{litReview}, this paper's pursuit and its corresponding contributions are organized as follows\edit{:}

\edit{\textsf{\textbf{Part I (Formalizing and Analyzing the Spectra of Synchronicity \& Interdependence):}}} First, we have two \edit{foundationally}-focused sections \edit{in which we develop and analyze a stochastic model of the customer-agent service interaction. Section~\ref{sec:model} defines the model, and Section~\ref{clusterSec} provides our core understanding of it. Specifically, in} Section~\ref{sec:model}, we define the Hawkes cluster model of the interaction at the level of individuals \citep[generalized from the form fit to contact center interaction data by][]{daw2021co}, and we also intuitively explain model parameters for the interdependence and synchronicity of the service. 
\edit{Then}, in Section~\ref{clusterSec}, we establish a novel  \edit{and} interpretable \edit{temporal} decomposition for this stochastic model through random parking functions, a family of probabilistic combinatorial objects\edit{, and we} formally prove equivalence to the standard Hawkes cluster definition (Theorem~\ref{equivThm})\edit{. We} then use this methodology to \edit{rigorously link the model to} the spectra of interdependence and synchronicity  \edit{and analyze their asymmetric extremes (Proposition~\ref{binSidesProp} and Theorem~\ref{convThm}, respectively)}. By consequence, we develop asymptotically tight upper and lower bounds for the mean cluster duration (Proposition~\ref{boundProp}).

\edit{\textsf{\textbf{Part II (Connecting the Individual-Level Model to System-Level Performance):}}} Following the \edit{modeling and methodological sections in Part I}, we have three sections which elucidate the (a)symmetry in individual-level \edit{service} interactions and relate this behavior to  system-level performance metrics. \edit{Section~\ref{nonMonSec} frames synchronicity in terms of concurrent or multi-tasked services, Section~\ref{symSec} contrasts the model and insights with prevailing approaches to concurrency, and Section~\ref{interSec} uncovers the impact of interdependence in these concurrent interactions.} \edit{Specifically}, in accordance with observations from the behavioral operations literature, \edit{Section~\ref{nonMonSec} contextualizes} the interaction-level synchronicity \edit{as dependent on the agent's overall concurrency level, a system-level quantity. In this section,} we prove general conditions under which the \edit{agent's} throughput \edit{will be} non-monotonic as a function of monotonic slowdown \edit{from their concurrent load}, characterizing the \editTwo{widely}-observed inverted-U shape of performance (Theorem~\ref{iuThm}). By consequence, we have that an optimal concurrency exists (Corollary~\ref{optKcor}), and we show how it can be found (Proposition~\ref{convexProp}). Section~\ref{symSec} \edit{is then} an interlude between our main managerial results\edit{. In} an insight that is more modeling than managerial\edit{,} we explain that Section~\ref{nonMonSec} was able to reveal non-monotonic system performance specifically because the \edit{model's agent-side slowdown} allows \edit{for} asymmetry between customer and agent\edit{. We show} that symmetric slowdowns cannot reproduce non-monotonicity (Propositions~\edit{\ref{symProp1} and~\ref{symProp2}})\edit{, revealing a \emph{service-computation contrast}.}
\edit{Finally}, in Section~\ref{interSec}, we revisit the other spectrum of the service interaction, interdependence, and prove that there exists a second non-monotonicity\edit{. \editTwo{That is}, we prove the existence of} a \editTwo{U}-shape that sits along the peaks of the concurrency-optimized throughput in each \editTwo{inverted-U}-shape (Theorem~\ref{uThm}). Hence, connecting the two dimensions, our insights show that the interdependence of the service \editTwo{interaction} dictates the efficacy of the operational improvements offered by controlling the synchronicity.

\edit{After these two primary parts, the main body of the paper concludes in Section~\ref{conclusion}} with a discussion of these results, possible extensions, and additional connections. All proofs are contained in the appendix, as are auxiliary \editTwo{theoretical results and} numerical experiments that both apply these findings to models with real-world parameter estimates and showcase the robustness of our insights.

\section{Motivation, Background, and Further Literature Review}
\label{litReview} 

Before proceeding with our stochastic service interaction model and the accompanying analysis of service asymmetry, let us provide additional context by surveying  literature that  inspires and guides this research. 

\subsection{Service Design, Customer Contact, and (A)Symmetry in Services}\label{sl:sd}


A central theme of this paper is that service is characterized by customer involvement. This is a core pillar of our stochastic modeling framework, and often we will invoke the customer's role as a distinguishing factor of service relative to, say, computation, manufacturing, or production. While we believe this perspective could be more well-represented in stochastic models of service, it has been widely considered in classic service theory.  
For example, \citet[pg. 193]{fuchs1968service} posited that the successful completion of a service relies on the ``knowledge, experience, and motivation'' of the customer, and \citet[pg. 331]{sampson2006foundations} proposed that dependence on the customer typifies service, asserting that the ``presence of customer inputs is a necessary and sufficient condition to define a production process as a service process.'' Building on this notion, the theory of ``customer contact'' proposed by \citet{chase1981customer,chase1978where} both recognizes service's dependence on the customer and proposes a spectrum of this dependence, ranging from scenarios in which the customer contributes significantly (high contact) to scenarios in which they are more passive (low contact). As discussed in the introduction, \citet{kellogg1995constructing} later refined this notion to measure both the cardinality (interdependence) and pace of contributions (synchronicity). We will explicitly show how these spectra are captured by our stochastic model.



The recognition of such spectra introduces degrees of freedom in both modeling and managing services, and a recent of stream of literature approaches this opportunity from a strategic design perspective. For example, \citet{roels2014optimal} offers several examples of interdependence, such as tax preparation, financial services, and education, and develops a model of joint production for a single-task service in which the customer and agent each strategize for the level of effort they exert in achieving the service goal. This leads to a product-process matrix for a firm's optimal design of co-produced services, which further informs how the interdependence of a service should be designed based on the degree of task standardization. Building on this, \citet{roels2021you} provides strategic behavioral explanations for where a given service may naturally exist on the interdependence spectrum, as informed by the incentives and stakes of the interaction. Also influential here is the multi-stage perspective taken by \citet{bellos2019should,bellos2021service}, in which the service process is modeled with a fixed number of steps and various settings of work allocation and complexity, and in which the customer's experience in one step has downstream influences on the experience in later steps. \citet{bellos2019should} considers a spectrum of routine to non-routine services and finds that the agent should have more control of the interaction when the service is more standard. \citet{bellos2021service} then generalizes this to a continuous decision of resource allocation, rather than a discrete task-offering choice, and  thus uncovers a richer understanding of the optimal design of effort.


\subsection{Evidence of Behavioral Phenomena Which Shape Services}\label{sl:bom}

Service theory has also recognized that behavior should be incorporated into the design and control of services \citep{chase2001want}. Indeed, evidence has amassed throughout the behavioral operations  management literature that the behavior of customers and agents has significant impact on the performance of services, often in ways that \edit{are not captured in and may even be contradicted by} the dynamics of hallmark stochastic service models.  Of particular interest here are observations of non-monotonic relationships between performance and load, which is by now well-documented across many service contexts and many definitions of those metrics \citep{Delasay2017LoadEffect}. For instance, in two different healthcare settings, \citet{Kc2009ImpactOperations} showed  that patient throughput would rise as the new patient rate (load) increases but then would decrease under prolonged periods of high load (overwork), leading to an inverted-U shape of throughput. Similar in spirit but quite different in setting, \citet{lee2021managing} observed an inverted-U relationship between sales and customer traffic in retail fitting rooms.



In this paper, we are specifically interested in load as defined by concurrency, meaning an agent multitasking across several simultaneous customers. The behavioral operations literature has documented non-monotonic performance under this structure as well. For instance, \citet{tan2014does} observes an inverted-U relationship between sales (measured in dollars per check) and concurrent workload by servers in multiple locations of a restaurant chain, and, likewise, \citet{Kc2013DoesDepartment} finds \edit{an} inverted-U shape of throughput (measured in patients per hour) as a function of the concurrent workload by physicians in an emergency department. Similarly, \citet{coviello2014time,coviello2015inefficiency} describe an inverted-U throughput (measured in cases completed per week) in the number of multitasked cases by judges in Milan. 

Critically, the empirical literature has also devoted great attention to distilling the underlying drivers of these observations and understanding such non-monotonicity on the level of individuals. For an example quite relevant to our analysis, in data from a chat-based contact center, \citet{Goes2017WhenMultitasking} finds that not only is the agent-side response time increasing in the concurrency, the marginal effect from additional concurrency is increasing as well. That is, the agent-side slowdown is \emph{super-linear in the concurrency}. 
Such phenomena may be supported by psychological studies that explain that productivity is lost in multitasking due to the additional effort needed to refocus on an interrupted task, which is known as cognitive sharing \citep[e.g.,][]{gladstones1989division,pashler1994dual,rubinstein2001executive}. 
These drivers find additional support in other operational studies.
For example, \cite{Kc2013DoesDepartment} hypothesizes that although the initial increase in service rate stems from more efficient use of physician time (treating one patient while waiting for the results from other patients to arrive),  the   eventual   decrease is due to the cognitive changeover costs from task switching. \cite{freeman2017gatekeepers} provides further evidence of multitasking‐related service slowdowns at high load, and \cite{berry2016past} observes over-treatment, where physicians order expensive low‐efficacy tests, possibly in order to postpone concurrent patients.

As this volume of evidence for unanticipated behavior has amplified, there has been a growing  response of stochastic models which incorporate or address these features. For example, several recent papers have used queueing models with state-dependent service rates to model behavior. For instance, \citet{dong2015service} investigates \edit{the impact of} load-induced (meaning total number in system of a modified  Erlang-A queue) service slowdown on staffing. Similarly, \cite{chan2014use} studies \edit{the effects of} state-dependent service speedup and returns on service performance. \citet{delasay2016modeling} analyzes the effect of both load and cumulative overwork in a similar manner to what was observed by \citet{Kc2009ImpactOperations}. \citet{cho2019behavior} studies both speedup and slowdown effects and analyzes the optimal staffing that results as a consequence of this profile. There has also been work on cases in which service rates can be selected strategically, such as in \citet{zhong2023behavior}, where both system load and payment can affect the service rate. Of course, behavior need not be captured just through service rates and queues; for instance, \citet{arlotto2014optimal} uses an infinite-armed bandit to model a population of workers who learn, and \citet{gurvich2015collaboration,gurvich2018collaboration} develop a fluid network perspective on collaboration and multitasking. Naturally, many other approaches exist for modeling behavior in services, and the interested reader can find several more in the survey from \citet{allon2018behavioral}. Our work aims to join this literature.





\subsection{Stochastic Models of Multitasking and Concurrency}\label{sl:sm}

As we discussed in the introduction, the prevailing queueing model of concurrency is the processor sharing service discipline \citep[e.g.,][]{kleinrock1967time,yashkov2007processor}. However, we have argued (and intend to formalize) that, without modifications, this approach may be better suited for its original context of computation than it is for services. As also  mentioned, the leading such modifications limit the total number of simultaneous customers and allow the service rate to depend on the state of the concurrency. These constructions are derived from messaging-based contact centers, where they offer a ``macroscopic model'' of the concurrent service system, capturing the system-level dynamics of many parallel agents that each can hold several simultaneous customer service conversations  \citep[][pg. 329]{luo2013staffing}. Naturally, given this perspective, many of the results in this stream develop  managerial insights and policies at the system-level through many-server limits. For instance, \citet{luo2013staffing} and \citet{tezcan2014routing} establish fluid limits of the Markov chain model to ascertain routing and staffing decisions, and \citet{cui2016approximations} and \citet{storm2023diffusion} broaden this scope by extending to diffusion limits of the models. \citet{legros2019scheduling} provides insight into routing policies outside of the asymptotic setting through a Markov decision process approach, and \citet{long2018customer} generalizes the multitasking model beyond its Markovian assumptions on the patience and service distributions.


 
 
 
 


Although this model and its associated stream of literature are the predominant stochastic process representations of concurrent services, we can point to two notable exceptions. First, the queueing model of case managers from \citet{Campello2017} also takes a macroscopic perspective through a Markov chain model of the full service system, but, rather than governing the impact of concurrency through an adaptation of the processor sharing discipline, each agent is essentially modeled as their own Erlang-R queue \citep[i.e.,][]{YomTov2014ErlangR}. That is, in addition to an external queue where customers wait to be matched to an agent, each agent has an internal queue where their (limited) concurrent customers have tasks served in first-come-first-serve fashion, with some probability of re-entering the internal queue (after some external delay) to have another task served. Relative to the aforementioned multi-server model, it is interesting to note that, on some level, this round-robin style processing in the internal queue is essentially what \citet{kleinrock1967time} sought to approximate or idealize through the processor sharing discipline, where a cycle of continuous rotation and infinitesimal progress among tasks is assumed for simplicity.\endnote{The re-entering customer's external delay is also notable relative to processor-sharing; \editTwo{see Section~\ref{conclusion} for further discussion}.} 

The second alternative stochastic model of concurrency comes from \citet{daw2021co}, which proposes a bivariate, marked Hawkes cluster model of the service interaction. By comparison to the preceding models, this perspective is \edit{more micro than macro}: the framework takes a single customer-agent interaction as the scope of the history-driven stochastic process. 
In some sense, this perspective is well-aligned with the principles of people-centric operations \citep{roels2021om}. 
Through applying and assessing the model on a contact center data set containing nearly 5 million messages across over 300,000 service conversations during one month's time, \citet{daw2021co} shows that modeling at this interaction-level indeed captures dynamics missed by simply representing the duration of service as some random variable\edit{, \editTwo{finding} that the times between messages in the interior of conversations exhibit the Hawkes process's history-driven behavior with statistical significance}. Moreover, the numerical experiments confirm the \edit{practical} importance of history dependence in the representation, as the Hawkes cluster model is also shown to have superior fit relative to phase-type service distributions, even when those models are granted state-dependence on \edit{the number of messages sent so far}. Some initial system-level managerial insights were drawn from this fit to data through a stylized adaptation of the routing static planning problem from \citet{tezcan2014routing}, but rigorous analysis of the individual-level model remains quite open. Such interaction-level analysis is the principal aim of the present paper, keeping in mind that the behavioral operations ideal of theoretical models is to be ``instrumental in deciding which  individual‐level  behavioral  anomalies  matter  and  whether  they  have  meaningful system‐level effects'' \citep[][pg. 358]{allon2018behavioral}. For example, by assuming that each agent's concurrency is limited up to some value, the aforementioned macroscopic stochastic models presume that something of a ``sweet spot'' exists in terms of optimally efficient multitasking. Moreover, this assumption implies that this ideal level can be found and known by management \editTwo{a priori}. 


  

\subsection{Endogenous Clusters in Hawkes Self-Exciting Stochastic Models}\label{sl:se}

Following the modeling and application to data in \citet{daw2021co}, this paper will be concerned with an endogenously driven cluster model, as derived from the original self-exciting point process proposed by \citet{hawkes1971spectra}. That seminal work introduced its namesake process through defining a stochastically time-varying intensity given by an integral over the history of the point process. \citet{hawkes1974cluster} then provided an alternate definition that connects intuitively to branching processes and cluster models. That is, \citet{hawkes1974cluster} distilled the Hawkes point process into an exogenous baseline stream (equivalent to a homogeneous Poisson process) and self-excited progeny streams.
These clusters then typify the ``self-exciting'' label, as they are an almost surely finite collection of points that are generated exclusively by the history of their own activity. Naturally, one can then define the duration of the cluster to be the time from the first point in the cluster to the last.
Phrased as a stochastic modeling objective, to understand how the throughput of the service is affected by the dimensions of interdependence and synchronicity is to understand how the Hawkes cluster duration depends on the service interaction parameters. 

Unfortunately, the cluster duration has been an elusive object. \citet[pg. 631]{moller2005perfect} remarked that this distribution is ``unknown even for the simplest examples of Hawkes processes.''  Only the mean of the duration would be needed to analyze the throughput, but even this has been out of reach \citep{chen2021perfect,graham2021regenerative}. 
Recently, \citet{daw2023conditional} made progress on this problem by establishing a novel adaptation of the random time change theorem, showing that the compensator transform of the cluster points will be conditionally uniform on a certain convex polytope when given the cluster size, and the distribution of the size is well-understood thanks to \citet{hawkes1974cluster}. This conditional uniformity then provides a deterministic recursive mapping from the vector of compensator points back to the original cluster times. In the case of the Markovian Hawkes process, the recursion collapses thanks to Cauchy's identity for the exponential function. Given the cluster size, the Markovian cluster duration can be seen to be distributionally equivalent to a sum of conditionally independent exponential random variables. 
However, this simplification requires the Hawkes cluster to be univariate with an exponential kernel, which is substantially more specific than the model we aim to study \edit{in this paper}. In particular, under such assumptions, we would not be able study the dimensions of service that we do here.
Outside of this special exponential setting, it is unclear if the recursive solution from \citet{daw2023conditional} can be as readily simplified. Explicitly \edit{obtaining} the general mean (and distribution) of the duration thus remains a challenging problem.


\section{Defining the Service Interaction Stochastic Model and Measures of (A)Symmetry}
\label{sec:model}


  


\edit{Let us now begin Part I of the paper, in which we develop and analyze the foundations of a stochastic model of service at the individual interaction level.}
\edit{We start} by defining this paper's focal representation of dynamically co-produced service as a two-sided Hawkes cluster model. 
As discussed in Section~\ref{litReview}, we seek a model of service that is micro- rather than macroscopic, meaning that it is focused on behavior at  the level of  individuals rather than at the level of the system. By comparison to simply representing the duration as some random variable, we \editTwo{claim} that this more granular framework strikes at the heart of service: service isn't formulaic or step-by-step like other classic domains of operations like manufacturing or production. The terms, contexts, and goals of the service interaction evolve as it progresses.
As the interaction unfolds, the two sides, customer and agent, each contribute to the service objective, and these then spur further contributions from both sides.

To ground this idea in a practical motivation, consider again the tax preparation example. Particularly in modern settings, this collaboration often occurs over a series of digital messages and virtual meetings through the tax preparation company's website. In the course of that correspondence, the customer may send documents and background details, and the agent may supply insights and forms. The incomplete information on each side creates a reciprocal question-and-answer cycle, where both parties may send each other questions to refine the scope and ascertain the goal, implying that new queries from one party can create additional tasks for the other. Hence, each contribution within the interaction may beget subsequent contributions, rendering the service exchange dependent on its own history. Furthermore, each contribution differs from the rest in the information it carries and the subsequent tasks it necessitates, and, thus, the number of points generated in response will likewise vary across contributions, injecting uncertainty into the process. This randomized ripple effect is the hallmark of the Hawkes point process and precisely what makes it ``self-exciting.'' To that end, the following process can be seen to be a \emph{behavioral} model, but rather than a model of strategy or rationality, it is a model of impulses, influences, and interaction.

\edit{Building from \editTwo{the application, calibration, and evaluation on real-world service data in} \citet{daw2021co}, we} specify this interaction model now in Definition~\ref{hcDef}\edit{.\endnote{\edit{\editTwo{Though} Definition~\ref{hcDef} shares similarities with \citet{daw2021co}, it also has important differences. For instance, all of the model variants fit to data in \citet{daw2021co} assume exponential response functions, which are left as general in Definition~\ref{hcDef}. Nevertheless, the most general model in \citet{daw2021co} also differs from Definition~\ref{hcDef} by treating the response functions as not only dependent on the \editTwo{responding} party, but also on the party which made the original contribution. 
Although analysis of \editTwo{that} fully bivariate model is challenging, in Appendix~\ref{numerSec}, we conduct numerical experiments to show that our insights extend to the broader, multi-directional model. 
Hence, Definition~\ref{hcDef} might be viewed as both a partial simplification and partial generalization of the \citet{daw2021co} model, where both directions of changes are meant to support the analytical pursuits of this paper.}\label{noteModelDiff}}} 


\begin{definition}[Two-Sided Hawkes Cluster Model of Service]\label{hcDef}
Assuming that the service is initiated by a time $\tau_0 = 0$ contribution, let $N_t$ with $N_0 = 1$ be the point process for the {number of contributions} up to time $t \geq 0$, where this point process is driven by a corresponding stochastic {service contribution rate} intensity $\mu_t$, defined 
\begin{align}
\mu_{t}
&=
\int_0^t
\left(
g_\mathsf{1}(t-s)
+
\eta \cdot g_\mathsf{2}\left(\eta \cdot (t-s) \right)
\right)
\mathrm{d}N_s
=
\sum_{i=0}^{N_{t} - 1} 
g_\mathsf{1}(t-\tau_i)
+
\eta \cdot
g_\mathsf{2}\left( \eta \cdot (t-\tau_i) \right)
,
\label{intensityDef}
\end{align}
\editTwo{and} where $\tau_i$ is the epoch for the $i$th contribution in the service interaction, and $\eta \in \mathbb{R}_+$. That is,
\begin{align}
\PP{N_{t+\delta} - N_t = n \mid \mathcal{F}_t} 
&
=
\begin{cases}
\mu_t \delta + o(\delta) & n = 1,\\
1 - \mu_t \delta + o(\delta) & n = 0,\\
o(\delta) & n > 1,
\end{cases}
\label{mainHawkesDef1}
\end{align}
where $\mathcal{F}_t$ is the natural filtration of the stochastic process.
Here, $g_\mathsf{1}: \mathbb{R}_+ \to \mathbb{R}_+$ and $g_\mathsf{2}: \mathbb{R}_+ \to \mathbb{R}_+$ are the Side $\mathsf{1}$ and Side $\mathsf{2}$ response functions, respectively, so that $g_\mathsf{1}(0)$ ($\eta g_\mathsf{2}(0)$) is the instantaneous impact of a given contribution on Side $\mathsf{1}$'s (Side $\mathsf{2}$'s) individual contribution rate, and, generally, $g_\mathsf{1}(u)$ ($\eta g_\mathsf{2}(\eta u)$) is its impact $u \geq 0$ time after its occurrence. Finally, we will suppose that the service ends at the natural duration of the cluster, $\tau = \tau_{N-1} = \inf\{t \geq 0\mid N_t = N\}$, with the size of the cluster, $N = \lim_{t \to \infty} N_t$, being the total number of contributions in the service (including the initial). Given this structure, let the points $\tau_i$ for $i \in \{0, 1, \dots, N-1\}$ be the \emph{Hawkes cluster model} of the contributions comprising the service interaction. \hfill \Halmos  
\end{definition}

In addition to the terms given in Definition~\ref{hcDef}, we may also use superscripts to refer to the same quantities specifically on one side of the service. For example, we will let $N^\mathsf{1}$ and $N^\mathsf{2}$ be the total number of Side $\mathsf{1}$ and Side $\mathsf{2}$ contributions, respectively, where the size of the service cluster is such that $N = N^\mathsf{1} + N^\mathsf{2} + 1$.\endnote{Without loss of generality, we assume that the service-initializing contribution at time 0 does not belong to either side.} For simplicity, we will also suppose that $g_\mathsf{1}(\cdot)$ and $g_\mathsf{2}(\cdot)$ are continuous functions, but this could be relaxed, if needed. Without loss of generality, we will let Side $\mathsf{1}$ represent the customer, and Side $\mathsf{2}$, the agent. \editTwo{Though it may not yet be obvious, Definition~\ref{hcDef} establishes $\eta$ as a hyperparameter for comparing the relative paces of these two sides across different model instances and different system-level dependencies. Hence, we refer to $\eta$ as the \emph{synchronicity parameter}, and we will soon formalize this concept.}


\begin{remark}[\editTwo{A Parsimonious Service Model Interpretation}]\label{hoRemark}
\editTwo{The intensity-oriented approach of Definition~\ref{hcDef} is consistent with the original establishment of self-exciting processes in \citet{hawkes1971spectra}, but the subsequent literature may offer a more intuitive construction for this service context. By \citet{hawkes1974cluster}, a Hawkes cluster model can be viewed from a branching-process lens: Definition~\ref{hcDef} can be equivalently stated as an endogenous collection of independent but inhomogeneous Poisson processes, in which each contribution point receives direct response contributions according to its own Poisson stream with rate $g_\mathsf{1}(u) + \eta g_\mathsf{2}(\eta u)$ for $u \geq 0$ time since that focal contribution occurred. By thinning these Poisson streams, we can further interpret this model as each contribution having a $g_\mathsf{1}(u)$ direct response process from the customer and a $\eta g_\mathsf{2}(\eta u)$ direct response process from the agent. Together, these interpretations highlight why Definition~\ref{hcDef} is a \emph{service interaction} model: the Hawkes cluster captures two sides interactively contributing to the service in response to their other contributions so far.}
\end{remark}

\subsection{\editTwo{Formalizing Performance Metrics and Key Parameters of the Interaction Model}}

While the focus of this section will be restricted to the interaction level, our aim is to eventually connect this individual-level behavior to system-level dynamics. 
\edit{\editTwo{As discussed in Section~\ref{questionSec}}, we are specifically interested in the system-level effects of concurrency or multi-tasking in service. We will let $\mathcal{K}$ denote the agent's concurrency level, meaning the number of customers that the agent serves simultaneously. Most commonly, $\mathcal{K}$ can be viewed to be a positive integer, but let us  generalize this to allow any $\mathcal{K} \in \mathbb{R}_+$, where non-integer concurrency can be thought of as customer-sharing by agents \citep[such as in the patient-to-nurse ratio discussed in][]{vericourt2011nurse,cho2019behavior}. 
}

\edit{As a foremost objective of this paper}, we seek to understand the \emph{throughput} or the \edit{agent's service rate}, 
\begin{align}
\frac{\mathcal{K}}{\E{\tau\left(\mathcal{K}\right)}}
,
\label{firstThroughput}
\end{align}
where the notation $\tau(\mathcal{K})$ reflects that fact that the service duration will depend on the concurrency level. \edit{In the case that $\mathcal{K} \in \mathbb{Z}_+$, one can further contextualize the micro-level service interaction model as being one of $\mathcal{K}$ parallel services held by the agent at the system-level, where the $\mathcal{K}$ different sequences of contributions for each customer-agent service pair can be thought of as $\mathcal{K}$ parallel Hawkes cluster models, which are conditionally independent given the concurrency. 
Accordingly, we will use the phrase ``agent-level'' to refer to measurements which span all service interactions held concurrently by one agent, such as the agent's throughput in \eqref{firstThroughput}. The agent-level perspective then becomes a bridge to the ``system-level'' perspective. If there is a single agent in the system, then these notions are the same; otherwise, system-level measurements are those that span all the system's agents. Following the micro-level modeling framework of this paper, we will primarily leave agent-to-system relations to be the subject of macro-level analyses, and we will instead focus on the interaction-to-agent bridge. Hence, most often, our discussions of system-level performance will be rooted in the context obtained at the agent-level.}

\edit{Eventually, in Part II of the paper, we will introduce an assumption that the synchronicity of the interaction (which we will soon show is captured through the parameter $\eta$) is a function of the concurrency, $\mathcal{K}$.} Before contextualizing this interaction model in the broader operation, though, we will first develop insight at the customer-agent \edit{interaction} level, with specific attention on the dimensions of interdependence and synchronicity within \edit{a single} service interaction. Hence, we set aside the dependence on $\mathcal{K}$ for now, and we will return to it in Section~\ref{nonMonSec} \edit{(and make it precise in Assumption~\ref{conA}, specifically)}.

\begin{figure}[ht]
\centering
\includegraphics[width=\textwidth]{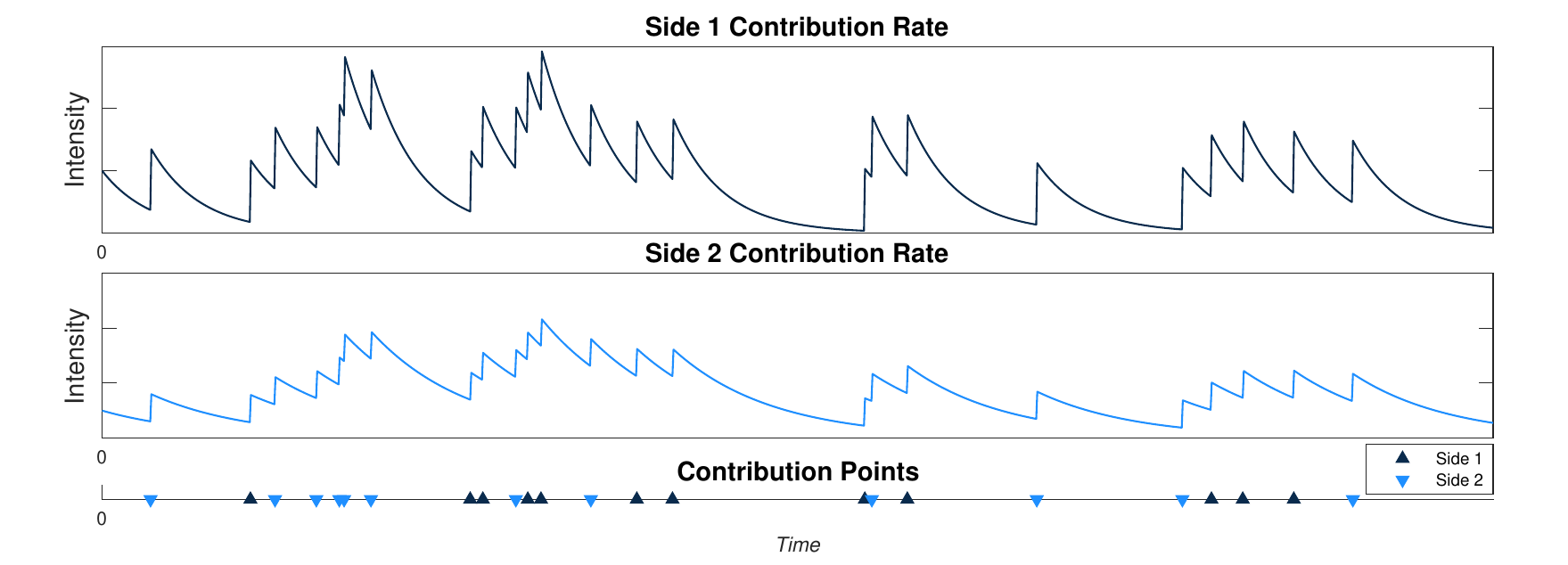}
\caption{A sample path of contribution points and the
corresponding contribution rates on the two sides.}\label{spFig}
\end{figure}

A visualization of one sample path of the two-sided Hawkes service model is given in Figure~\ref{spFig}. Here, the contribution rate given in Equation~\eqref{intensityDef} is shown in two sub-intensities, one for Side $\mathsf{1}$ and one for Side $\mathsf{2}$. As suggested by the motivating examples, we can see in Figure~\ref{spFig} that, upon each new contribution point within the service, there is a reciprocal increase in the rate at which each side makes new contributions. However, because the model differentiates the behavior on the two sides of service, each side has its own impact. \editTwo{We} can see that, in this example, the instantaneous impact of each contribution (the size of the jumps in the intensity) is larger on Side $\mathsf{1}$ than on Side $\mathsf{2}$, but the rate of decay is also faster on Side $\mathsf{1}$. Hence, in this \editTwo{case}, Side $\mathsf{1}$ is more likely to react quickly to a new contribution, but the effect of each contribution lingers longer on Side $\mathsf{2}$.

We can also observe two key performance metrics of the service interaction within Figure~\ref{spFig}. First, the duration of the service, $\tau$, is the time from the first contribution point until the last, so it is marked by the right-most triangle on the time axis. In stochastic modeling terminology, this is the duration of the Hawkes process cluster. Then, the total number of triangles along the horizontal axis in Figure~\ref{spFig} serves as a measurement of the total taskload within the service exchange. This cluster size, $N$, counts the total number of contributions in the service \edit{(including the initial contribution at time 0)}.


The service process depicted in Definition \ref{hcDef} \edit{also encapsulates two important dimensions of the customer-agent interaction.} 
The first of these is readily observable: $\eta$, which appears in the Side $\mathsf{2}$ response functions, governs the synchronicity between the customer and agent, meaning the comparison of the relative time scales on each side of the service. Then, the second spectrum is actually contained in a pair of parameters just beyond the surface of the definition. Letting $\rho_\mathsf{1}$ and $\rho_\mathsf{2}$ be defined $\rho_\mathsf{1} = \int_0^{\edit{\infty}} g_\mathsf{1}(t) \mathrm{d}t$ and $\rho_\mathsf{2} = \int_0^{\edit{\infty}} g_\mathsf{2}(t) \mathrm{d}t$, we will see that this pair governs the interdependence of the customer and agent, meaning the distribution of production between the two sides.\endnote{\editTwo{Note that $\int_0^\infty \eta g_\mathsf{2}(\eta t) \mathrm{d}t = \rho_\mathsf{2}$ regardless of the value of $\eta$. This confirms our intuition that $\rho_\mathsf{1}$ and $\rho_\mathsf{2}$ are time-agnostic metrics.}} 
Hence, we will focus on $\eta$, $\rho_\mathsf{1}$, and $\rho_\mathsf{2}$ as the parameters capturing the (a)symmetry of the service, as $\eta$ answers ``who sets the pace?'', and $\rho_\mathsf{1}$ and $\rho_\mathsf{2}$ answer ``who does the work?''

Formally, we associate $\eta$, $\rho_\mathsf{1}$, and $\rho_\mathsf{2}$ with the service spectra through the following definitions:
\begin{definition}[Synchronicity]\label{defSync}
    Let $\eta \in \mathbb{R}_+$ be the \emph{synchronicity parameter} of the Hawkes service model (Def.~\ref{hcDef}): extreme $\eta$ ($\eta \rightarrow 0$ or $\eta\rightarrow \infty$) connotes \emph{asynchrony} in the service interaction, and moderate $\eta$ ($\eta \approx 1$) connotes \emph{synchrony}.
    \hfill \Halmos 
\end{definition}

\begin{definition}[Interdependence]\label{defInter}
    Let the pair $(\rho_\mathsf{1}, \rho_\mathsf{2}) \in [0,1]^2$ be the \emph{interdependence parameters} of the Hawkes service model (Def.~\ref{hcDef}): disparate $\rho_\mathsf{1}$ and $\rho_\mathsf{2}$ ($\rho_\mathsf{1} \gg \rho_\mathsf{2}$ or $\rho_\mathsf{1} \ll \rho_\mathsf{2}$) implies \emph{self-production}, and comparable $\rho_\mathsf{1}$ and $\rho_\mathsf{2}$ ($\rho_\mathsf{1} \approx \rho_\mathsf{2}$) implies \emph{co-production}. 
    \hfill \Halmos 
\end{definition}


To motivate these claims, let us turn to what might be the most widely used tool of analysis for Hawkes processes, the \citet{hawkes1974cluster} cluster decomposition. 
\editTwo{First, through the perspective granted by Remark~\ref{hoRemark},} we can \editTwo{immediately} see that the mean number of direct Side $\mathsf{1}$ responses per contribution is $\rho_\mathsf{1}$, and likewise on Side $\mathsf{2}$, $\rho_\mathsf{2}$. 
The larger (smaller) $\rho_s$ is, the more (fewer) \emph{total} direct responses we expect from Side $s \in \{\mathsf{1}, \mathsf{2}\}$ for any given contribution.

\editTwo{However}, the mean total \editTwo{response} count per contribution only tells us so much about the service interaction; in particular the $\rho_\mathsf{1}$ and $\rho_\mathsf{2}$ service parameters do not infer any insights about the pace. For this, we turn to $\eta$. 
\edit{One can think of $\eta$ as modulating the agent's time scale, akin to a change of variable for $t$.}
Given that there is a Side $\mathsf{1}$ response to a given initial point, we can further invoke \editTwo{the Remark~\ref{hoRemark} lens} to see that the mean time at which it occurs is $\int_0^\infty t g_\mathsf{1}(t) \mathrm{d}t / \rho_\mathsf{1}$; for Side $\mathsf{2}$, this is $\int_0^\infty t \eta g_\mathsf{2}(\eta t) \mathrm{d}t / \rho_\mathsf{2} = \int_0^\infty \edit{t} g_\mathsf{2}(\edit{t}) \mathrm{d}\edit{t} / (\eta \rho_\mathsf{2})$. For each $s \in \{\mathsf{1}, \mathsf{2}\}$, we will refer to the respective quantity as the expected Side $s$ response offset time.
By noticing that the mean Side $\mathsf{2}$ response time is proportional to $1/\eta$, we can see how this parameter can imply synchrony or asynchrony between the two sides. 
If $\eta \approx 1$, then \edit{each side processes the service history} at approximately the same pace. On the other hand, if $\eta \gg 1$ then the agent processes information faster than the customer does, and if $\eta \ll 1$ then the agent operates more slowly than the customer does.





\begin{figure}[ht]
\centering
\includegraphics[width=\textwidth]{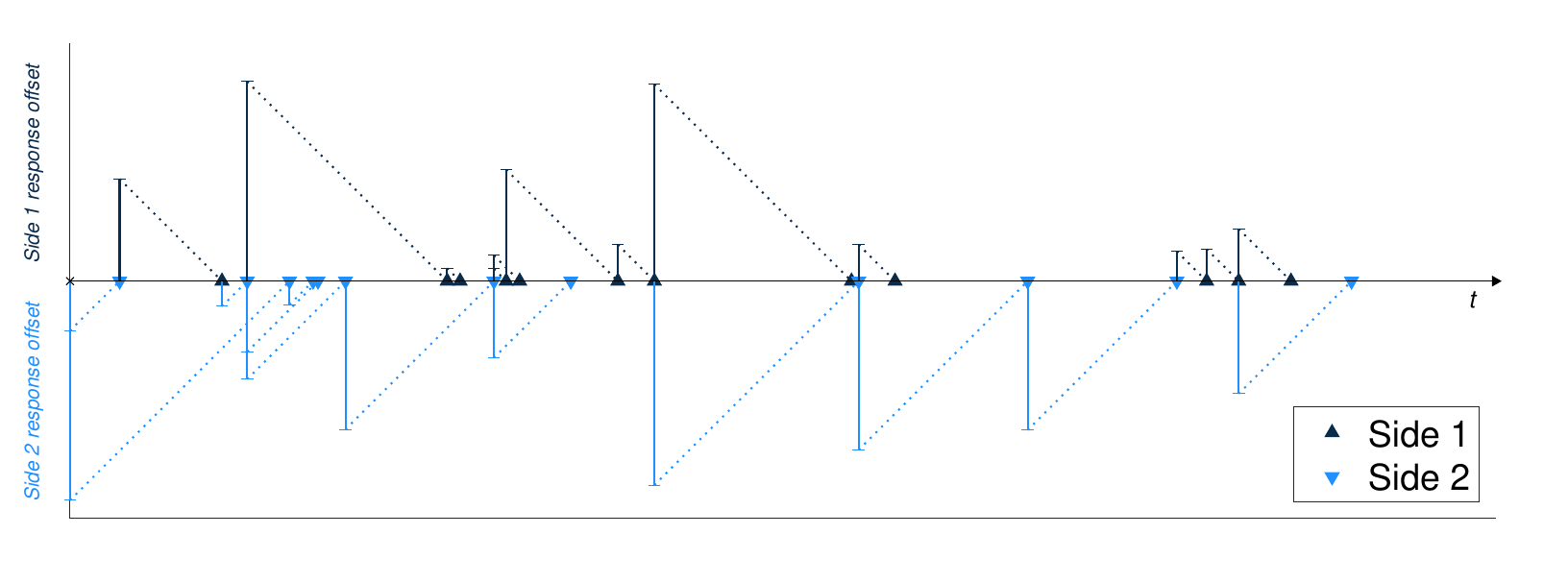}
\caption{A ``telephone wire'' diagram of the contribution points within the model of the service interaction, which shows the response structure and the associated offsets. 
The vertical lines show the delay between prompt and response(s), and the diagonal lines trace each point to the prior contribution to which it responds.}\label{telephoneFig}
\end{figure}

Drawing inspiration from \editTwo{the \citet{hawkes1974cluster} decomposition} and from the figures in \citet{eick1993physics}, let us introduce an alternate visualization of the service interaction. In Figure~\ref{telephoneFig}, we plot the same sample path as in Figure~\ref{spFig}, now in a layout that we will refer to as a ``telephone wire'' diagram. Instead of displaying the contribution points along with the associated jumps in contribution rate intensities as in Figure~\ref{spFig}, Figure~\ref{telephoneFig} shows the response structure within the interaction. That is, each contribution point is traced to the prior point to which it responds, and thus each response offset time can be gleaned both horizontally along the time axis and vertically along the response offset axis.
The ``telephone poles,'' meaning the vertical lines, show the time elapsed from the originating point and \edit{the} associated response, and the diagonal ``wires'' map each vertical offset height to the corresponding response.
Given the randomness in the inhomogeneous Poisson response streams, the number of poles varies across the points; some contributions beget multiple direct replies, but others receive none at all. For example, counting left-to-right, the third contribution triangle receives three direct responses: the sixth, seventh, and eighth points.
In the case that a point receives multiple responses from the same side, the poles stand single-file, so that the difference in the heights of, say, the first and second poles is the difference in the respective response offsets, rather than the length of the second response offset itself.

This telephone wire display also lets us quickly discern the degree of synchronicity and interdependence within the realized service interaction. First, we can see that the number of tasks is evenly split among the two sides. Not counting the initial point for either side, Figure~\ref{spFig} shows a co-productive arrangement, with both parties contributing 12 points each. Then, like how Figure~\ref{spFig} shows faster decay on Side $\mathsf{1}$, we can see that the response offset triangles are, generally, slightly larger on Side $\mathsf{2}$ than on Side $\mathsf{1}$. Still, the offsets are not overly different, so we might describe this interaction as exhibiting at most mild asynchrony. Predominantly, it shows a synchronous service exchange.

To finalize the scope of the service model, let us establish a set of assumptions which will ensure that the interaction actually ends, eventually.  

\begin{assumption}[Stability \edit{in Service}]\label{stabAssum}
Let $\rho = \rho_\mathsf{1} + \rho_\mathsf{2} < 1$, and let $\int_0^\infty t g_s(t) \mathrm{d}t < \infty$ for each $s \in \{\mathsf{1},\mathsf{2}\}$. 
\end{assumption}




\begin{proposition}\label{stabProp}
If and only if Assumption~\ref{stabAssum} holds, the service duration will be finite almost surely and have finite mean: $\PP{\tau < \infty} = 1$ with $\E{\tau} < \infty$ for every $\eta \in \mathbb{R}_+$.
\end{proposition}


Having a finite mean duration is centrally important for the focus of this work, as the principal performance metric we study is the throughput of a concurrent service system. Hence, Proposition~\ref{stabProp} provides justification for us to study $\tau$ \editTwo{at the individual-level and the throughput given by $\tau$ at the system-level}. 

\subsection{Formalizing the Shape Characterizations of Performance}

To be able to properly describe the shape of the throughput or any other performance metric, let us first classify certain structures which connect our analysis to the observations of the empirical literature. \editTwo{To} close this section, \edit{we now} define a pair of contrasting function classes. First, in Definition~\ref{uDef}, we provide analytic analogs for two of the conceptual shapes of interest, \editTwo{U-shaped and inverted-U-shaped}.


\begin{definition}[Asymptotically Non-Monotonic {[$\mathcal{N}$]}]\label{uDef}
{ For $\mathcal{D} \subseteq \mathbb{R}$ convex, let a \editTwo{positive} function $f:\mathcal{D} \to \mathbb{R}_+$ be called \textit{asymptotically non-monotonic with $\U$-shape} ($f \in \mathcal{N}_\U$) {[\textit{asymptotically non-monotonic with $\IU$-shape} ($f \in \mathcal{N}_\IU$)]} if $\lim_{x\to \inf(\mathcal{D})}\edit{f(x)}=\lim_{x\to\sup(\mathcal{D})} f(x) = \infty$ {[$0$]}.} 
\end{definition}

Naturally, the function class $\mathcal{N}_\U$ is a formalization of a U-shape, whereas $\mathcal{N}_\IU$ is likewise a formalization of an inverted-U-shape. In some senses, these definitions are weaker or more general of conditions than concavity or convexity with the same asymptotes: $f$ need not be convex (concave) to be in $\mathcal{N}_\U$ ($\mathcal{N}_\IU$). Furthermore, $f$ belonging to either $\mathcal{N}_\U$ or $\mathcal{N}_\IU$ need not be strictly unimodal. However, it is guaranteed that, if a function belongs to either of these classes, then it attains a global optimum on the interior of its domain. Furthermore, for arguments approaching the extremes of the domain, there will always be a more extreme outcome. In this sense, \editTwo{these formalized} $\U$ or $\IU$ shapes might be classified as \emph{strong}, and there could be a corresponding weaker classification in the spirit of Definition~\ref{uDef} but with finite asymptotes. Our analysis will focus on the strong $\U$ and $\IU$ shapes in Definition~\ref{uDef}, but we will also occasionally comment on where it may be possible to relax assumptions and still establish existence of these shapes in a weaker form. 




To contrast with $\mathcal{N}_\U$ and $\mathcal{N}_\IU$, let us also introduce a partial converse to the \editTwo{asymptotic} non-monotonicity of those classes. In Definition~\ref{nonuDef}, we define a class of functions that are similarly strong at the extremes of the domain, but the limiting values do not match.

\begin{definition}[Asymptotically Monotonic {[$\mathcal{M}$]}]\label{nonuDef}
{ For $\mathcal{D} \subseteq \mathbb{R}$ convex, let a \editTwo{positive} function $f:\mathcal{D} \to \mathbb{R}_+$ be called \textit{asymptotically monotonic} ($f \in \mathcal{M}$) if $\lim_{x\to \inf(\mathcal{D})} \edit{f(x)} = 0$ and $\lim_{x\to\sup(\mathcal{D})} f(x) = \infty$, or if $\lim_{x\to \inf(\mathcal{D})} \edit{f(x)} = \infty$ and $\lim_{x\to\sup(\mathcal{D})} f(x) = 0$.} \hfill \Halmos 
\end{definition}

Much like functions in $\mathcal{N}_\U$ or $\mathcal{N}_\IU$ need not be convex or concave, the functions in $\mathcal{M}$ need not be strictly monotonic. However, for any given positive value of a function belonging to this class, there exists an argument at which point the function achieves a larger value than the given one, and the same is true for smaller values. Hence, whether the goal is to maximize or minimize, if the objective function belongs to $\mathcal{M}$, the performance within the interior can always be improved by a decision closer to the boundaries.\endnote{\edit{Nevertheless, it is important to not lose sight of the fact that these characterizations in Definitions~\ref{uDef} and~\ref{nonuDef} are broad and most prominently oriented around the asymptotic extremes: one can notice that an ``M'' shape would actually belong to $\mathcal{N}_\IU$, a ``W'' shape would belong to $\mathcal{N}_\U$, and both an ``N'' or \editTwo{an} ``inverted N'' would belong to $\mathcal{M}$. We show how our modeling framework can be used to analyze more specific notions of non-monotonicity in Appendix~\ref{hSec}.\label{mwnNote}}}


\section{Analyzing and Decomposing the Hawkes Cluster Stochastic Model}\label{clusterSec}




\edit{In this second section within Part I of the paper}, we establish the foundational analysis of the core stochastic models in this paper, leading to methodology \editTwo{for answering} the service asymmetry questions \editTwo{we ask} throughout the rest of the paper. Specifically, Section~\ref{decompSec} deepens the connection between Hawkes processes and parking functions through a novel alternate definition of the cluster model, and Section~\ref{limitSec} employs this definition to illuminate the limiting behavior of the model at the extremes of its parameters.






\subsection{A \editTwo{Novel} Parking Function Decomposition for the Cluster Model of Service \edit{Interaction}}\label{decompSec}

In the course of obtaining \edit{a} transformation which yields conditional uniformity, \citet{daw2023conditional} finds a surprising connection between the time-changed cluster points and uniformly random parking functions. For the sake of self-containment of the paper, let us now define those discrete combinatorial objects. Centrally related to many discrete and algebraic structures \citep[e.g., see the many connections in][]{yan2015parking}, (deterministic) parking functions can be defined through a simple condition:

\begin{definition}[Parking Function]\label{pfDef}
{For $k \in \mathbb{Z}_+$, $\pi \in \mathbb{Z}_+^k$ is a \emph{parking function} of length $k$ if and only if it is such that, when sorted $\pi_{(1)} \leq \dots \leq \pi_{(k)}$, $\pi_{(i)} \leq i$ for each $i \in \{1, \dots, k\}$.
} \hfill \Halmos 
\end{definition}

Then, for $\mathsf{PF}_k$ as the set of all parking functions of length $k \in \mathbb{Z}_+$, a uniformly random parking function of length $k$ is simply some $\pi$ drawn from $\mathsf{PF}_k$ with equal weightings for each possible $\pi$. There are $(k+1)^{k-1}$ parking functions of length $k$, and sampling from these with uniform probability $(k+1)^{-k+1}$ can be easily done via an elegant rotation argument \citep[see, e.g.,][for modern probabilistic explorations of parking functions]{diaconis2017probabilizing,kenyon2023parking}. The idea is thus \citep[phrased here in the style of][\edit{but not taken in direct quotation}]{diaconis2017probabilizing}: 
\begin{myquote}
Draw $k$ i.i.d.~random variables that are uniformly random on $1, \dots, k+1$, say $\tilde \pi_i$ for $i \in \{1, \dots, k\}$. Then, ``park'' this sequence one-by-one on \edit{a ``circle'' according to the $\mathbb{Z}/(k+1)\mathbb{Z}$ factor group}, assigning space $\tilde \pi_i$ to car $i$ if cars 1 through $i-1$ have not already occupied it and otherwise placing car $i$ in the next available space after $\tilde \pi_i$, restarting at space 1 if needed. By the pigeonhole principle, there will be precisely one space left empty. If this empty space is $k+1$, then $\tilde \pi$ is a parking function, and, if not, rotate the circle (by updating $\tilde \pi_i $ to $\tilde \pi_i + 1 \mod k+1$ for each $i$) until $k+1$ is no longer taken. 
\end{myquote}
\noindent
This rotational parking \edit{concept} is often referred to as ``Pollak's circle argument,'' as recorded by Pollak's contemporaries \citep[e.g.,][]{riordan1969ballots}.

In the Hawkes cluster setting, \citet{daw2023conditional} finds that parking functions can be used to decompose the convex polytope on which the compensator transformed points (meaning, the integral of the intensity evaluated at the point process epochs) are themselves conditionally uniformly distributed. Specifically, a randomly sampled parking function will correspond to a sub-region of the polytope, and, in this sub-polytope, the Hawkes compensator points can be reduced to an ordering of mutually independent standard uniform random variables with affine shifts.
Through this tractable partition, the compensator points become more interpretable and amenable to analysis. However, such an analysis still suffers from \editTwo{opacity:} it can be unclear how to interpret insights from the compensator space in the original time space of the cluster, particularly so outside of the Markovian setting.


To ameliorate this, we now seek to uncover a connection from parking functions to Hawkes processes exclusively in the time space, i.e.,~without random time change arguments or compensator transforms. Instead of invoking such transformations, our approach will be  first to leverage the relationship between (close relatives of) these combinatorial objects and branching processes, and then to extend this to Hawkes clusters via \citet{hawkes1974cluster}, \editTwo{\`a la Remark~\ref{hoRemark}}. 
The inspiration for this approach begins with \citet{devroye2012simulating}, which shows that a Pollak-type rotation argument produces Cayley trees that are uniformly random when conditioned on their total size (number of nodes). When not conditioned on the total size, \citet{devroye2012simulating} motivates these trees as the lineage structures within a branching process's progeny, where the offspring distribution is Poisson. This already hints at a connection to \editTwo{the cluster model as viewed through Remark~\ref{hoRemark}:} the relationship between Cayley trees and parking functions is well-known \citep[e.g.,][]{yan2015parking}. Hence, from these Cayley trees we can find parking functions in the same space as the cluster itself.

The commonality among \edit{parking functions, Cayley trees, and branching processes} leads us to define a cluster model from the threads of all three. Bringing in the notion of the two sides from Definition~\ref{hcDef}, we propose the following parking function cluster now in Definition~\ref{pfcDef}.




\begin{definition}[Parking Function Two-Sided Cluster]\label{pfcDef}
{ For kernels $\gamma_1: \mathbb{R}_+ \to \mathbb{R}_+$ and $\gamma_2: \mathbb{R}_+ \to \mathbb{R}_+$ with $\gamma_1(x) = g_\mathsf{1}(x)$ and $\gamma_2(x) = \eta g_\mathsf{2}(\eta x)$ for each $x \geq 0$,
construct the two-sided cluster of points $0 = \tau_0 < \tau_1 < \dots < \tau_{N-1}$ as follows:
\begin{enumerate}[i)]
\item Let $N \sim \mathsf{Borel}(\rho)$ be the size of the cluster, where $\rho = \rho_\mathsf{1} + \rho_\mathsf{2}$.
\item For $\pi \in \mathsf{PF}_{N-1}$ as a uniformly random parking function of length $N-1$, let $\kappa_i(\pi) = \left|\{j \mid \pi_j = i\}\right|$ \edit{for each $i \in \{1, \dots, N-1\}$} be the number of points descending directly from point $i-1$.
\item Let $s_i \stackrel{\mathsf{iid}}{\sim} \mathsf{Bern}\left(\rho_\mathsf{1}/\rho\right)$ for each $i \in \{1, \dots N-1\}$ be indicator variables recording whether point $i$ was generated on Side $\mathsf{1}$ ($s_i = 1$) or Side  $\mathsf{2}$ ($s_i = 0$).
\item For each $i \in \{1, \dots, N-1\}$, let $\Delta_i$ be independently generated according to the density $\gamma_{1}(\cdot)/\rho_{\mathsf{1}}$ if $s_i = 1$ and $\gamma_{2}(\cdot)/\rho_{\mathsf{2}}$ if $s_i = 0$, and set $T_i = T_{\pi_{(i)}-1} + \Delta_i$, with $T_0 = 0$.
\end{enumerate}
Finally, let $\tau_i = T_{(i)}$ for each $i \in \{0, 1, \dots, N-1\}$ be the \emph{parking function cluster}.} \hfill \Halmos 
\end{definition}

Inherently, this cluster definition is in the lineage of the perspective given by \citet{hawkes1974cluster}, as its construction centers around the responses or offspring from each point. However, unlike the time-forward branching structure used by \citet{hawkes1974cluster}, in which the definition iterates forward from the initial time point, Definition~\ref{pfcDef} uses \emph{end-state conditioning} like \citet{devroye2012simulating} and \citet{daw2023conditional}. That is, Definition~\ref{pfcDef} works backwards from the size of the cluster, meaning the end-of-time total count of epochs, and then generates these times conditionally. The uniformly random parking function (with length determined by the size) becomes the key intermediate step.  This vector refines the size information to further include the tree of responses, meaning the number of points that are in response to each point before. This may bring to mind the telephone wire diagram in Figure~\ref{telephoneFig}, where the parking function contains the number of telephone poles sprouting from each point. Then, given the number of responses for a given point, the side of each response reduces to an independent coin flip. After all the time-agnostic structure (size, response pattern, and sides) is established, the times follow as the final step of the definition, completing the backwards approach to forming a cluster.







As the connections among parking functions, branching trees, and Hawkes processes might suggest, this end-state conditioned  construction can be seen to be equivalent to the time-forward Hawkes derivation.

\begin{theorem}\label{equivThm}
The Hawkes cluster (Definition~\ref{hcDef}) and the parking function cluster (Definition~\ref{pfcDef}) are distributionally equivalent.
\end{theorem}

In many ways, Definition~\ref{pfcDef} serves as an addendum to the \citet{hawkes1974cluster} cluster definition \editTwo{invoked in Remark~\ref{hoRemark}}, where the conditioning approach leverages the branching process interpretation of its hallmark non-stationary Poisson response streams. 
The proof of Theorem~\ref{equivThm} essentially follows from this exact idea, empowered by Poisson thinning. Using the independence of those inhomogeneous streams and the total number of offspring for each point through the parking function, obtaining the final times (before being placed in true chronological order) reduces to finding the offsets through inverse transform sampling with the non-stationary rates.

\begin{remark}[\editTwo{Hawkes Cluster Models Yield Service Trees, Forward and Backward}]
\editTwo{Whether viewed from the backward-looking Definition~\ref{pfcDef} or the forward-looking Definition~\ref{hcDef} (or Remark~\ref{hoRemark}), a key underlying idea to this micro-level model of service is that each customer-agent interaction contains a tree of contributions. Definition~\ref{hcDef} generates these ``service trees'' organically from roots to leaves, whereas Definition~\ref{pfcDef} samples them in full. Because  Definition~\ref{hcDef} offers interpretation whereas Definition~\ref{pfcDef} enables analysis, the benefit of Theorem~\ref{equivThm} is precisely that these oppositely angled perspectives are equivalent.}
\end{remark}





By uncovering the parking function in the Hawkes cluster without compensator transformations, Definition~\ref{pfcDef} and Theorem~\ref{equivThm} both deepen and explicate \citet{daw2023conditional}'s claim that \emph{parking functions are the hidden spines of Hawkes processes}. 
\edit{Notice that, relative to the compensator-based connection found in \citet{daw2023conditional}, Theorem~\ref{equivThm} provides a decomposition of the Hawkes cluster in the original temporal context of the model.}
Certainly more interpretable and arguably more elegant, the manner in which the probabilistic combinatorial \edit{connection} is obtained here will empower the remaining pursuits in this work. That is, beyond any conceptual intrigue to the parking function construction, the value of Definition~\ref{pfcDef}  for service modeling is that the chronological details are ignored until the very last step. Recalling the well-documented challenges of analyzing $\tau$, there is value in isolating the dependence on time, as we will show\edit{, and this perspective supports our performance analysis of the service interaction model}.
Given Theorem~\ref{equivThm}, we will henceforth work in the probability space aligned with Definition~\ref{pfcDef}, and, in particular, conditioning on the latent parking function will facilitate direct proofs of the stochastic modeling results of interest. 

\begin{remark}[\edit{Temporal and Compensator-Based Parking Functions Need Not Agree}]\label{agreeRemark}
\edit{Although \citet{daw2023conditional} identified the first connection between Hawkes processes and parking functions (and the present paper is certainly indebted to that novel recognition), the connection therein actually is not the same as the one now obtained in Theorem~\ref{equivThm}. \editTwo{The} two notions are meaningfully distinct not only in interpretation, but also in realization at the sample path level. Furthermore, we can observe that these two distinct connections offer complementary benefits through their respective strengths and weaknesses. In Appendix~\ref{agreeSec}, we prove that, for any Hawkes cluster sample path, there is a strictly positive probability that the temporal parking function of Theorem~\ref{equivThm} will differ in value (irrespective of order) from the compensator-based parking function found in \citet{daw2023conditional}, meaning that the two concepts meaningfully disagree. Additionally, we numerically demonstrate this rarity of agreement through simulation, and we discuss how the two objects each hold distinct practical and analytical value. \editTwo{(See Remark~\ref{pfUseRemark} at the close of Appendix~\ref{agreeSec} for discussion of the contrasting uses.)}} 
\end{remark}

\subsection{Analyzing the Spectra of (A)Symmetry Within the Two-Sided Cluster}\label{limitSec}




Let us now leverage the perspective from Definition~\ref{pfcDef} to analyze how the cluster model depends on the \edit{service spectra} parameters: $\rho_\mathsf{1}$, $\rho_\mathsf{2}$, and $\eta$. We begin with interdependence, characterizing the conditional distribution of \editTwo{contributions} among the two sides, given the total number of \editTwo{contributions overall}.


\begin{proposition}\label{binSidesProp}
Let $N=n+1$ for $n \in \mathbb{N}$. Then, $N^s \sim \mathsf{Bin}\left(n,  \rho_s \slash \rho \right)$ for each $s \in \{\mathsf{1}, \mathsf{2}\}$.
\end{proposition}

The conditional binomial distribution of contributions across the sides follows intuitively from steps (i) and (iii) from Definition~\ref{pfcDef}. Proposition~\ref{binSidesProp} now formally justifies the motivations for Definition~\ref{defInter}'s assertion that $\rho_\mathsf{1}$ and $\rho_\mathsf{2}$ control the interdependence spectrum. In fact, we can now clearly understand \emph{co-production} as existing in the middle of the range from \emph{agent self-production} to \emph{customer self-production}. For example, consider a fixed $\rho = \rho_\mathsf{1} + \rho_\mathsf{2}$, and let us evaluate the spectrum given by $\rho_\mathsf{1} \in (0, \rho)$. If $\rho_\mathsf{1}$ is near 0, then Proposition~\ref{binSidesProp} shows that it is quite likely that the agent (Side $\mathsf{2}$) makes most, if not all, the contributions in the service, which is a \edit{``}service factory\edit{''} or agent self-production setting. On the other hand, if $\rho_\mathsf{1}$ is near $\rho$, then the customer (Side $\mathsf{1}$) is likely to provide the lion's share of the contributions, creating self-service  or customer self-production. Between these extremes, if $\rho_\mathsf{1} \approx \rho / 2$ (or, equivalently, $\rho_\mathsf{1} \approx \rho_\mathsf{2}$), then the contributions are evenly divided among the two agents, and this symmetry between the two sides yields service co-production. Naturally, these qualitative classifications can be further refined (e.g., ``moderate agent self-production'') as needed to describe the service setting at hand, and this spectrum of interdependence can be quantified on the range of possible values of $\rho_\mathsf{1}$ (or $\rho_\mathsf{2}$).   




From the parking function decomposition and the resulting Proposition~\ref{binSidesProp}, we can further recognize that this  service interaction model exhibits a \emph{\edit{conservation} of work} property.

\begin{remark}[\editTwo{On} Conservation of Work]\label{workRem}
Because $\rho$, $\rho_\mathsf{1}$, and $\rho_\mathsf{2}$ do not depend on $\eta$, we see that the synchronicity may alter the agent's pace, but it does not change the interdependence: the total taskload distribution between the two sides is independent of $\eta$. This reflects the fact that the service runs until its natural duration, meaning the time of the last contribution.
Because all contributions are made and the work in the service is conserved regardless of the customer and agent's respective speeds, this model may be most appropriate for services of higher quality, formality, or importance, where customers may be more likely to remain engaged. For example, customers have a vested stake in the success of the tax preparation services discussed in Section~\ref{intro}, and thus they should be likely to respond to the tax agent's contributions, regardless of the agent's pace. By comparison, there may be less sensitive services in which customers simply may renege from the interaction entirely if the agent does not respond sufficiently quickly. \editTwo{Such a setting} may be better \edit{modeled and} managed through \emph{systematic closure}. \edit{Under systematic closure, if the natural closure point cannot be observed \citep[meaning the customer does not ``say goodbye,'' as was observed by][]{ascarza2018some}, the service platform must deploy some policy to determine when to mark the service as concluded. For example, the system may automatically close the service after some period of customer inactivity.} \edit{In this service interaction modeling framework, \cite{daw2021co} proposed to model the space of possible systematic closure policies as the collection of almost surely finite stopping times defined on the filtration of the Hawkes cluster model. One can quickly notice that \editTwo{this} conservation of work property  need not hold for any arbitrary stopping time: a deterministic closure time $t$ (which is a trivially adapted stopping time) will \emph{not} exhibit this property}. Fitting the scope of this work, we remain focused on the natural duration setting.\endnote{Nevertheless, in Appendix~\ref{numerSec}, we offer numerical demonstration that the managerial insights that follow are robust to both natural and systematic closure. We reserve formal analysis of systematic closure for future research.}
\end{remark}

Now, while the interdependence spectrum offers valuable \edit{time-agnostic} perspective, we still seek chronological insights into the service interaction. 
Turning to the dependence on $\eta$, let us consider the model through its limits at the two asynchronous extremes. For this, we will use Definition~\ref{pfcDef} to  construct two special case models, and we will emphasize the pair of  durations resulting in these two special cases.

First, let us take $\bar \tau^\mathsf{1}$ as the cluster duration in which, when sampling each $\Delta_i$ at step (iv) of Definition~\ref{pfcDef}'s parking-function-based construction, the density of $\Delta_i$ remains $\gamma_\mathsf{1}(\cdot)/\rho_\mathsf{1}$ if $s_i = 1$, but if $s_i = 0$, then $\Delta_i$ is simply deterministically set to 0: $\PP{\Delta_i = 0 \mid s_i = 0} = 1$. Note that, in this special case, the model definition no longer depends on $\eta$; hence, the distribution of $\bar \tau^\mathsf{1}$ does not depend on $\eta$ either. Then, for the second special case, we mirror the first. Let us take $\bar \tau^\mathsf{2}$ as the cluster duration in which, again at step (iv) of Definition~\ref{pfcDef}, each $\Delta_i$ is sampled from density $\gamma_2(\cdot)/\rho_\mathsf{2}$ with $\eta = 1$ if $s_i = 0$ (so that the density is $\gamma_2(\cdot)/\rho_\mathsf{2} = g_\mathsf{2}(\cdot)/\rho_\mathsf{2}$) and otherwise is set to  0 if $s_i = 1$: $\PP{\Delta_i = 0 \mid s_i = 1} \edit{= 1}$. \editTwo{As} in the first special case, because the value of $\eta$ has been specified to be 1 in this second construction, the distribution of $\bar \tau^\mathsf{2}$ does not depend on $\eta$, either.






\editTwo{These two special cases represent the service settings in which one side always responds infinitesimally fast, leaving the duration entirely dependent on the pace of the other side}. In this sense, let us refer to $\bar \tau^\mathsf{1}$ as the service duration under \emph{$\mathsf{1}$-paced asynchrony} (meaning Side $\mathsf{2}$ is effectively instantaneous), and likewise let $\bar \tau^\mathsf{2}$ be the service duration under \emph{$\mathsf{2}$-paced asynchrony} (meaning Side $\mathsf{1}$ is effectively instantaneous).

\begin{theorem}\label{convThm}
As $\eta \to \infty$, $\tau \to \bar \tau^\mathsf{1}$ almost surely. As $\eta \to 0$, $\eta\tau \to \bar \tau^\mathsf{2}$ almost surely.
\end{theorem}

It may be of broader stochastic modeling interest to recognize that $\bar \tau^\mathsf{1}$ and $\bar \tau^\mathsf{2}$ are durations of \emph{marked Hawkes clusters}, where the marks are Borel distributed with mean $1/(1-\rho_\mathsf{2})$ and $1/(1-\rho_\mathsf{1})$, respectively. \edit{(}This assertion is formalized in Corollary~\ref{barMarks} within Appendix~\ref{markedSec}, which contains extended discussion and analysis of the marked cluster setting.\edit{)} Moreover, from Theorem~\ref{expMarkedDist} in Appendix~\ref{markedSec}, we can see that there are broad parameter settings in which $\bar \tau^\mathsf{1}$ and $\bar \tau^\mathsf{2}$ may be tractable to characterize explicitly but $\tau$ may not be, such as when $g_\mathsf{1}(\cdot)$ and $g_\mathsf{2}(\cdot)$ are both exponential functions. Hence, through Theorem~\ref{convThm} we can reduce the duration object into simpler components.
To provide intuition for this and for the convergence in Theorem~\ref{convThm} in general, let us again invoke the telephone wire diagrams.

\begin{figure}[ht]
\centering
\subfigure[$\eta=0.01$, $N=21$]{
\includegraphics[width=0.17\textwidth]{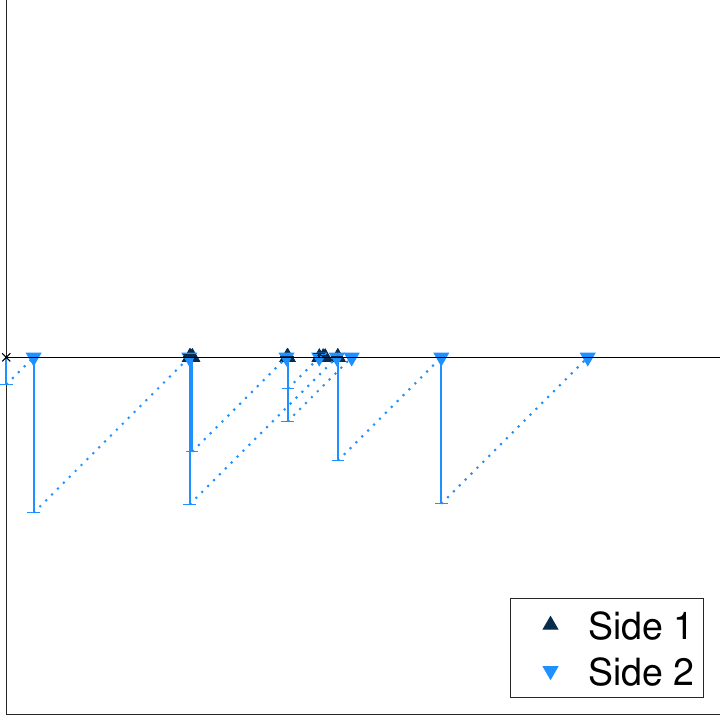} \label{fig:conv2-100-20}
} 
\subfigure[$\eta=0.1$, $N=21$]{
\includegraphics[width=0.17\textwidth]{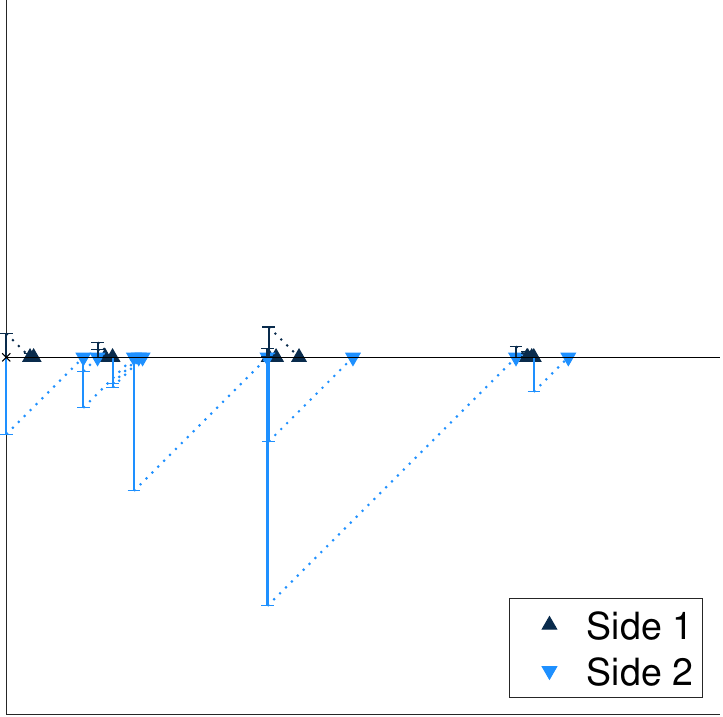} \label{fig:conv2-10-20}
} 
\subfigure[$\eta=1$, $N=21$]{
\includegraphics[width=0.17\textwidth]{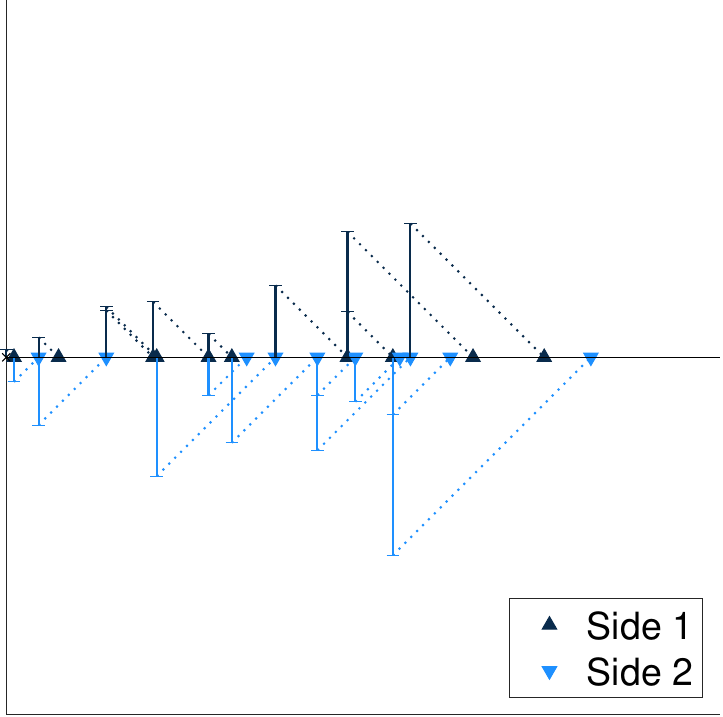} \label{fig:conv-1-20}
}
\subfigure[$\eta=10$, $N=21$]{
\includegraphics[width=0.17\textwidth]{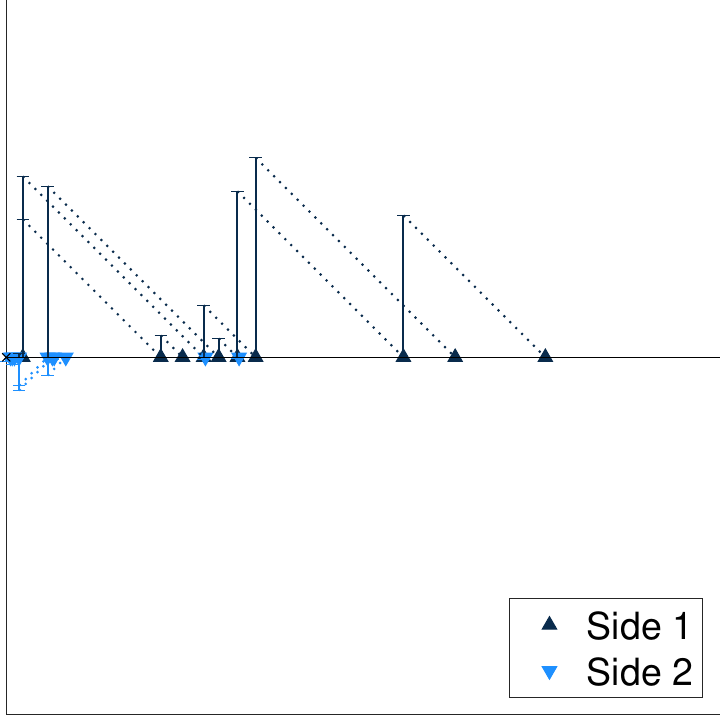} \label{fig:conv-10-20}
} 
\subfigure[$\eta=100$, $N=21$]{
\includegraphics[width=0.17\textwidth]{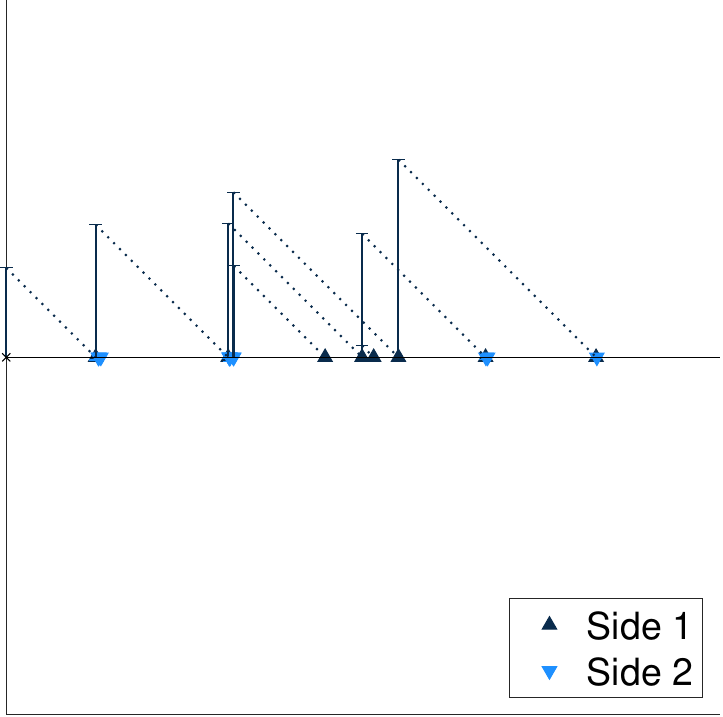} \label{fig:conv-100-20}
} 
\subfigure[$\eta=0.01$, $N=201$]{
\includegraphics[width=0.17\textwidth]{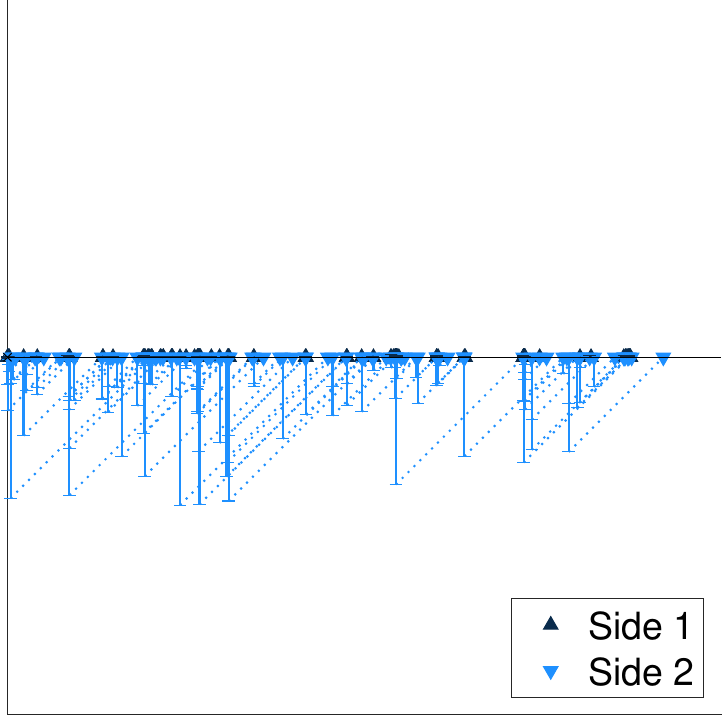} \label{fig:conv2-100-200}
} 
\subfigure[$\eta=0.1$, $N=201$]{
\includegraphics[width=0.17\textwidth]{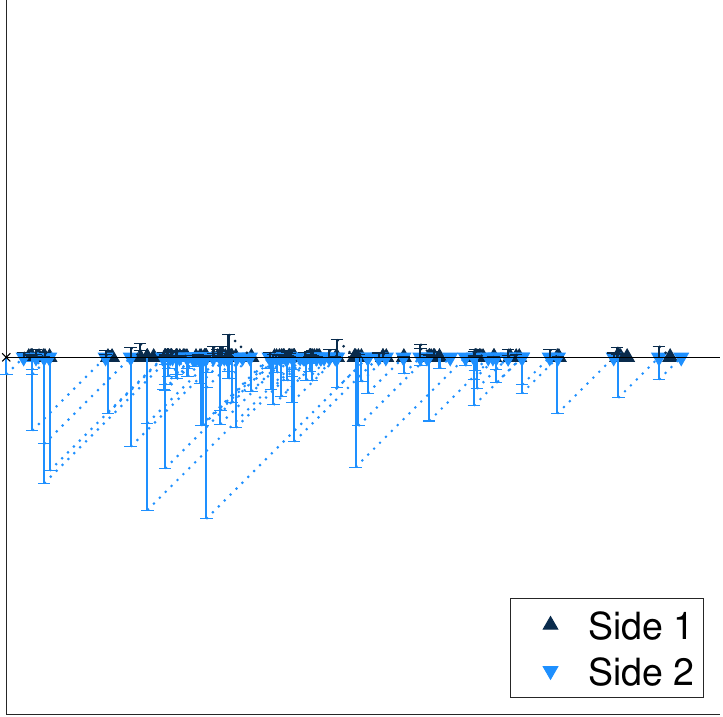} \label{fig:conv2-10-200}
} 
\subfigure[$\eta=1$, $N=201$]{
\includegraphics[width=0.17\textwidth]{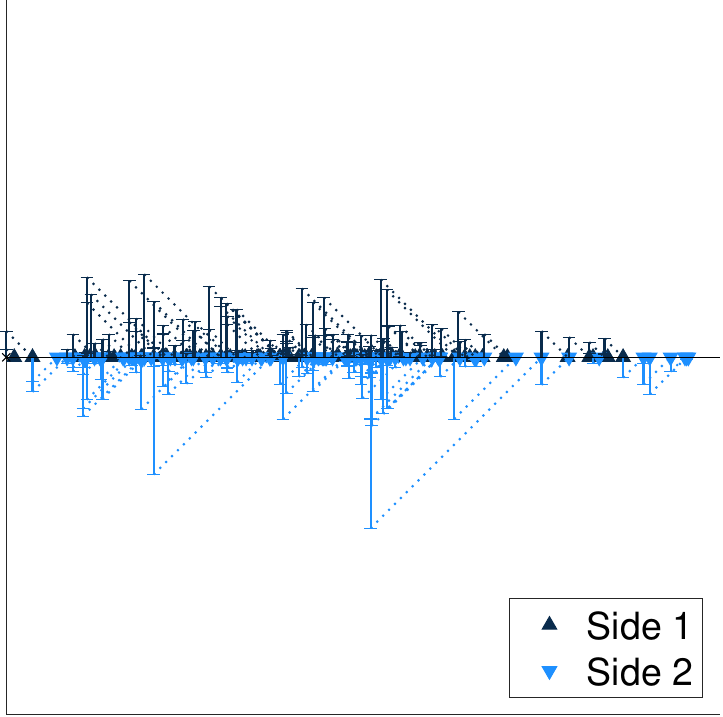} \label{fig:conv-1-200}
}
\subfigure[$\eta=10$, $N=201$]{
\includegraphics[width=0.17\textwidth]{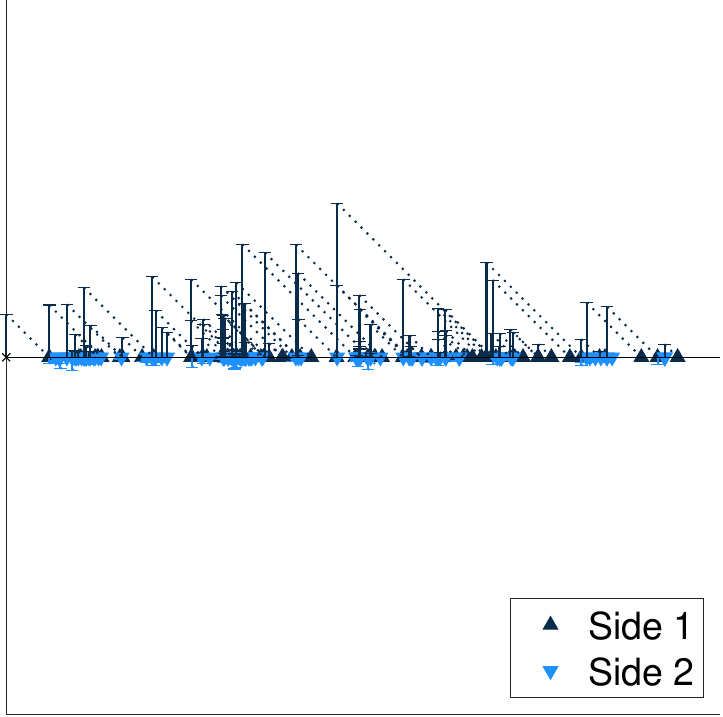} \label{fig:conv-10-200}
} 
\subfigure[$\eta=100$, $N=201$]{
\includegraphics[width=0.17\textwidth]{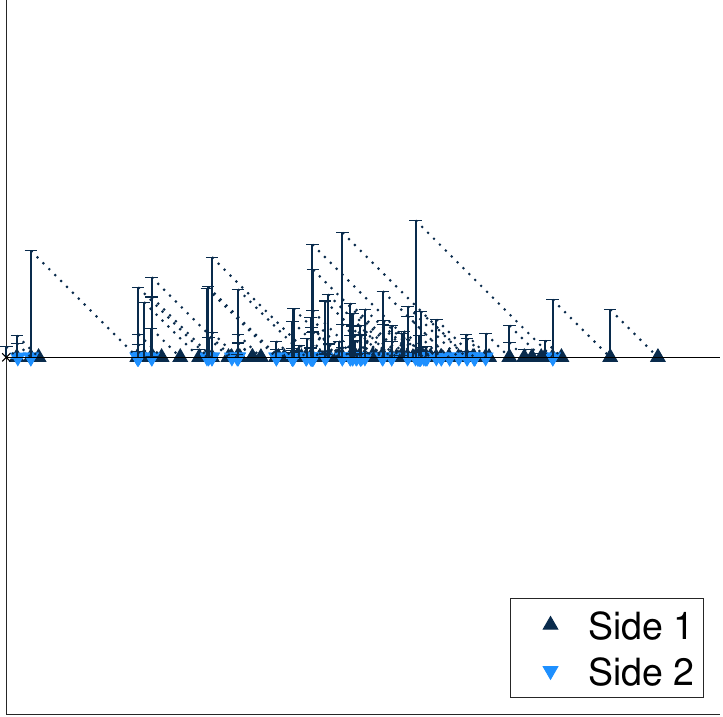} \label{fig:conv-100-200}
} 
\caption{Demonstration of the convergences from Theorem~\ref{convThm} in telephone wire diagrams. Here, ${\eta}$ increases from left-to-right in each row. Every sample path is generated independently, and each in the first row is conditioned on having 20 follow-ups, while each in the second row is conditioned on having 200.}
\label{fig:convAll}
\end{figure}

In Figure~\ref{fig:convAll}, we plot \edit{the telephone wire diagrams for} two collections of mutually independent cluster model sample paths. All axes are plotted on relative scales. Furthermore, the two \editTwo{sides'} underlying response kernels are equivalent up to $\eta$, and thus exactly equal at $\eta = 1$. In the first collection, shown in Figures~\ref{fig:conv2-100-20} through~\ref{fig:conv-100-20}, the clusters are conditioned to have 21 total contributions (including the initial), and the clusters in the second collection, Figures~\ref{fig:conv2-100-200} through~\ref{fig:conv-100-200}, are each conditioned to have 201. In each collection, $\eta$ increases by an order of magnitude from one diagram to the next. In the middle, at the synchronous setting of $\eta = 1$ in Figures~\ref{fig:conv-1-20} and~\ref{fig:conv-1-200}, one can see the offset heights are roughly the same size on the two sides. As $\eta$ increases or decreases, though, one can see that one side dominates the other in length, \edit{yet} the other side does not disappear. That is, particularly in the $N=21$ collection, one can simply count the wires to recognize that there are not 20 poles on either side. Recalling that $\eta$ has no bearing on the interdependence, we can observe that all of these diagrams show service co-production, and contributions are in fact equally likely to be made by either side, as implied by the equivalence of the response kernels. In the extreme cases of the synchronicity parameter, at $\eta = 10^{-2}$ and $\eta = 10^2$ in Figures~\ref{fig:conv2-100-20},~\ref{fig:conv-100-20},~\ref{fig:conv2-100-200}, and~\ref{fig:conv-100-200}, we can see that one side's poles are of effectively zero height, leaving the other side's responses to essentially entirely dictate the pace of service. 


For the purposes of our intended study, it is actually not so important that the duration converges specifically to the duration of marked Hawkes clusters in these asynchrony limits. Instead, what is most important for our present goals is that these limiting objects, $\bar \tau^\mathsf{1}$ and $\bar \tau^\mathsf{2}$, have no dependence on $\eta$. As we will now show in Proposition~\ref{boundProp}, through the same decomposition-based techniques as used in Theorem~\ref{convThm}, we can bound the elusive $\E{\tau}$ through simple affine functions of $\eta$. In fact, by consequence of Theorem~\ref{convThm}, these inequalities are asymptotically tight.

\begin{proposition}\label{boundProp}
For any $\eta \geq 0$, the mean service duration $\E{\tau}$ is bounded above and below by
\begin{align}
\left(\E{\bar \tau^\mathsf{1}} \vee \frac{\E{\bar \tau^\mathsf{2}}}{\eta} \right)
\leq
\E{\tau} 
\leq 
\E{\bar \tau^\mathsf{1}} + \frac{\E{\bar \tau^\mathsf{2}}}{\eta} 
,
\label{boundPropEq}
\end{align}
where $\E{\bar \tau^\mathsf{1}}, \E{\bar \tau^\mathsf{2}} > 0$ are the mean durations under $\mathsf{1}$-paced and $\mathsf{2}$-paced asynchrony, respectively.
\end{proposition}

Recalling our interpretation that $\bar \tau^\mathsf{1}$ and $\bar \tau^\mathsf{2}$ are the durations of service if Sides $\mathsf{2}$ and $\mathsf{1}$ are respectively instantaneous, we can then extend this intuition to the bounds in Proposition~\ref{boundProp}. Through this lens, we can see that the lower bound represents an idealized form of \emph{simultaneous} work, and the upper bound is similarly a simplified form of \emph{successive} work. That is, the lower bound implies that the true mean duration is as least as long as the duration needed for Side $\mathsf{1}$ to complete all their tasks, $\E{\bar \tau^\mathsf{1}}$, and also at least as long as the duration needed for Side $\mathsf{2}$ to do all their tasks, $\E{\bar \tau^\mathsf{2}}/\eta$ (where the normalization by $\eta$ adjusts the Side $\mathsf{2}$ pace \edit{in} $\bar \tau^\mathsf{2}$ to match that in $\tau$). Analogously, the upper bound is then the sum of these two, which is as if one side does all their tasks first and then the other does all their tasks after. 

\edit{As is shown in the telephone wire diagrams in Figures~\ref{telephoneFig} and~\ref{fig:convAll}}, reality is somewhere between these extremes. Unlike the upper bound, there is some overlap among successive triangles, but, unlike the lower bound, there is also some waiting incurred in the back-and-forth between sides. From that perspective, the idea underlying Proposition~\ref{boundProp}'s bounds can be thought of as rearranging the triangles in the telephone wire diagram, or, analogously, switching the hidden parking function spines of the clusters through step (ii) of Definition~\ref{pfcDef}. However, as $\eta$ approaches the limiting asynchrony established by Theorem~\ref{convThm}, the pace is dictated by a spine corresponding to only one side: one can see that the upper and lower bounds in~\eqref{boundPropEq} coincide as $\eta \to 0$ and as $\eta \to \infty$.



Proposition~\ref{boundProp} also implies a practically important distinction of this interaction model relative to common models of the service duration. Because each one-sided asynchrony duration is of positive mean length, Equation~\eqref{boundPropEq} shows that $\E{\tau}$ is strictly positive not only at the asynchronous extremes, but also throughout the range of possible $\eta \geq 0$. We view this as a reflection of a central component in this paper's modeling philosophy. By comparison to conventional approaches, this Hawkes cluster model specifically captures the existence of two sides within the service interaction, and it distinguishes one side's contributions from the other's. Hence, Proposition~\ref{boundProp} follows as a natural consequence. Even if one side always contributes at an instantaneous speed,  if the other side doesn't match that rapid pace, then the mean duration overall won't be instantaneously fast either. In its essence, this \edit{finding} is indebted to the fact that the \edit{micro-level} model distinguishes the customer from the agent. In the remainder of this paper, we will show how the (a)symmetry between these two sides of the service shapes its performance overall.






\section{Non-Monotonic System Performance from Monotonic \editTwo{Agent} Slowdown}\label{nonMonSec}




Having now developed this understanding of the interaction along the service asymmetry spectra, let us return to the original operational goal \editTwo{as framed in Section~\ref{questionSec}}: understanding \edit{the operational impact of} concurrency. 
Because the goal of this section is to understand how system\edit{-level} performance, specifically the agent's throughput, depends on the agent's concurrency, let us  update our notation so that the service duration is a function of the concurrency (as alluded to following Definition~\ref{hcDef}). That is, for $\tau(\mathcal{K})$ as the natural duration of a service given that the agent has $\mathcal{K}$ simultaneous customers, we aim to understand the throughput, $\mathcal{K}/\E{\tau(\mathcal{K})}$. \edit{Hence, this section begins Part II of the paper, in which we connect the understanding we have developed at the individual-level to the performance we seek to characterize at the system-level.}

To connect concurrency to the dimensions of service we have analyzed in Section~\ref{clusterSec} and to expand the model from the interaction-level to an agent- or system-level context, let us likewise map concurrency to the behavioral phenomenon of agent-side slowdown.
\editTwo{As reviewed in Section~\ref{sl:bom}, both the empirical operations literature and general psychological studies have found that} concurrent customers \edit{in} multitasked services \editTwo{induce} \emph{slowdown effects} on the agent. Intuitively, this means that, if the agent serves a higher number of simultaneous customers, their speed in each individual service will be slower than when they have a lower concurrency. Recalling from the discussions in Sections~\ref{sec:model} and~\ref{clusterSec} that the mean agent response time is proportional to $1/\eta$, we will capture the slowdown effect of concurrency by defining \editTwo{the interaction model's synchronicity parameter ($\eta$)} as the inverse of a non-decreasing function of the concurrency.

\begin{assumption}[Synchronicity as a Function of Concurrency]\label{conA}
Let the synchronicity $\eta(\mathcal{K})$ be determined by the agent's concurrency level $\mathcal{K}$ through $\eta(\mathcal{K}) = 1/h(\mathcal{K})$, where the Side $\mathsf{2}$ slowdown function,   $h: \mathbb{R}_+ \to \mathbb{R}_+$, is such that $h(\cdot)$ is non-decreasing, $h(0) = 0$, $h(1) = 1$, and $\lim_{x \to \infty} h(x) = \infty$. 
\end{assumption}

Assumption~\ref{conA}, \edit{which we will  take for granted throughout Part II of the paper, provides} a general form of the observed and intuitive structure of the concurrency. As $\mathcal{K}$ increases, the agent's mean response time increases, because $1/\eta(\mathcal{K}) = h(\mathcal{K})$ and $h(\cdot)$ is non-decreasing. Hence, we refer to $h(\cdot)$ as the Side $\mathsf{2}$ or agent-side slowdown function. Note that \emph{agent-side} is contextualized within a given service interaction, whereas \emph{agent-level} refers to all of the interactions held concurrently by one agent. The remaining (mild) conditions on $h(\cdot)$ can be easily interpreted. First, in the hypothetical extreme in which the agent has infinitely many customers, they should take infinitely long to respond in any individual service interaction: $\lim_{x \to \infty} h(x) = \infty$. On the other hand, if the agent has a vanishingly small fraction of customers, we assume that they should be able to respond instantaneously: $h(0) = 0$. Finally, to set the scale, we assume that $h(1) = 1$ so that one customer is essentially the ``standard'' concurrency (but this \edit{precise value} is simply for the sake of interpretation, and it is not necessary for the technical results). Naturally, $h(\cdot)$ can be tailored to the specific problem and behavioral environment, as appropriate.\endnote{\edit{See Section~\ref{hSec} of the Appendix for an example of such problem-specific tailoring.}}

\edit{With the conditions of Assumption~\ref{conA} in hand, the research question of this section (and, more broadly, Part II of the paper) is thus: what conditions, if any, on the monotonic agent-side slowdown function can produce the non-monotonic system-level performance that has been broadly observed empirically?}

Let us emphasize that Assumption~\ref{conA} defines $h(\cdot)$ as a \emph{monotonic effect on the pace of agent-side contributions}.  Aligned with the observations of the empirical operations and general psychological literature, $h(\cdot)$ reflects the fact that, as the agent's concurrency workload rises, their average pace should slow. This model of the agent-side behavior is both flexible and parsimonious, as we impose no other universal conditions on the slowdown effect other than its non-decreasing growth from instantaneous under zero work ($h(0) = 0$) to overloaded under ceaseless work ($\lim_{x \to \infty} h(x) = \infty$).  

\begin{remark}[\edit{Relative vs.~Absolute Agent-Side Slowdown}]\label{slowdownRemark}
\edit{Let us note that Assumption~\ref{conA} does not imply that the agent's slowdown must hold in a normalized or relative sense, but only in a strictly absolute sense. For the sake of example, suppose that the agent presently has two concurrent customers and each customer just made a contribution at time 0. Then, in each interaction, the agent's expected response time is proportional to $h(2)$. The assumptions on $h(\cdot)$ do \emph{not} require that the agent's time to respond to each of the two customers must be at least twice as long as their time to respond if they were serving just one customer (i.e., $h(2) \leq 2 h(1)$ and $h(2) > 2 h(1)$ can each satisfy the slowdown assumptions), but they do imply that, focusing on \editTwo{an arbitrary} one of the two concurrent services without loss of generality, the expected agent response time to the focal customer cannot be faster than the expected response time if that customer was their \emph{only} customer (i.e., $h(2) < h(1)$ does not satisfy the assumptions).\endnote{\edit{\editTwo{If} $g_\mathsf{1}(\cdot)$ and $g_\mathsf{2}(\cdot)$ are both exponential, \editTwo{this remark's} arguments \editTwo{extend} to the order statistics of the random variables, meaning that $h(2)/2$ is proportional to the expected time of the \emph{first} agent response across the two simultaneously started interactions.}}
Equivalently said, our modeling approach does allow the agent to possibly have a faster average rate of \emph{overall contribution} in $k > 1$ parallel services than they would in one service on its own ($h(k)/k < h(1)$ can be possible), but the agent cannot do $k$ things at once faster on average than one thing on its own ($h(k) < h(1)$ is not possible). Hence,  non-decreasing $h(\cdot)$  assumes \emph{absolute slowdown}, but it does not make the stronger assumption of \emph{relative slowdown}: $h(x)/x$ need not be non-decreasing even though $h(x)$ is.}
\end{remark}



Now, by the construction of the service interaction model, both the customer and the agent have a role in setting the overall pace of the service. Recalling the one-sided asynchronous durations from Proposition~\ref{boundProp}, ${\bar \tau^\mathsf{1}}$ and ${\bar \tau^\mathsf{2}}$, let us now define two counterparts to the focal agent-level throughput: the guaranteed service rate and the idealized service rate. Let the \textit{guaranteed service rate} $\mathcal{G}: \mathbb{R}_+ \to \mathbb{R}_+$ be defined
\begin{align}
\mathcal{G}(\mathcal{K})
=
\frac{\mathcal{K}}{
\E{\bar \tau^\mathsf{1}} + \E{\bar \tau^\mathsf{2}} h(\mathcal{K})
}
,
\label{gDefEq}
\end{align}
and, then, let the \textit{idealized service rate} $\mathcal{I}: \mathbb{R}_+ \to \mathbb{R}_+$ be defined
\begin{align}
\mathcal{I}(\mathcal{K})
=
\frac{\mathcal{K}}{
\left(\E{\bar \tau^\mathsf{1}} \vee \E{\bar \tau^\mathsf{2}} h(\mathcal{K})\right)
}
.
\label{iDefEq}
\end{align}
Immediately from Proposition~\ref{boundProp}, we can use $G(\mathcal{K})$ and $\mathcal{I}(\mathcal{K})$ to bound the  throughput at concurrency $\mathcal{K}$.

\begin{corollary}\label{rateBound}
The guaranteed and idealized service rates are lower and upper bounds, respectively, for the agent's throughput:
\begin{align}
\mathcal{G}(\mathcal{K})
\leq
\frac{\mathcal{K}}{\E{\tau(\mathcal{K})}}
\leq
\mathcal{I}(\mathcal{K})
,
\end{align}
for every concurrency level $\mathcal{K}$.
\end{corollary}



Naturally, the intuition for Corollary~\ref{rateBound}, and thus also the intuition for the names for $\mathcal{G}(\cdot)$ and $\mathcal{I}(\cdot)$, is identical to that of Proposition~\ref{boundProp}. The guaranteed service rate, meaning the slowest that the throughput will be, occurs when the agent works with each customer in successive fashion. Similarly, the idealized service rate, or the fastest that the throughput can be, occurs when the parties work simultaneously. \editTwo{Of course, as exemplified in the telephone wires of  Figures~\ref{telephoneFig} and~\ref{fig:convAll}}, reality lies somewhere in the middle: at some times, there may be overlap among the successive contributions, and at other times there will be no overlap at all. 

This leads us now to a sequence of our main and most general results for the concurrency-driven throughput, in which we prove the existence of an $\IU$-shaped throughput and demonstrate its managerial implications. In Theorem~\ref{iuThm}, we begin by providing first-principles conditions for this metric to exhibit an $\IU$-shape. 

\begin{theorem}\label{iuThm}
If  ${h(x)}\slash {x} \to \infty$ as $x \to \infty$, then the throughput is $\IU$-shaped:
$$
\frac{\mathcal{K}}{\E{\tau(\mathcal{K})}} \in \mathcal{N}_{\IU}
.
$$ 
However, 
if there exists $c < \infty$ such that $h(x) \slash x \to c$ as $x \to \infty$, then the throughput will not belong to $\mathcal{N}_\IU$.
\end{theorem}


Building from the bounds on the duration obtained by the parking function decomposition of the interaction model, the proof of Theorem~\ref{iuThm} leverages that Corollary~\ref{rateBound}'s bounds on the throughput are asymptotically tight as $\mathcal{K} \to 0$ and as $\mathcal{K} \to \infty$. Thus, if $\mathcal{G}(\cdot)$ and $\mathcal{I}(\cdot)$ each belong to the class of asymptotically non-monotonic functions that tend to zero at each asymmetric extreme ($\mathcal{N}_\IU$), then so will the throughput. 


We can recognize that these limit-based conditions on $h(\cdot)$ come with a natural intuition in the service context. First, if $h(\cdot)$ is super-linear ($h(x) \slash x \to \infty$ as $x \to \infty$), then the agent's slowdown effect is \textit{stronger than time}. In this case, the agent's pace is slowed down at a magnitude \edit{that} exceeds the number of customers gained, as is incurred from the aforementioned behavioral drivers like task juggling \citep{coviello2014time,bray2015multitasking}, switch-over costs \citep{Kc2013DoesDepartment,gurvich2020collaboration}, or cognitive load \citep{pashler1994dual,rubinstein2001executive}. 
This also aligns with observations from the empirical operations literature, such as the finding by \citet[pg. 3035]{Goes2017WhenMultitasking} that the ``marginal effect of more multitasking on response delays should be increasing.'' 
On the other hand, if $h(\cdot)$ is either asymptotically linear or sub-linear ($h(x) \slash x \to c$ as $x \to \infty$), then the agent's slowdown effect is \textit{equal to} or \emph{weaker than time}. This means that the agent's time to complete one contribution while handling $k$ customers may be proportional to (or faster than) the product of $k$ and the time needed for one contribution while handling only one customer. We would expect this to be most relevant for services with large amounts of task redundancy across interactions, so that, as the concurrency grows, there will always be an efficiency gain to handling additional customers \edit{\citep[such as \editTwo{through synergistic multi-brand verticals} of concurrent customer streams, as demonstrated in the data of][]{batt2024multitasking}}.


Hence, Theorem~\ref{iuThm} implies that if the slowdown from $h(\cdot)$ is sufficiently strong, the system will find non-monotonic performance from monotonic agent-side slowdown. 
\editTwo{In immediate operational consequence, 
Corollary~\ref{optKcor} recognizes} that if the throughput is $\IU$-shaped, the agent's concurrency can be optimized.

\begin{corollary}\label{optKcor}
If  ${h(x)}\slash {x} \to \infty$ as $x \to \infty$, then there exists a globally optimal concurrency level at which the throughput is maximized.
\end{corollary}

If the conditions that $h(\cdot)$ is at most linear are made slightly more specific, then we can offer a full monotonic converse to the non-monotonic results of Theorem~\ref{iuThm}. \edit{Specifically, the partial converse to Theorem~\ref{iuThm} can be stratified into two settings: $h(\cdot)$ being either sub-linear or (asymptotically) linear. Let us state a corollary for each case. If $h(\cdot)$ is sub-linear, then the throughput will be asymptotically monotonic, as shown in Corollary~\ref{monoCor}. If $h(\cdot)$ is asymptotically linear,} we can refine the \edit{conditions of the} second half of Theorem~\ref{iuThm} so that the throughput converges \emph{up to} \edit{some level as $\mathcal{K}$ grows large}, meaning that \edit{there is always a faster throughput at some higher concurrency level}.  We provide these conditions in Corollary~\ref{noOptKcor}.

\begin{corollary}\label{monoCor}
\edit{If $h(x)/x \to 0$ as $x \to \infty$, the throughput is asymptotically monotonic\editTwo{,}
$
{\mathcal{K}}/{\E{\tau(\mathcal{K})}}
\in 
\mathcal{M}
$\editTwo{,}} 
\editTwo{with $\lim_{\mathcal{K} \to \infty}\mathcal{K}/{\E{\tau(\mathcal{K})}} = \infty$.}
\end{corollary}

\begin{corollary}\label{noOptKcor}
    If $h(x)/x \to c$ as $x\to\infty$ and $h(x) \slash x~\editTwo{>}~c$ for all $x$ such that $h(x) \geq \E{\bar \tau^\mathsf{1}} \slash \E{\bar \tau^\mathsf{2}}$\edit{, then} 
for every $\mathcal{K}$ there exists some $\mathcal{K}' > \mathcal{K}$ such that $\mathcal{K}'/{\E{\tau(\mathcal{K}')}} > \mathcal{K}/{\E{\tau(\mathcal{K})}}$.
\end{corollary}

\edit{Both of these corollaries make a stronger statement than simply saying there is not an $\IU$-shape according to Definition~\ref{uDef}: each implies an unambiguous lack of even a partial $\IU$-shape. Equivalently said, the important managerial conclusion from both Corollary~\ref{monoCor} and~\ref{noOptKcor} is that, in either the sub-linear setting or this specific linear setting, there will be no optimal concurrency level. Hence, under general conditions of the agent-side slowdown, Corollaries~\ref{optKcor},~\ref{monoCor}, and~\ref{noOptKcor} reveal a contrast in system-level performance (and, accordingly, managerial decision making) that is characterized by the agent's individual-level features.}



While this level of generality around $h(\cdot)$ is often desirable, let us demonstrate that as the structure on $h(\cdot)$ is made even just slightly more explicit, we can substantially refine our insights regarding the shape of the throughput. For simplicity of expression, let us here forward add the following technical assumption. 

\begin{assumption}\label{twiceAssump}
The agent-side response slowdown function $h(\cdot)$ is twice-differentiable.
\end{assumption}

Under Assumption~\ref{twiceAssump}, not only do we receive simple conditions on $h(\cdot)$ that yield non-monotonic throughput, we can also characterize this shape of the performance through explicit partition of the range of concurrency. For example, under strict convexity \citep[which would align \editTwo{with} the increasing marginal effect of more multitasking seen by][]{Goes2017WhenMultitasking}, Proposition~\ref{convexProp} casts the $\IU$-shape of the throughput across two easily computed partition points: increasing below both, peaking between, and decreasing after.

\begin{proposition}\label{convexProp}
If $h(\cdot)$ is strictly convex, then the throughput attains its global maximum within the interval $[\underline{\mathcal{K}},\overline{\mathcal{K}}]$, where $\mathcal{K}^*$ is the unique solution to
\begin{align}
\frac{
\E{\bar \tau^\mathsf{1}}
}{
\E{\bar \tau^\mathsf{2}}
}
=
\mathcal{K}^* h'(\mathcal{K}^*) - h(\mathcal{K}^*)
,
\label{convEq}
\end{align}
with $\underline{\mathcal{K}} = \E{\bar \tau^\mathsf{1}} \slash \left( \E{\bar \tau^\mathsf{2}} h'(\mathcal{K}^*)\right)$ and  $\overline{\mathcal{K}}$ as the unique solution to $h'(\mathcal{K}^*) = h(\overline{\mathcal{K}})/\overline{\mathcal{K}}$.
\end{proposition}

\subsection{Illustrative Slowdown Function: Polynomial ($h(\mathcal{K}) = \mathcal{K}^{\,^\sigma}$)}\label{polySec}

To showcase these results in full clarity, let us close this section with \editTwo{one specific form of the slowdown function}. \edit{This evincive setting of polynomial $h(\cdot)$ will be a recurring example throughout the remainder of the paper, so let us properly describe it.}

\edit{For the sake of illustration}, let us \editTwo{temporarily} assume that $h(\mathcal{K}) = \mathcal{K}^\sigma$ for some $\sigma > 0$. At any such $\sigma$, we see that the agent's rate of contributions slows as the concurrency increases: $\mathcal{K}^\sigma$ is an increasing function for any $\sigma > 0$. However, the context and pace of the increase changes with this parameter. First, if $\sigma = 1$, then the agent-side slowdown is analogous (but not equivalent, as we will soon discuss) to standard processor sharing, in the sense that the agent can perfectly divide their attention across their $\mathcal{K}$ concurrent customers, splitting the unit rate into $\mathcal{K}$ even parts. Then, $\sigma > 1$ would correspond to added delays, such as from switch-over costs or cognitive load, \editTwo{which} hamper the agent's efficiency and induce a response rate that is slower \editTwo{than} the unit rate divided the concurrency. On the other hand, $\sigma < 1$ corresponds to the case where the agent actually becomes more efficient as the scale increases. 

However, the discussion above only captures the agent-side effects \emph{within the interaction}, and it does not capture the overall performance \editTwo{at the system-level. Instead,} the guaranteed and idealized service rates provide us insight into these broader perspectives. Under this subsection's assumptions on the agent-side slowdown function, \editTwo{$\mathcal{G}(\cdot)$ and $\mathcal{I}(\cdot)$ become
\begin{align}
\mathcal{G}(\mathcal{K})
=
\frac{\mathcal{K}}{
\E{\bar \tau^\mathsf{1}}
+
\E{\bar \tau^\mathsf{2}} \mathcal{K}^{\,^\sigma}
}
\qquad
\text{ and }
\qquad
\mathcal{I}(\mathcal{K})
=
\frac{\mathcal{K}}{
\left(
\E{\bar \tau^\mathsf{1}}
\vee
\E{\bar \tau^\mathsf{2}} \mathcal{K}^{\,^\sigma}
\right)
}
.
\end{align}}
As one can see, these bounding functions agree at the asymmetric extremes ($\mathcal{K} \to 0$ and $\mathcal{K} \to \infty$), and this is what illuminates the shape of the throughput.


\begin{proposition}\label{polyOptProp}
Let $h(x) = x^\sigma$. If $\sigma > 1$, then the throughput is $\IU$-shaped and will attain its global maximum on the interval 
\begin{align}
(1 - 1\slash\sigma)\mathcal{K}^* 
\leq 
\mathcal{K} 
\leq 
\sigma^{1/(\sigma-1)}\mathcal{K}^*
,
\end{align}
where
\begin{align}
\mathcal{K}^*
=
\left(
\frac{\E{\bar \tau^\mathsf{1}}}{(\sigma - 1) \E{\bar \tau^\mathsf{2}}}
\right)^{{1}\slash{\sigma}}
.
\label{polyOptKstarEq}
\end{align}
However, if $\sigma \leq 1$, then the throughput is not $\IU$-shaped; instead, it is asymptotically monotonic and there does not exist an optimal concurrency.
\end{proposition}

Through Proposition~\ref{polyOptProp}, the results of this section can be summarized into one unified statement for the context of \editTwo{this example: if $\sigma > 1$, the $\IU$-shape exists and there is a globally optimal concurrency level for the agent, and if $\sigma \leq 1$, there will be no $\IU$-shape and there is a larger concurrency with better throughput.} 

\section{Symmetric Slowdowns Cannot Reproduce Non-Monotonicity\edit{: A Service-Computation Contrast}}\label{symSec}




Throughout the paper so far, we have compared and contrasted the perspective of this service interaction model with the framework of processor sharing, the classic queueing theoretic model of multitasking, particularly so in computation contexts. In this \editTwo{interlude within} \edit{Part II's study of the system-level impact of concurrency}, we focus on the contrast \edit{between these micro- and macro-modeling frameworks}, highlighting the difference between the insights of our proposed model of service asymmetry with the symmetric slowdown implied by processor sharing. By symmetric slowdown, we mean that both Side $\mathsf{1}$ and Side $\mathsf{2}$ time scales are modulated by $\eta$, or, equivalently in the service context, both the customer and agent contribution speeds are modulated by \editTwo{a} function of the concurrency $\mathcal{K}$. \edit{As we will demonstrate, symmetric slowdown yields meaningfully different system-level dynamics and performance metrics by comparison to the asymmetric slowdown specifically applied to the agent side (Side $\mathsf{2}$), and we will explain why this reveals a \editTwo{contrast in modeling desiderata for services and modeling desiderata for computation contexts. Hence, we will refer to this observation as a} \emph{service-computation contrast}.}

In this section only, let us consider a service \edit{interaction} model in which Side $\mathsf{1}$ and Side $\mathsf{2}$ are slowed down symmetrically as a function of the concurrency. Applying $\hat{\cdot}$ markers to denote the symmetric setting, let us update the interaction model so that the contribution rate from~\eqref{intensityDef} is now given by
\begin{align}
\hat \mu_t
&=
\int_0^t \left( \frac{1}{h(\mathcal{\mathcal{K}})} \cdot g_\mathsf{1}\left( \frac{1}{h(\mathcal{\mathcal{K}})} \cdot (t - s) \right) + \frac{1}{h(\mathcal{\mathcal{K}})} \cdot g_\mathsf{2}\left(\frac{1}{h(\mathcal{\mathcal{K}})} \cdot (t-s)\right)\right) \mathrm{d}\hat N_s
\label{symModelDef}
,
\end{align}
where, by comparison to the asymmetry in Definition~\ref{hcDef}, Equation~\eqref{symModelDef} applies the concurrency-dependent synchronicity term $\eta(\mathcal{K}) = 1 / h(\mathcal{K})$ to both of the response functions, instead of \edit{only applying it} to $g_\mathsf{2}(\cdot)$. Let \editTwo{$\hat \tau(\mathcal{K}) = \hat \tau_{\hat N - 1}$} denote the duration of service in this symmetric setting \editTwo{with concurrency $\mathcal{K}$}. Through the parking function decomposition \edit{from} Definition~\ref{pfcDef}, we can quickly notice that $\hat \tau(\mathcal{K}) = \hat \tau(1) \cdot h(\mathcal{K})$ \edit{for every $\mathcal{K}$}, because every offset timing random variable will be multiplied by $h(\mathcal{K})$. Equivalently stated, Equation~\eqref{symModelDef} implies that $h(\mathcal{K})$ functions like a universal deterministic time change. 

\begin{remark}[\editTwo{Processor Sharing Is a Symmetric Slowdown}]\label{psRemark}
\edit{Let us compare the $\hat \tau(\mathcal{K}) = \hat \tau(1) \cdot h(\mathcal{K})$ relationship to the dynamics of service duration under processor sharing as discussed in, for example, section 2.2 of \citet{long2018customer}. In \citet{long2018customer}, the system-level queueing model employs limited, state-dependent processor sharing as its service discipline over general service durations by dividing each exogenous service duration random variable by a state-dependent rate that is strictly decreasing with the agent's concurrency level. The only difference relative to $\hat \tau(\mathcal{K}) = \hat \tau(1) \cdot h(\mathcal{K})$ is that here we have converted the decreasing rate divisor into an increasing time-modulation multiplier. In the words of \citet[][pg. 1678]{long2018customer} with our notation substituted in [$\cdot$], ``if a customer always receives service from an agent at level [$\mathcal{K}$], then [$\hat \tau(1) \cdot h(\mathcal{K})$] signifies the required service time for this customer.'' Hence,  the symmetric slowdown in \eqref{symModelDef} is exactly the same as state-dependent processor sharing: a time change modulation applied universally across all of the service interaction, rather than just on one side. To further relate this to the original form of processor sharing as proposed by \citet{kleinrock1967time}, one could simply take $h(\mathcal{K}) = \mathcal{K}$.}
\end{remark}

\edit{Following \editTwo{directly} from the construction in \eqref{symModelDef}}, we can recognize that the throughput under symmetric slowdown will be \edit{simply proportional to the ratio of the concurrency and the slowdown function\editTwo{:}}
\begin{align}
\edit{\frac{\mathcal{K}}{\E{\hat \tau(\mathcal{K})}}
= } 
\frac{
\mathcal{K}
}{
\E{\hat \tau(1)} h(\mathcal{K})
}
.
\label{symRateDef}
\end{align}
 Thus, we do not need to obtain equivalents to the guaranteed and idealized service rates; \editTwo{we can analyze the shape of the throughput without the need to bound or even explicitly compute $\E{\hat \tau(1)}$}. 
 
 Immediately, we find contrast to the results of Section~\ref{nonMonSec}. By comparison to Theorem~\ref{iuThm}'s finding that the throughput is $\IU$-shaped for super-linear $h(\cdot)$ or the partial converse in  Corollary~\ref{monoCor} that sub-linear $h(\cdot)$ will yield asymptotic monotonicity, Proposition~\ref{symProp1} shows that, if the derivative of the symmetric slowdown function is asymptotically monotonic (in any form), the throughput will likewise be asymptotically monotonic under symmetric slowdown.

\begin{proposition}\label{symProp1}
\edit{If $h'(\cdot) \in \mathcal{M}$, then the throughput under symmetric slowdown will also be asymptotically monotonic:
\begin{align}
\frac{\mathcal{K}}{\E{\hat \tau(\mathcal{K})}} \in \mathcal{M}.
\end{align}
Hence, there is no optimal concurrency level under symmetric slowdown of the customer and agent.}
\end{proposition}


\edit{Though this asymptotic characterization adheres to the themes set across the rest of the paper, we actually need not lean upon such limiting analysis in this symmetric setting. In contrast to Proposition~\ref{convexProp}, we now show in Proposition~\ref{symProp2} that under convexity \emph{or} concavity of the symmetric slowdown function, the resulting throughput will be monotonic globally, rather than just asymptotically.}

\begin{proposition}\label{symProp2}
\edit{If $h(\cdot)$ is either convex or concave, then the throughput under symmetric slowdown will be  truly monotonic:
\begin{enumerate}[i)]
\item If $h(\cdot)$ is convex, then ${\mathcal{K}}/{\E{\hat \tau(\mathcal{K})}}$ will be non-increasing for all $\mathcal{K}$.
\item If $h(\cdot)$ is concave, then ${\mathcal{K}}/{\E{\hat \tau(\mathcal{K})}}$ will be non-decreasing for all $\mathcal{K}$.
\end{enumerate}
In either case, there is no optimal concurrency level under symmetric slowdown of the customer and agent.}
\end{proposition}


The slowdown function $h(\mathcal{K}) = \mathcal{K}^\sigma$ can again  illustrate \edit{the ideas behind Propositions~\ref{symProp1} and~\ref{symProp2}. In this case, the symmetric-slowdown throughput becomes}
\begin{align}
\frac{
\mathcal{K}
}{
\E{\hat \tau(1)} \mathcal{K}^{\,^\sigma}
}
=
\frac{
\mathcal{K}^{\,^{1-\sigma}}
}{
\E{\hat \tau(1)} 
}
,
\end{align}
and this \edit{function} has first derivative proportional to $(1-\sigma)\slash \mathcal{K}^{\,^\sigma}$. Hence, the sign only depends on $\sigma$. If $\sigma < 1$, then performance always increases with the concurrency; if $\sigma = 1$, then all concurrencies are alike; if $\sigma > 1$, then performance always decreases with the concurrency. \edit{By contrast to this monotonic throughput for any $\sigma$ under symmetric slowdown, recall that in Proposition~\ref{polyOptProp}, we found that the throughput is $\IU$-shaped for $\sigma > 1$ if the slowdown is localized to the agent's side.}


\edit{\editTwo{In light of Remark~\ref{psRemark}}, the key takeaway \editTwo{of this section} is that the conditions for non-monotonic throughput identified through this service interaction model will instead produce monotonic throughput if used in a processor-sharing-type model. We feel that this insight is  more modeling than managerial.} Although there could perhaps be settings in which the service could attempt to distract, entertain, or otherwise slow the customer's contribution rate to match the concurrency-burdened agent's (where then monotonic throughput would be advantageous), this is likely not practical in most service settings.  Instead, we believe the first-order benefit of \edit{Propositions~\ref{symProp1} and~\ref{symProp2}} is the explanation \edit{they} offer for why processor-sharing-based stochastic models \edit{of service} would have to make non-monotonic throughput assumptions explicitly. 

\edit{As we claimed at the start of this section, we view the distinction between asymmetric and symmetric slowdown as a contrast between modeling services and modeling computation. In computation contexts such as the evenly divided machine capacity originally modeled by \citet{kleinrock1967time}, all present jobs share time on the processor equally, so that the machine's throughput remains constant for any amount of multi-tasking. Even when generalized to be state-dependent or limited, processor sharing inherently supposes that the speed \emph{of the system} is evenly divided among all the present jobs. By consequence, that classic modeling approach presumes that any slowdown (such as from load) is applied across the full service duration: $\hat \tau(\mathcal{K}) = \hat \tau(1) h(\mathcal{K})$.} 

\edit{If the context is truly just jobs or other inanimate tasks being processed in parallel, then such a modeling approach makes sense. However, by contrast to those computation settings, \editTwo{this paper's service interaction framework  inherently proposes} that the customer and agent should instead be disentangled and \editTwo{modeled} separately, so that any effects (from load or otherwise) can be applied \emph{asymmetrically}. Intuitively, in services and unlike in computation, even if the agent was instantaneously fast, the customer still plays a role in the service interaction, and thus the duration of the service will never be instant. Moreover, the duration of the service interaction is not merely a multiple of the duration of the agent's work. Because the classic processor-sharing discipline is at the macro- or system-level,  this fact would be missed by that queueing theoretic model. Thus, we view this contrast as justification of this interaction framework (and, more broadly, this micro-level approach) as a model specifically \emph{for services}.}



\section{A Second Dimension of Customer-Agent (A)Symmetry}\label{interSec}





By consequence of the model's implicit conservation of work noted in Remark~\ref{workRem}, focusing on the operational impacts of concurrency has meant that we have thus far only considered the customer-agent synchronicity. However, in Section~\ref{sec:model}, we have shown that this model captures not only this spectrum of time (a)symmetry but also the spectrum of work allocation, or the interdependence. \edit{Here in this final section of Part II of the paper, we} will now inspect how this second, often managerially-designed dimension impacts the performance of concurrent services. 

To begin, let us return to our illustrative  polynomial agent-side slowdown function, $h(\mathcal{K}) = \mathcal{K}^\sigma$, and let us suppose that $\sigma > 1$. By Proposition~\ref{polyOptProp}, we know that in this setting, the throughput is $\IU$-shaped, and, \edit{therefore}, there exists a \edit{global} optimal concurrency at which point the throughput is maximized. \editTwo{Furthermore,} this maximal throughput is bounded below by the peak of the guaranteed service rate, 
\begin{align}
\mathcal{G}^*
=
\mathcal{G}(\mathcal{K}^*)
&=
\left(
\frac{\E{\bar \tau^\mathsf{1}}}{(\sigma - 1) \E{\bar \tau^\mathsf{2}}}
\right)^{\frac{1}{\sigma}}
\frac{\sigma - 1}
{\sigma \E{\bar \tau^\mathsf{1}}}
=
\frac{(\sigma - 1)^{1 - \frac{1}{\sigma}}}{\sigma \E{\bar \tau^\mathsf{1}}^{1-\frac{1}{\sigma}} \E{\bar \tau^\mathsf{2}}^{\frac{1}\sigma}}
,
\end{align}
and above by the peak of the idealized service rate. This maximized idealized service rate is
\begin{align}
\mathcal{I}^*
=
\frac{1}{
\E{\bar \tau^\mathsf{1}}^{1-\frac{1}{\sigma}}
\E{\bar \tau^\mathsf{2}}^{\frac{1}{\sigma}}
}
,
\end{align}
and this occurs at $\mathcal{K} = \left(\E{\bar \tau^\mathsf{1}}\slash\E{\bar \tau^\mathsf{2}}\right)^{1\slash \sigma}$, at which point the two terms within \edit{the maximum in the denominator of~\eqref{iDefEq}} are equal. Now, we can recognize that if $\E{\bar \tau^\mathsf{1}} \to 0$ or $\E{\bar \tau^\mathsf{2}} \to 0$, both $\mathcal{G}^*$ and $\mathcal{I}^*$ will grow without bound. By immediate consequence, if the one-sided asynchrony duration tends to 0 for either party in the service, the maximal throughput will tend to infinity. 

To connect this observation to interdependence, let us invoke the idea behind the telephone wire diagrams of Figures~\ref{telephoneFig} and~\ref{fig:convAll}. If, say, $\rho_\mathsf{1}$ goes to 0, then we expect the triangles on the top half of these plots to vanish, and thus the mean $\mathsf{1}$-paced asynchrony duration would likewise go to 0. Equivalently, if $\rho_\mathsf{2} \to 0$, we would find that $\E{\bar \tau^\mathsf{2}} \to 0$. These arguments can be quickly formalized through the parking function decomposition, as Proposition~\ref{binSidesProp} can be used to imply that 
\editTwo{$
\E{\bar \tau^s}
\leq
{\rho_s}\E{\Delta} /({1 - \rho}) 
$} 
for each side $s \in \{\mathsf{1}, \mathsf{2}\}$, where $\Delta$ is distributed according to the density $\gamma_s(\cdot)$ given in Definition~\ref{pfcDef}.
Then, drawing inspiration from the frameworks of service design, let us suppose that the total average taskload is fixed, but the service platform can control the average amount of tasks per side. Again, by Proposition~\ref{binSidesProp}, this means that we can let $\rho$ be held constant, and  the mean number of customer contributions will be $\rho_\mathsf{1}/(1 - \rho)$, \edit{whereas} the mean number of agent contributions will be $\edit{\rho_\mathsf{2}/(1-\rho) = (\rho - \rho_\mathsf{1})/(1-\rho)}$. 

Hence, a second main \editTwo{service asymmetry} result emerges: as $\rho_\mathsf{1}$ tends to either of its extremes, 0 (meaning agent self-production) or $\rho$ (meaning customer self-production), we find a $\U$-shape of the concurrency-optimized throughput as a function of the interdependence.
In Theorem~\ref{uThm}, we generalize this insight.

\begin{theorem}\label{uThm}
Let $\rho \in (0,1)$ \edit{be fixed}. If $h(\cdot)$ is strictly convex with $h'(0) = 0$, then the concurrency-maximized throughput is $\U$-shaped as a function of $\rho_1$, i.e. 
$$
\frac{
\mathcal{K}(\rho_\mathsf{1})
}{
\E{\tau(\mathcal{K}(\rho_\mathsf{1}))}
}
\in
\mathcal{N}_\U
,
$$
where the concurrency $\mathcal{K}(\rho_\mathsf{1})$ maximizes the throughput at the given $\rho_\mathsf{1}$.
\end{theorem}

\begin{figure}[htb]
\centering
  \begin{minipage}[b]{.38\linewidth}
\includegraphics[width=\linewidth, trim=0cm 0cm 1.25cm 0cm, clip=true]{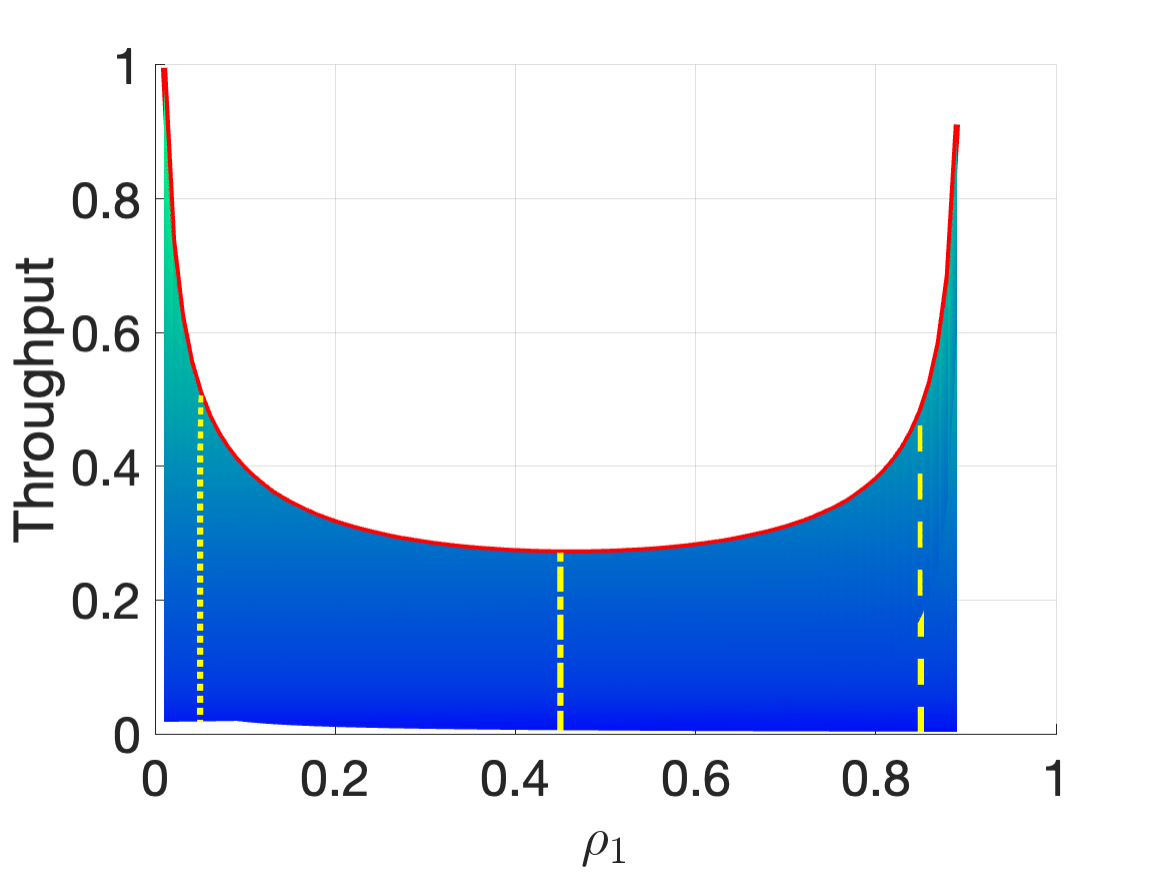}
\includegraphics[width=\linewidth, trim=0cm 0cm 1.25cm 0cm, clip=true]{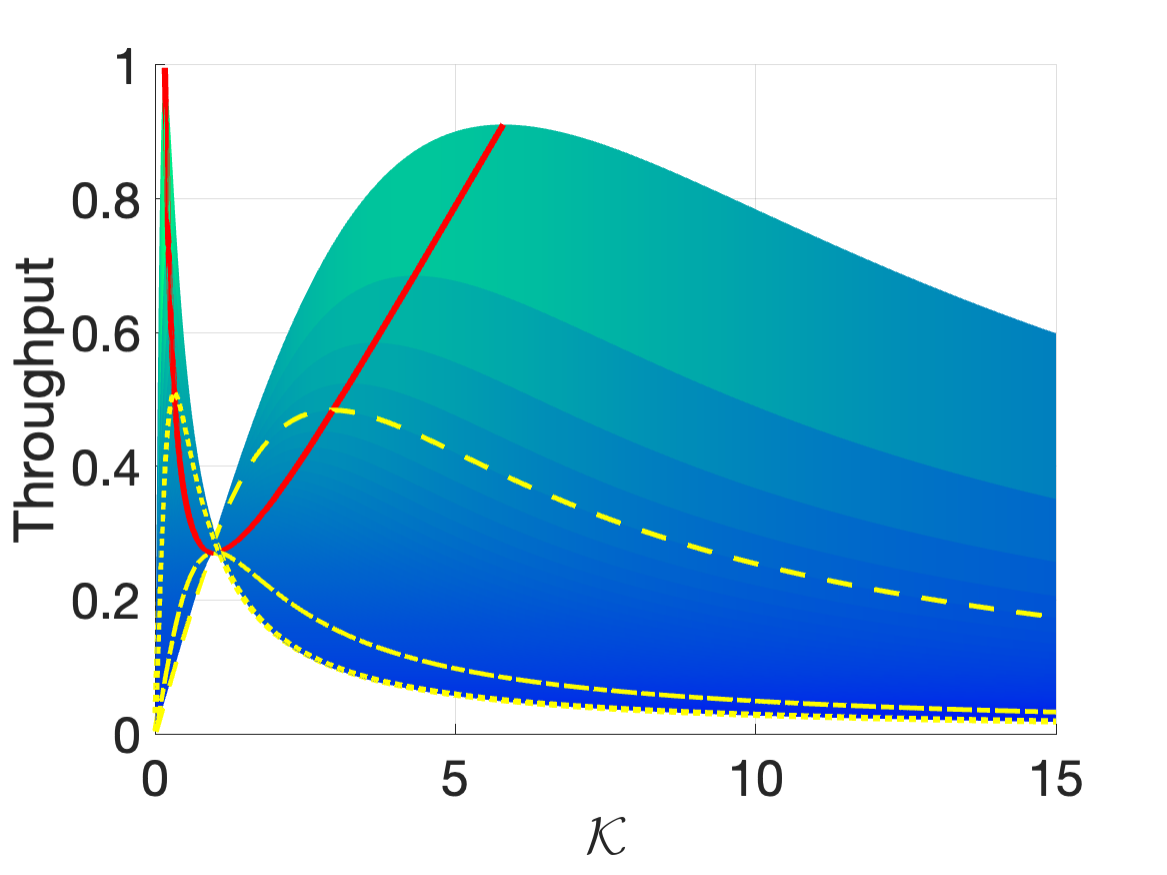}
\end{minipage}
\hfill
\begin{minipage}[t]{.6\linewidth}
\includegraphics[width=\linewidth, trim=0.3cm -1.8cm 0.3cm 0cm, clip=false]{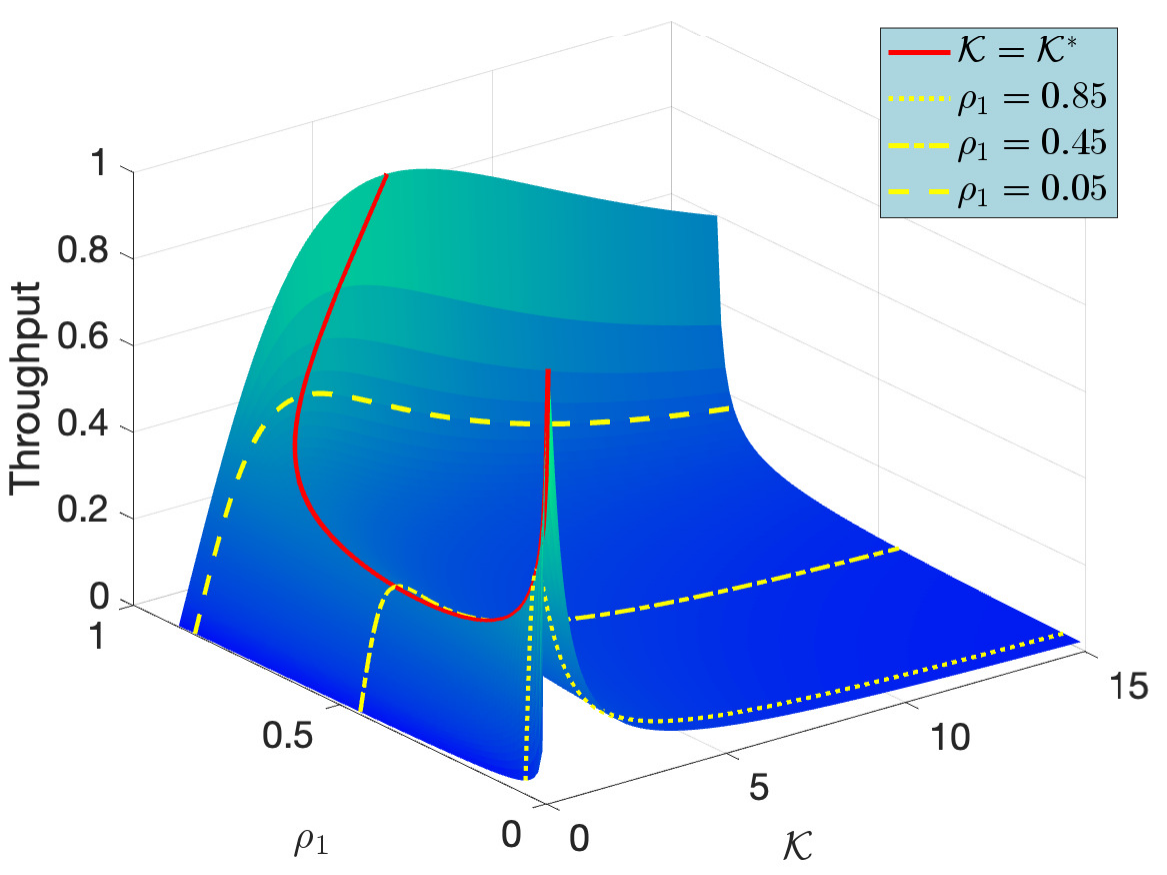}
\end{minipage}%
\caption{Demonstration of the interdependence-driven $\U$-shape along the top of the synchronicity-driven (as a function of the concurrency) $\IU$-shape \edit{(right), with $\rho_\mathsf{1}$- and $\mathcal{K}$-\editTwo{perpendicular} perspectives to emphasize the $\U$ and $\IU$'s, respectively (left)}.
\edit{In each plot, the surface $\mathcal{G}(\cdot)$ is shown for $g_\mathsf{1}(x) = \rho_\mathsf{1} e^{-x}$, $g_\mathsf{2}(x) = (\rho - \rho_\mathsf{1}) e^{-x}$, $h(x) = x^{2.05}$, and $\rho = 0.9$.}}
\label{uiuFig}
\end{figure}

Theorem~\ref{uThm} reveals that, tracing the maximal throughput at the crest of the concurrency-driven, synchronicity-based $\IU$-shape found in Theorem~\ref{iuThm} (and specified for these assumptions in Proposition~\ref{convexProp}) as a function of the interdependence, we find a $\U$-shape along the axis from agent self-production to customer self-production, with co-production occupying the min-max in the middle of this range. We visualize these spectra of asymmetry in  Figure~\ref{uiuFig}.

\edit{Figure~\ref{uiuFig} shows the shape of the throughput (through only the guaranteed service rate, for the sake of representational simplicity) as functions of $\rho_\mathsf{1}$ and $\mathcal{K}$, which respectively capture the interdependence and (as a function of the concurrency) the synchronicity of the interaction. First, because $\rho = 0.9$ is held fixed, $\rho_\mathsf{1} \in (0, 0.9)$ captures the spectrum from agent self-production \editTwo{($\rho_\mathsf{1} = 0$)} to customer self-production \editTwo{($\rho_\mathsf{1} = 0.9$)}. Then, by Step (iv) of Definition~\ref{pfcDef}, the $\Delta$ offset distributions in this example become $\mathsf{Exp}(1)$ for Side $\mathsf{1}$ and $\mathsf{Exp}(1/h(\mathcal{K}))$ for Side $\mathsf{2}$, where $h(\mathcal{K}) = \mathcal{K}^{2.05}$. Hence, at $\mathcal{K} = 1$, the customer and agent are synchronous, as $\mathcal{K}$ grows larger than 1, the interaction tends toward asynchrony, with the pace dictated by the relatively slower agent, and for fractional load $\mathcal{K} < 1$ the interaction also becomes asynchronous, but instead paced by the relatively slower customer. }
\edit{Figure~\ref{uiuFig} contains three separate plots: one ``main'' plot (on the right) that shows the three-dimensional surface from a perspective of both $\rho_\mathsf{1}$ and $\mathcal{K}$, one $\rho_\mathsf{1}$-isolating plot that restricts the view \editTwo{by showing} the throughput \editTwo{surface at a $\mathcal{K}$-perpendicular angle} (on the top left), and one $\mathcal{K}$-isolating plot \editTwo{by showing} the throughout \editTwo{surface at a $\rho_\mathsf{1}$-perpendicular angle} (on the bottom left). To both provide context and emphasize the $\U$ and $\IU$ shapes, each plot in Figure~\ref{uiuFig} also shows the same four lines across the throughput surface. First, there are three yellow contour curves that trace the change with $\mathcal{K}$ for three respectively fixed values of $\rho_1$, each of which can be seen to be $\IU$-shaped through the $\mathcal{K}$-isolating perspective on the bottom left. Then, there is one red line for $\mathcal{K} = \mathcal{K}^*$, which traces along the peaks of the $\IU$'s (as showcased on the bottom left) and is itself $\U$-shaped (as showcased on the top left).\endnote{{For further perspective on the $\U$ and $\IU$ shapes, 
see \url{https://faculty.marshall.usc.edu/Andrew-Daw/asymmetryUIU.html} 
for a rotation of this diagram.}}}

Managerially, these coinciding $\U$ and $\IU$ reveal that asymmetry of interdependence (either customer or agent self-production) is where the operational improvements of optimal concurrency will be most effective. Equivalently said, Theorem~\ref{uThm} shows that it is \emph{symmetry} of interdependence (co-production) where the optimized concurrency offers the least improvement. 
\edit{This phenomenon can be readily observed in Figure~\ref{uiuFig}: the top left plot highlights that the exact nadir of the concurrency-optimized red curve occurs at $\rho_\mathsf{1} = \rho/2 = 0.45$, which is perfect co-production.} 

Figure~\ref{uiuFig} also suggests a robustness of the optimal concurrency for services that rely predominantly on customer contributions ($\rho_\mathsf{1}$ near $\rho$), and a fragility of this same quantity when the service is largely agent driven ($\rho_\mathsf{1}$ near 0). \edit{For example, the dashed yellow contour curve (for $\rho_\mathsf{1} = 0.85$) is relatively flat and its throughout does not change quickly with $\mathcal{K}$, whereas the dotted yellow contour curve (for $\rho_\mathsf{1} = 0.05$) is quite sharp and its throughput drops significantly for $\mathcal{K}$ away from $\mathcal{K}^*$.}\endnote{This also may be more easily observed by rotating Figure~\ref{uiuFig}: 
see \url{https://faculty.marshall.usc.edu/Andrew-Daw/asymmetryK.html}} 
This relative flatness atop the $\IU$ for large optimal concurrency is formalized for the polynomial slowdown function case by Proposition~\ref{polyOptProp}, in which the width of the ``peaking'' range is proportional to $\mathcal{K}^*$ itself.  Hence, Figure~\ref{uiuFig} shows that there is a higher degree of flexibility in setting the agent's concurrency level when the service \editTwo{features} significant customer self-production.




\section{\edit{Discussion and} Conclusion}\label{conclusion}





Following \citet{daw2021co}'s data-driven insight that modeling the service duration simply as a random variable may be too coarse, in this paper, we have analyzed the associated Hawkes cluster model of service. \edit{First, in Part I of the paper}, we have shown that this individual-level model captures both the synchronicity and the interdependence within the customer-agent interaction. This was built upon a \edit{novel temporal} decomposition methodology we establish for Hawkes clusters, in which parking functions are revealed to be hidden spines within these chronologies, encoding the pattern of responses. This \edit{decomposition also} led to asymptotically tight bounds for the mean service duration, which naturally translate to bounds on the agent's throughput. \edit{In Part II of the paper, we} contextualized these individual dynamics in terms of the system-level performance by modeling the synchronicity as a function of the agent's concurrency. In doing so, we obtained general conditions for the throughput to be $\IU$-shaped as a function of the concurrency, and this implies that there exists an optimal concurrency level. Turning to interdependence, we have found that this second dimension of the service interaction actually prescribes the degree of improvement achieved by optimizing the concurrency. \editTwo{Our $\U$-shape} results suggest that services co-produced by the customer and agent will have less room to improve than services characterized as either agent or customer self-production.



\edit{To close our presentation of this service interaction model and its analysis in the main body of this paper, let us  discuss broader implications, extensions, and contributions stemming from it in a series of brief expoundings on particular relevant themes.}

\edit{\textbf{Implications for service operations stochastic modeling:}} Relative to prior stochastic models, we believe that the two most salient differences of our approach are that we have distinguished the roles of the customer and agent from one another and that we have distilled the service interaction into a series of endogenously-driven contributions from these two sides. The former contrasts our approach with the processor sharing service discipline, as the slowdowns in our service model are asymmetric, and thus they are not like the symmetric slowdowns that match well to processor sharing's origins in computation contexts\edit{, yielding what we have deemed a \emph{service-computation contrast}}. Hence, we believe our \edit{micro-level} model and results work well in conjunction with the literature on macroscopic stochastic models of concurrent service systems. In particular, our insights augment and enlighten the ways in which those macroscopic models generalize processor sharing to be limited and state-dependent, as we provide agent-specific conditions that validate the presumed structures. Turning to the micro-level perspective and modeling approach built on the self-exciting cluster, let us note that this framework may also hold value in neighboring arenas, such as shared decision making in healthcare \citep[e.g.,][]{tuncalp2023should}. \edit{Nevertheless, we believe that one of the foremost conclusions from our results is the ability of this interaction framework as a model \emph{for services}.}


\edit{To that end, in the broader context of service modeling, one of this work's more conceptual contributions is that we have shown that the model from \citet{daw2021co} can be used to obtain theoretical insights about service interactions at a micro-level, particularly so in ways that have eluded or otherwise required specific assumptions from the prevailing macro- or system-level approaches to modeling services. Furthermore, we have also provided a methodological framework to analyze these general Hawkes cluster models of the service interactions, which \citet{daw2021co} posed as an important open area. Rather than simply closing one problem, we view this paper as both showing how questions in this area can be addressed and providing a potential blueprint for future service interaction pursuits.}


\edit{\textbf{Interaction modeling beyond Hawkes clusters:} Building from the potential of this micro-level service modeling perspective as discussed above, we can also recognize that there may be paths to future works that model interactions with stochastic models other than the Hawkes. At its most general, the idea of modeling services at the interaction level is to consider a finite point process for the contribution epochs within each customer-agent service exchange. In this paper, as justified by the application to data in \citet{daw2021co}, we have taken this point process to be the history-driven or self-exciting Hawkes process cluster, but it may be of both scholarly interest and practical use to consider either general point processes or other specific variants. Similarly, through the ideas granted by the equivalence in Theorem~\ref{equivThm}, one could also consider \editTwo{oother tree-based models of service interactions built upon other probabilistic combinatorial objects.} }

In light of the relative novelty of the Hawkes cluster model compared to the service models literature, we \edit{also} recognize that it is natural to ask how dependent our \editTwo{system-level} results are on this particular choice of \editTwo{individual-level model}. 
\edit{In general, we are highly interested in analyzing how the Hawkes cluster compares and contrasts with other stochastic models of service. However, it is important to notice that} the confluence of our twin main managerial results --- the conditions for the $\IU$ and $\U$ in Theorems~\ref{iuThm} and~\ref{uThm}, respectively --- followed, perhaps unexpectedly, from the Hawkes cluster. \edit{It remains to be seen if other point processes \editTwo{or other service trees} could produce these results if used within this service interaction modeling framework.}

\edit{\textbf{Connecting interaction-level and system-level models:}}
\edit{Even granted the discussions above on the merits of this micro-level model, we can recognize that it may be quite detailed to construct such intra-service point-process models}, and this specification may feel a bit rich for some macro-level questions. For instance, perhaps one is interested in a broad, offline decision, like setting the staffing level in a concurrent service. In such a case, it could be sufficient to employ one of the bounds from Corollary~\ref{rateBound} to approximate the service rate in a state-dependent Markov chain model of the system. \edit{As we discuss in Section~\ref{hSec} in the Appendix, we believe that the guaranteed service may be a fairly promising approximation for the throughput, and one could simply use $\mathcal{G}(\mathcal{K})$ as the service rate at concurrency state $\mathcal{K}$.} Under the right conditions on $h(\cdot)$, this should still produce the $\IU$-shape (although not necessarily the interdependence $\U$) through the concurrency, and this may be enough for staffing-type questions. \edit{In fact, it is quite possible that embedding these approximations in a Markov chain model is sufficiently detailed for many system-level questions, as the guaranteed service rate does effectively capture the existence of both customer and agent service contributions through the presence of both $\E{\bar{\tau}^\mathsf{1}}$ and $\E{\bar{\tau}^\mathsf{2}}$ (which could perhaps be simply treated as deterministic model parameters if the marked Hawkes clusters are to be entirely avoided, though Appendix~\ref{markedSec} advances the analytical toolkit for those models). Furthermore, the guaranteed service rate applies the slowdown only to the agent side, and we have argued that this asymmetric slowdown is essential for modeling concurrent services. In this way, one can construct more tractable system-level models while preserving some of the core interaction-level insights identified here.} 

One may also wonder whether other models can produce similar results under similar conditions. As we have shown in Section~\ref{symSec}, the symmetric slowdown implied by models employing processor sharing will not be able to yield the $\IU$ without modifying the structure specifically to assume such a shape, and that is because this service discipline does not distinguish customer from agent. By comparison, we have pointed out the queueing model of case managers from \citet{Campello2017} as a notable exception among the macroscopic models, as this Markov chain model features both task completion times that depend on the agent and return times that depend on the customer. Although \citet{Campello2017} did not consider the agent's rate as dependent on their concurrency, it may be possible that introducing this behavior could allow the model to yield non-motonicity under certain conditions. In general, we are quite interested in analyzing how the Hawkes cluster compares and contrasts with other stochastic models of service in future work.


\edit{\textbf{Extending beyond monotonic agent-side slowdown and asymptotic characterizations:} A core research goal of this paper was to identify conditions under which non-monotonic performance could arise from monotonic agent-side slowdown. Though we have maintained this scope within the narrative of the paper, our results do not \editTwo{actually all require that the agent's slowdown is globally monotonic}. For instance, the proof of Theorem~\ref{iuThm} does not rely on Assumption~\ref{conA}'s stipulation that $h(\cdot)$ is universally non-decreasing. In fact, using the asymptotic characterizations of this paper, readers may notice that the proof of Theorem~\ref{iuThm} immediately carries through in the more general setting that $h(\cdot) \in \mathcal{M}$ with $h(0) = 0$.}

\edit{Furthermore, although we have aimed to maintain generality in the model where possible, let us add that it may be possible to produce more precise shape results using similar techniques. For example, assuming a more specific structure for $h(\cdot)$ may be able to yield some of the more complex shapes of performance seen in the literature, such as an N-shape like what is observed by \citet{berry2016past}. In Section~\ref{hSec} of the Appendix, we both demonstrate this numerically and formalize conditions under which this type of non-asymptotic non-monotonicty will occur in this model.}









\edit{\textbf{Possible directions for empirical research:}} As discussed in Section~\ref{litReview}, part of the motivation for this paper has been the myriad empirical works which find real-world dynamics that violate canonical stochastic models of service. This has led us to revisit the underlying assumptions in these queueing models, and, in the spirit of a feedback loop between analytical and empirical work in behavioral operations management \citep{allon2018behavioral}, we are now quite interested in seeing if these results offer any questions worth asking in experiments or data. \edit{For example, though there may be a natural intuition to the $\U$-shape of Theorem~\ref{uThm} --- as once said by Peter Drucker, ``one either meets or one works'' --- to the best of our knowledge, there have not been empirical results that have demonstrated the $\U$-shape of optimized throughput as a function of interdependence. If there are indeed not yet any such results, we are very curious how, or even if, this could be tested. In collecting data directly from practice, it may be challenging to verify that the concurrency has been truly optimized to maximize throughput at each inspected value of interdependence, so it seems that testing such a question would rely at least partially on lab experiments.} 

\edit{More generally}, we hope that our work can offer greater understanding of when multitasking may create non-monotonicity in practice, as our theoretical results could provide guidelines for when one would expect such shapes to be reproduced. \edit{Because our system-level results are all essentially dependent on $h(\cdot)$, these findings imply that, in some sense, to understand the shape of overall system performance, it may be sufficient for data to be collected simply on the agent in a vacuum. To that end, these agent-isolating insights may be of help in efforts to translate service data into models, which \citet{ding2024express} has recently identified as potentially ripe for faulty generalizations.}

\edit{\textbf{Possible directions for modeling next-generation services:}} Of course, present technological advances may add a wrinkle to one of the background premises of this paper, in that services are interactions between two \emph{people}. Indeed, many firms are already facing the  question of if (and how) artificial intelligence (AI) could be incorporated in the design of their service delivery, possibly in place of a human agent \citep{dada2023omnichannel}. However, given that the roles of customer and agent would remain distinct even if one side is assisted or outright conducted by AI, our modeling approach should still apply. In fact, it would be interesting to study how the response and slowdown functions differ between agents who are people and those that are AI, and to contrast the resulting system performance. Given that the relevant empirical literature is nascent but undoubtedly growing, we look forward to pursuing this subject in future work. 






\edit{\textbf{Possible directions in stochastic modeling methodology:}}
Finally, to close on a methodological note, let us remark that Theorem~\ref{equivThm}'s equivalence of Definitions~\ref{hcDef} and~\ref{pfcDef} may well hold independent interest for Hawkes process modeling and analysis. It may be possible to generalize this parking-function-based cluster to construct alternate definitions of more general Hawkes models. For example, the \citet{hawkes1974cluster} decomposition does extend to Hawkes processes with higher dimensions of self- and mutual-excitement, and such an equivalence result could be of use in analyzing many-sided generalizations of the interaction model. We do, in fact, prove an equivalence result beyond what is stated here in the main body of the paper; in  \edit{Appendix~\ref{markedSec}}, we establish an alternate definition for marked Hawkes clusters, which requires a generalization from uniformly random parking functions to a weighted distribution over Dyck paths. This supports our analysis of $\bar \tau^\mathsf{1}$ and $\bar \tau^\mathsf{2}$ in the study of the spectrum of interdependence. \edit{As briefly discussed in Remark~\ref{agreeRemark}, we have also shown in Appendix~\ref{agreeSec} that the parking functions used in this temporal decomposition are provably distinct at the sample path level from those identified by \citet{daw2023conditional} through compensator-based transformations. } We look forward to trying to advance this decomposition methodology for a broader family of history-driven stochastic models in future research.



\section*{Acknowledgements}

The authors are grateful for the generous support of this work by the National Science Foundation Division of Civil, Mechanical and Manufacturing Innovation  [Award \#2441387] (A. Daw) and the United States-Israel Binational Science Foundation [Award \#2022095] (A. Daw, G. B. Yom-Tov).

\theendnotes

\bibliographystyle{informs2014} 
\bibliography{Bibliography.bib}

\newpage

\setcounter{page}{1}

\edit{
\begin{center}
\begin{large}
\textit{Electronic Companion to}\\
\textsf{Asymmetries of Service: Interdependence and Synchronicity}\\
\end{large}
Andrew Daw and Galit Yom-Tov
\end{center}
}

\begin{APPENDIX}{} 

This appendix contains the proofs of the statements in the main body, along with supporting technical results and auxiliary numerical demonstrations. In Section~\ref{smProofs}, we provide the proofs for the stochastic modeling and methodology results presented in \edit{Part I of the paper (Sections~\ref{sec:model} and~\ref{clusterSec})}. Then, Section~\ref{sbProofs} contains the proofs of the results regarding the service dynamics and behavior presented in \edit{Part II (Sections~\ref{nonMonSec},~\ref{symSec}, and~\ref{interSec})}. Next, in Section~\ref{markedSec}, we provide additional technical results driving the analysis and decomposition of marked Hawkes cluster models, which support the results and proofs of the main body. \edit{Section~\ref{agreeSec} is then focused on the parking functions hidden within Hawkes clusters, specifically on the distinction between the temporal and compensator-based concepts as claimed in Remark~\ref{agreeRemark}. In Section~\ref{hSec}, we show how more specific assumptions on $h(\cdot)$ can produce similarly specific non-monotonic throughput shapes, such as an N-shape.} Finally, in Section~\ref{numerSec}, we conduct simulation experiments that demonstrate our main analytical results and suggest that they extend to a more general family of models.

\section{Proofs of Parking Function Decomposition and Stochastic Model Dynamics}\label{smProofs}

\edit{As fitting for Part I of the paper, this first section of proofs focuses on understanding the individual-level interaction model and elucidating the ways that the interaction is shaped by the parameters for interdependence and synchronicity. The core methodological idea here (which thus supports the rest of the paper) is the shift from the intensity-based Definition~\ref{hcDef} to the end-state conditioning and combinatorial interpretations given by Definition~\ref{pfcDef}. This equivalence is established in Theorem~\ref{equivThm}, which itself draws upon analytical tools for marked, univariate Hawkes clusters, which are developed in \editTwo{Appendix~\ref{markedSec}}.}

\subsection{\editTwo{Discussion and} Proof of Proposition~\ref{stabProp}}

\proof{Proof.}
By recognizing that the response process is equivalent to a branching process with Poisson offspring distribution, we have that $N \sim \mathsf{Borel}(\rho)$, and, moreover, $\PP{N < \infty} = 1$ for all $\rho \leq 1$ \citep[e.g., Ex.~6.1.10 in][]{roch2015modern}. By, \edit{e.g.,} \citet{haight1960borel,daly2019borel}, the mean number of contribution points is $\E{N} = 1/(1-\rho)$, which would imply $\E{\tau} = \infty$ if $\rho = 1$. Hence, we must have $\rho < 1$. 

\editTwo{With} the added assumption that the response offsets each have finite mean, say $\E{S^\mathsf{1}} \edit{= \int_0^\infty t g_\mathsf{1}(t) \mathrm{d}t / \rho_\mathsf{1}}$ and $\E{S^\mathsf{2}} \edit{= \int_0^\infty t g_\mathsf{2}(t) \mathrm{d}t / (\eta \rho_\mathsf{2})}$, \editTwo{we can now show that $\E{\tau} < \infty$.} 
\editTwo{Starting with $\tau$ expressed as a telescoping sum of inter-contribution-point gaps, $\tau = \tau_{N-1} = \sum_{i=1}^{N-1} \tau_i - \tau_{i-1}$, we can first obtain an almost sure upper bound on $\tau$ in the style of the ``total birth time'' bound used in \citet[Prop. 3]{chen2021perfect}. In the context of this proof only (but in foreshadowing of results to follow), let $\pi(i) \in \{0, \dots, i - 1\}$ be defined for each $i \in \{1, \dots, N-1\}$ as the prior point to which point $i-1$ responds. Immediately, we have that $\tau_i - \tau_{i-1} \leq \tau_i - \tau_{\pi(i)}$ for each $i$. By the \citet{hawkes1974cluster} decomposition of Hawkes clusters, we also know that each $\tau_i - \tau_{\pi(i)}$ is equivalent to an arrival time of a non-stationary Poisson process with time-varying rate $g_\mathsf{1}(u) + \eta g_\mathsf{2}(\eta u)$. Moreover, from the commutativity of the sum, we are agnostic to the chronological sequencing of these gaps, and we can thus leverage the fact that the unordered collection of gap random variables is equivalent to $N-1$ mutually independent samples from the density $(g_\mathsf{1}(\cdot) + \eta g_\mathsf{2}(\eta ~\cdot))/\rho$. Hence, together with Wald's identity, we have
\begin{align}
\E{\tau}
\leq
\E{\sum_{i=1}^{N-1} \tau_i - \tau_{\pi(i)}}
=
\E{N - 1}\left(\frac{\rho_\mathsf{1}}{\rho}\E{S^\mathsf{1}} + \frac{\rho_\mathsf{2}}{\rho}\E{S^\mathsf{2}}\right)
=
\frac{\rho_\mathsf{1}\E{S^\mathsf{1}} + \rho_\mathsf{2}\E{S^\mathsf{2}}}{1-\rho}
.
\label{birthBound}
\end{align}
Thus, if $\E{S^\mathsf{1}}$ and $\E{S^\mathsf{2}}$ are both finite, then we have that $\E{\tau} < \infty$ as well.}

\editTwo{To complete the proof and show that the finiteness of $\E{S^\mathsf{1}}$ and $\E{S^\mathsf{2}}$ is not only sufficient but also necessary, one can simply recognize that $\E{\tau} \geq \E{\tau_1 \mid N > 1} \PP{N> 1}$. Through cluster decomposition arguments analogous to those above, we can express this lower bound as
$$
\E{\tau}
\geq
\E{\tau_1 \mid N > 1} \PP{N > 1}
=
\left(\frac{\rho_\mathsf{1}}{\rho}\E{S^\mathsf{1}} + \frac{\rho_\mathsf{2}}{\rho}\E{S^\mathsf{2}}\right)
\left(
1 - e^{-\rho}
\right)
,
$$
and thus if either $\E{S^\mathsf{1}}$ and $\E{S^\mathsf{2}}$ are not finite, then $\E{\tau}$ will not be either.}
\hfill\Halmos\endproof

\editTwo{Let us note that Assumption~\ref{stabAssum} aligns with well-known general Hawkes point process stability conditions \citep{bremaud1996stability}, but here we also leverage the within-cluster setting to further establish that the mean duration is finite and to state the condition as both necessary and sufficient (whereas the typical phrasing is just that $\rho < 1$ is sufficient). 
Additionally, let us comment that, though it has been valuable analytically, the bound in Equation~\eqref{birthBound} is not particularly tight. Based on evaluating the expression numerically and comparing to simulations, this upper bound appears to be well-above the mean, which makes sense given that it is based on an almost sure bound on $\tau$. It is also important that we note that the expression in \eqref{birthBound} is agnostic to $\eta$, which showcases the need for our analysis to follow.}

\subsection{\editTwo{Discussion and} Proof of Theorem~\ref{equivThm}}

\proof{Proof.}
We will prove this in the order of the steps in Definition~\ref{pfcDef}, \edit{in} which each successive step will condition on all those prior. Naturally, the proof of (i) starts with the cluster-perspective of the Hawkes that originates in \citet{hawkes1974cluster} \citep[one can see, e.g., Definition 2.2 or Definition 2 in][respectively, for contemporary phrasings]{koops2017infinite,chen2021perfect}. \editTwo{This decomposition was the basis of the interpretation of the model in Remark~\ref{hoRemark}.} Because this perspective yields that each point has its own independent descendant stream (across both sides) according to a Poisson process with time-varying rate $g_\mathsf{1}(u) + \eta g_\mathsf{2}(\eta u)$ for $u$ time elapsed since its own occurrence, it is immediately clear that $N \sim \mathsf{Borel}(\rho)$. Hence, we proceed to (ii).

Following Lemma~\ref{dyckLemma} with $m_i=1$ for every $i \in \{0, \dots, N-1\}$, we have that the number of descendant points (or follow-up contributions in the service model language) in the Hawkes cluster model is equivalent to the vector $\left(\kappa_1(\vec{\pi}), \kappa_2(\vec{\pi}), \dots, \kappa_{N-1}(\vec{\pi})\right)$, where $\vec{\pi}$ is an $(N-1)$-step Dyck path randomly selected according to the distribution
\begin{align*}
\PP{ \vec{\pi} = d}
=
\frac{
\prod_{i=1}^{N-1} \frac{1}{\kappa_i(d)!}
}{
\sum_{\delta \in \mathsf{D}_n}
\prod_{j=1}^{N-1}
\frac{
1
}{
\kappa_{j}(\delta)!
}
}
.
\end{align*}
Let us now recognize that $\vec \pi$ is precisely a uniformly random parking function sorted increasingly. Clearly, conditioned on the same length, the sample spaces of Dyck paths and of sorted parking functions are the same: both collections contain all the vectors in $\mathbb{Z}_+^{N-1}$ which meet the requirement that each of their elements is no larger than its index \citep[e.g.,][]{yan2015parking}. To see that the unsorted $\vec \pi$ is indeed uniformly random among all parking functions, we simply need to observe that there are exactly $\prod_{i=1}^{N-1} {\kappa_i(d)!}$ ways to de-order \edit{any} Dyck path $d$ (and each of these is a parking function).

Moving to (iii), here we can petition directly to the \citet{hawkes1974cluster} perspective and the independent, non-stationary Poisson streams that it reveals. That is, recall that each Hawkes cluster point has its own offspring stream with time-varying rate $g_\mathsf{1}(u) + \eta g_\mathsf{2}(\eta u)$ for $u$ time since it occurred, and this implies that a given point will generate a $\mathsf{Pois}(\rho_\mathsf{1})$-distributed number Side $\mathsf{1}$ descendants and a $\mathsf{Pois}(\rho_\mathsf{2})$ number on Side $\mathsf{2}$. By Poisson thinning, the probability that a given descendent is on Side $\mathsf{1}$ is $\rho_\mathsf{1}/\rho$. Furthermore, because the offspring streams do not depend on the parent type, thinning further implies that the side outcomes are mutually independent and identically distributed across points.

Similarly, for (iv), we can turn to the generalization of conditional uniformity for non-stationary Poisson processes, while still leveraging the independent Poisson streams of the \citet{hawkes1974cluster} perspective.
Because the side $s_{\editTwo{\ell}} \in \{\mathsf{1},\mathsf{2}\}$ Poisson stream from point $\ell$ will follow rate $\gamma_{s_\ell}(u)$, conditional on there being $m$ such follow-up points, the offset times have joint distribution equivalent to that of the order statistics for $m$ independent samples from the cumulative distribution function $\PP{\Delta \leq x} = \int_0^x \gamma_{s_\ell}(u) \mathrm{d}u / \rho_{s_\ell}$ \citep[see, e.g.,][section 2.1]{daley2003introduction}. 
Thus, by ignoring the order of the responses, each offset in the Hawkes cluster is independently drawn according to the density given in Definition~\ref{pfcDef}.

To conclude the proof of equivalence of the clusters, we can recognize that by Definition~\ref{hcDef}, the Hawkes cluster points are indexed in increasing order of time, and this is precisely what is implied in the conclusion of Definition~\ref{pfcDef}. Hence, by analyzing the output of Definition~\ref{hcDef}, we have arrived at the output of Definition~\ref{pfcDef}, confirming their equivalence, as in each case, the collection of epochs fully characterizes the cluster.
\hfill\Halmos
\endproof

\editTwo{As can be seen in the proof above, steps (ii) and (iii) essentially exist in parallel in Definition~\ref{pfcDef}, and the order of the two could be switched without changing the result. In fact, step (iii) isn't even fully needed. One could compress it into step (iv) and draw according to the density $(\gamma_1(\cdot) + \gamma_2(\cdot))/\rho$, but keeping the sides distinct will be valuable in the analysis of asymmetry to follow. \edit{Given any of these implementations, another benefit of Theorem~\ref{equivThm} is that the steps of Definition~\ref{pfcDef} can now also be seen to be a straightforward simulation procedure for the general Hawkes clusters of Definition~\ref{hcDef}.}}

\subsection{Proof of Proposition~\ref{binSidesProp}}

\proof{Proof.}
Following Theorem~\ref{equivThm}, we can establish this through use of Definition~\ref{pfcDef}, specifically through (iii). Given that $N = n+1$, the placement of each contribution on Side $\mathsf{1}$ or $\mathsf{2}$ is an independent $\mathsf{Bern}(\rho_\mathsf{1}/\rho)$ variable, and this immediately yields the stated result.
\hfill\Halmos\endproof

\subsection{Proof of Theorem~\ref{convThm}}

\proof{Proof.}
For $s \in \{\mathsf{1},\mathsf{2}\}$ and $x \geq 0$, let $G_s(x) = \int_0^x g_s(u)\mathrm{d}u$. Then, the $\Delta_i$ offset random variables given for the contribution times in Definition~\ref{pfcDef} have cumulative distribution function $G_{1}(x) / \rho_\mathsf{1}$ if $s_i = 1$ and $G_2(\eta x) / \rho_\mathsf{2}$ if $s_i = 0$. Using the idea of inverse transform sampling, we can recognize that there exists an i.i.d. sequence of standard uniform random variables $U_1, \dots, U_{N-1}$, such that 
\begin{align}
\Delta_i &= 
\begin{cases}
G_{\mathsf{1}}^{-1}(\rho_\mathsf{1} U_i) & \text{ if $s_i = 1$,}\\ 
\frac{1}{\eta}G_\mathsf{2}^{-1}(\rho_\mathsf{2} U_i)& \text{ if $s_i = 0$.}
\end{cases}
\label{deltaDefEq}
\end{align}
Also through Definition~\ref{pfcDef}, we can recognize that because the duration of the service is given by the maximum of the contribution times, this can be written as the maximum of the epochs before they are sorted. That is, 
\begin{align}
\tau
&=
\max_{0 \leq i \leq N-1} T_i
=
\bigvee_{1 \leq i \leq N-1} \left(T_{\pi_{(i)}-1} + \Delta_i\right)
\label{tauDecompEq}
.
\end{align}
By convention, let $\bigvee$ over an empty set be equal to 0. Now, let us observe that neither the parking function $\pi$ nor the cluster size $N$ have any dependence on the synchronicity parameter $\eta$; only the offset variables do. Furthermore, $\Delta_i$ depends on $\eta$ only if $s_i = 0$. Hence, through the parking function construction of the cluster, we can see that, outside of these timing variables, the response structure and the sides of each contribution will be unchanged by changes to $\eta$.

Let us first consider the limit as $\eta \to \infty$. By Equation~\eqref{deltaDefEq}, if $s_i = 1$, then $\Delta_i$ will remain constant throughout the limit, whereas if $s_i = 0$ then $\Delta_i \to 0$ almost surely. By consequence, if $s_i = 0$, $T_i \to T_{\pi_{(i)}-1}$ almost surely as $\eta \to \infty$. However, this does not mean that the Side $\mathsf{2}$ contributions vanish. Because all Side $\mathsf{2}$ contributions are instantaneous in the limit, we can recognize each Side $\mathsf{1}$ contribution incurs a full progeny exclusively composed of Side $\mathsf{2}$ points. More formally, let us recognize that the $T_{\pi_{(i)}-1}$ to which $T_i$ for $s_i = 0$ converges may also be a Side $\mathsf{2}$ contribution point, and, if so, the same can be said for the point to which $T_{\pi_{(i)}-1}$ converges. Recalling the \citet{hawkes1974cluster} branching structure used in Definition~\ref{pfcDef}, we have that the total number of Side $\mathsf{2}$ descendants for each Side $\mathsf{1}$ \edit{ancestor} is $\mathsf{Borel}(\rho_\mathsf{2})$ distributed. Hence, this instantaneous collection of a  $\mathsf{Borel}(\rho_\mathsf{2})$ number of points becomes a random mark at the given Side $\mathsf{1}$ point. Because Equation~\eqref{tauDecompEq} specifies $\tau$ as the maximum over these times, we thus find the  duration under $\mathsf{1}$-paced asynchrony in the limit: $\tau \to \bar \tau^\mathsf{1}$ almost surely as $\eta \to \infty$. 

Then, the limit of $\eta\tau$ as $\eta \to 0$ will follow by similar arguments. Upon passing the $\eta$ coefficient through the maximum, we can see that, by comparison to the limit as $\eta \to \infty$,  it is the $s_i = 0$ case that remains constant. If $s_i = 0$, $\eta \Delta_i$ will be equal to $G_\mathsf{2}^{-1}(\rho_\mathsf{2}U_i)$ for any $\eta$. On the other hand, if $s_i = 1$, we now find that $\eta \Delta_i \to 0$ almost surely as $\eta \to 0$. Hence, if $s_i = 1$, $\eta T_i$ collapses down to $\eta T_{\pi_{(i)}-1}$, which itself converges down to the point from which it descended if $s_{\pi_{(i)}-1} = 1$, and so on. \edit{Thus}, only the Side $\mathsf{2}$ offsets remain in the limit, and by analogous arguments to the previous paragraph, we have that $\eta \tau \to \bar \tau^\mathsf{2}$ almost surely as $\eta \to 0$.
\hfill\Halmos\endproof

\subsection{\editTwo{Discussion and} Proof of Proposition~\ref{boundProp}}

As the technique in the following proof suggests, we believe this pair of bounds could be elevated to almost sure, if so desired, through careful construction of a shared probability space. However, as aligned with the goals of this paper, we restrict our attention to the mean bounds for the sake of simplicity.

\proof{Proof.}
Let us begin by proving the lower bound, which we show as two individual inequalities. From Equation~\eqref{tauDecompEq} in the proof of Theorem~\ref{convThm}, let us recall that 
$
\tau
=
\bigvee_{1 \leq i \leq N-1} T_i$ where $T_i = T_{\pi_{(i)}-1} + \Delta_i$. Because each $\Delta_i$ is an independent, positive random variable as defined in Equation~\eqref{deltaDefEq}, we can see that with probability 1 for every $i$, $\Delta_i \geq \Delta_i'$ where $\Delta_i' = G^{-1}_\mathsf{1}(\rho_\mathsf{1} U_i) \mathbf{1}\{s_i = 1\}$. Lower bounding the mean duration by $\E{\bar \tau^\mathsf{1}}$ follows immediately. For the bound by $\E{\bar \tau^\mathsf{2}}/ \eta$, let us define $\Delta_i'' = \frac{1}{\eta}G^{-1}_\mathsf{2}(\rho_\mathsf{2} U_i) \mathbf{1}\{s_i = 0\}$, and by analogous arguments we can see that $\Delta_i \geq \Delta_i''$ with probability 1 for every $i$.



Now, for the upper bound on the mean duration, let us notice that $\Delta_i = \Delta_i' + \Delta_i''$ for every $i$. Drawing inspiration from this, let us introduce an alternate sequence of times and offsets that runs through the parking function spine twice; first over the Side $\mathsf{1}$ offsets and then the Side $\mathsf{2}$'s. That is, let us define $\tilde T_0, \tilde T_1, \dots, \tilde T_{2N-2}$ as follows. For $i \in \{1, \dots, N-1\}$, let $\tilde T_i = \tilde T_{\pi_{(i)}-1} + \Delta_i'$, just like in the $\bar \tau^\mathsf{1}$ construction. Then, for $i \in \{N, \dots, 2N-2\}$, let $\tilde T_i = \tilde T_{\pi_{(i-N+1)}+N-1} + \Delta_{i-N+1}''$. By construction, this restarts the descendant sequence after the conclusion of the first and instead uses the Side $\mathsf{2}$ offsets on this second pass. Let $\tilde \tau = \bigvee_{1\leq i\leq 2N-2}\tilde T_i$. This structure commutes: the analogous construction starting with Side $\mathsf{2}$ instead of Side $\mathsf{1}$ will yield the same $\tilde \tau$. From this symmetry, we see that $\tau \leq \tilde \tau$ almost surely, as the true final offset must occur at least as late in (at least one of the two orderings of) the repeated construction as it does in the original. Finally, we can recognize that because this repeated construction iterates through one side and then through the other, we have $\tilde \tau = \bar \tau^\mathsf{1} + \bar \tau^\mathsf{2}$, and this completes the proof.
\hfill\Halmos\endproof

\section{Proofs of System Behavior and Phenomena in Service Asymmetries}\label{sbProofs}

\edit{Following the thematic expansion in Part II of the paper, this second section of the appendix is concerned with the proofs of the connection from the individual-level interaction model to the system-level performance metrics. Thus, by comparison to the proofs in Appendix~\ref{smProofs}, the proofs in this section all take the concurrency dependence of the synchronicity for granted as posed in Assumption~\ref{conA}. On top of those conditions, the proofs of Proposition~\ref{convexProp},~\ref{polyOptProp},~\ref{symProp1}, and~\ref{symProp2} and Theorem~\ref{uThm} additionally invoke the existence of first and second derivatives of $h(\cdot)$ via Assumption~\ref{twiceAssump} (whereas Assumption~\ref{conA} need not imply differentiability, and the proofs of Theorem~\ref{iuThm} and Corollaries~\ref{monoCor} and~\ref{noOptKcor} rely only upon Assumption~\ref{conA}).}

\subsection{Proof of Theorem~\ref{iuThm}}

\proof{Proof.}
From Definition~\ref{uDef}, to show that the existence of the $\IU$-shape, we must show that the throughput tends to 0 as $\mathcal{K}\to 0$ and as $\mathcal{K} \to \infty$. Before proceeding with either $h(x)/x$ case, let us observe that the guaranteed and idealized service rates both go to 0 as $\mathcal{K} \to 0$. That is, regardless of the behavior of the limit of $h(\mathcal{K})/\mathcal{K}$ as $\mathcal{K} \to 0$, we know that $\E{\bar \tau^\mathsf{1}} / \mathcal{K} \to \infty$ as $\mathcal{K} \to 0$. Hence, $\lim_{\mathcal{K}\to 0} \mathcal{G}(\mathcal{K}) = \lim_{\mathcal{K}\to 0} \mathcal{I}(\mathcal{K}) = 0$ \edit{because}
$$
\edit{\lim_{\mathcal{K} \to 0} 
\frac{\mathcal{K}}{\E{\bar \tau^\mathsf{1}} + \E{\bar \tau^\mathsf{2}} h(\mathcal{K})}
=
\lim_{\mathcal{K} \to 0} 
\left(
\frac{\mathcal{K}}{\E{\bar \tau^\mathsf{1}}}
\wedge
\frac{\mathcal{K}}{\E{\bar \tau^\mathsf{2}} h(\mathcal{K})}
\right)
=
0
,}
$$
and by Corollary~\ref{rateBound} we have that $\lim_{\mathcal{K}\to 0} \mathcal{K}/\E{\tau(\mathcal{K})} = 0$ \edit{accordingly}. We are thus left to consider the limit as the concurrency grows large.

Let us start with the case that $h(x)/x \to \infty$.  Given this assumption, we can see that both
$$
\edit{\lim_{\mathcal{K} \to \infty}}
\frac{\E{\bar \tau^\mathsf{1}}}{\mathcal{K}} + \frac{\E{\bar \tau^\mathsf{2}} {h(\mathcal{K})}}{\mathcal{K}}
\edit{=}
\edit{\lim_{\mathcal{K} \to \infty}}
\left(
\frac{\E{\bar \tau^\mathsf{1}} }{\mathcal{K}} \vee \frac{\E{\bar \tau^\mathsf{2}} h(\mathcal{K})}{\mathcal{K}}
\right)
\edit{=} \edit{\infty}
,
$$
and thus $\lim_{\mathcal{K}\to \infty} \mathcal{G}(\mathcal{K}) = \lim_{\mathcal{K}\to \infty} \mathcal{I}(\mathcal{K}) = 0$, completing the proof of the $\IU$-shape in this case. Turning to the case that $h(x)/x \to c$, we can quickly find through analogous arguments that both 
$$
\edit{\lim_{\mathcal{K} \to \infty}}
\frac{\E{\bar \tau^\mathsf{1}}}{\mathcal{K}} + \frac{\E{\bar \tau^\mathsf{2}} {h(\mathcal{K})}}{\mathcal{K}}
\edit{=}
\edit{\lim_{\mathcal{K} \to \infty}}
\left(
\frac{\E{\bar \tau^\mathsf{1}} }{\mathcal{K}} \vee \frac{\E{\bar \tau^\mathsf{2}} h(\mathcal{K})}{\mathcal{K}}
\right)
\edit{=} \edit{c\E{\bar \tau^\mathsf{2}}}
,
$$ and hence the $\IU$ does not exist in this case.
\hfill\Halmos
\endproof

\subsection{\edit{Proof of Corollary~\ref{monoCor}}}

\proof{Proof.}
\edit{To show that the throughput belongs to $\mathcal{M}$, we must show that it tends to 0 at one extreme of its domain and towards infinity at the other. As done in the proof of Theorem~\ref{iuThm}, we will prove this by showing that $\mathcal{G}(\mathcal{K})$ and $\mathcal{I}(\mathcal{K})$ each satisfy these extremal conditions. First, as $\mathcal{K}$ shrinks to 0, we can quickly see that
$$
\lim_{\mathcal{K} \to 0} 
\frac{\mathcal{K}}{\E{\bar \tau^\mathsf{1}} + \E{\bar \tau^\mathsf{2}} h(\mathcal{K})}
=
\lim_{\mathcal{K} \to 0} 
\left(
\frac{\mathcal{K}}{\E{\bar \tau^\mathsf{1}}}
\wedge
\frac{\mathcal{K}}{\E{\bar \tau^\mathsf{2}} h(\mathcal{K})}
\right)
=
0
,
$$
and, as seen in Theorem~\ref{iuThm}, this simply relies on the fact that $h(x) \to 0$ as $x \to 0$. Then, at the other extreme, we find that 
$$
\lim_{\mathcal{K} \to \infty} 
\frac{\mathcal{K}}{\E{\bar \tau^\mathsf{1}} + \E{\bar \tau^\mathsf{2}} h(\mathcal{K})}
=
\lim_{\mathcal{K} \to \infty} 
\left(
\frac{\mathcal{K}}{\E{\bar \tau^\mathsf{1}}}
\wedge
\frac{\mathcal{K}}{\E{\bar \tau^\mathsf{2}} h(\mathcal{K})}
\right)
=
\infty
,
$$
which now follows from the assumption that $h(x)/x \to 0$ as $x \to \infty$.}
\hfill\Halmos\endproof

\subsection{\editTwo{Discussion and} Proof of Corollary~\ref{noOptKcor}}

\proof{Proof.}
\edit{To prove this result, we will show that both the guaranteed and idealized service rates converge up to the same asymptotic value. Specifically, we will show that $\lim_{\mathcal{K} \to \infty} \mathcal{G}(\mathcal{K}) = \lim_{\mathcal{K} \to \infty} \mathcal{I}(\mathcal{K}) = 1/(c\E{\bar \tau^\mathsf{2}})$ and that $\mathcal{I}(\mathcal{K})~\editTwo{<}~1/(c\E{\bar \tau^\mathsf{2}})$ for all finite $\mathcal{K}$; the latter will immediately imply that $\mathcal{G}(\mathcal{K})~\editTwo{<}~1/(c\E{\bar \tau^\mathsf{2}})$ for all $\mathcal{K}$ as well. By consequence of these claims, we will have that the throughput itself must converge up to the same value as the concurrency tends to infinity.}

\edit{For the limits, we can readily observe that the guaranteed service rate converges to
$$
\lim_{\mathcal{K} \to \infty} \mathcal{G}(\mathcal{K})
=
\lim_{\mathcal{K} \to \infty} \frac{1}{\E{\bar \tau^\mathsf{1}}/\mathcal{K} + \E{\bar \tau^\mathsf{2}} h(\mathcal{K})/\mathcal{K}}
=
\frac{1}{c \E{\bar \tau^\mathsf{2}}}
,
$$
and, similarly for the idealized service rate, we also find that
$$
\lim_{\mathcal{K} \to \infty}
\mathcal{I}(\mathcal{K})
=
\lim_{\mathcal{K} \to \infty}
\left(
\frac{\mathcal{K}}{\E{\bar \tau^\mathsf{1}}}
\wedge
\frac{\mathcal{K}}{\E{\bar \tau^\mathsf{2}} h(\mathcal{K})}
\right)
=
\lim_{\mathcal{K} \to \infty}
\frac{\mathcal{K}}{\E{\bar \tau^\mathsf{2}} h(\mathcal{K})}
=
\frac{1}{c \E{\bar \tau^\mathsf{2}}}
.
$$
Now, let us show that $\mathcal{I}(\mathcal{K})~\editTwo{<}~1/(c\E{\bar \tau^\mathsf{2}})$. For this, let us analyze the two terms within the minimum operator (as written directly above) separately, starting with $\mathcal{K}/(\E{\bar{\tau}^\mathsf{2}} h(\mathcal{K}))$. Because we have that $h(x) \slash x~\editTwo{>}~c$ for all $x$ such that $h(x) \geq \E{\bar \tau^\mathsf{1}} \slash \E{\bar \tau^\mathsf{2}}$, we thus have
$$
\frac{\mathcal{K}}{\E{\bar \tau^\mathsf{2}} h(\mathcal{K})}
\mathbf{1}\left\{ 
h(\mathcal{K}) \geq \frac{\E{\bar \tau^\mathsf{1}}}{\E{\bar \tau^\mathsf{2}}}
\right\}
~\editTwo{<}~
\frac{1}{c \E{\bar \tau^\mathsf{2}} }
\mathbf{1}\left\{ 
h(\mathcal{K}) \geq \frac{\E{\bar \tau^\mathsf{1}}}{\E{\bar \tau^\mathsf{2}}}
\right\}
,
$$
and thus $1/(c\E{\bar \tau^\mathsf{2}})$ \editTwo{strictly} upper bounds $\mathcal{I}(\mathcal{K})$ when $h(\mathcal{K}) \geq \E{\bar \tau^\mathsf{1}} \slash \E{\bar \tau^\mathsf{2}}$. Then, let us further observe that by the continuity of the minimum operator, we also can bound
$$
\sup_{\mathcal{K} : h(\mathcal{K}) < \frac{\E{\bar \tau^\mathsf{1}}}{\E{\bar \tau^\mathsf{2}}}} \mathcal{I}(\mathcal{K})
\leq
\frac{\tilde{\mathcal{K}}}{\E{\bar \tau^\mathsf{2}} h(\tilde{\mathcal{K}})}
,
$$
where $\tilde{\mathcal{K}}$ is the switch point such that $\mathcal{K} = \inf\{\mathcal{K} \geq 0 ~|~ h(\mathcal{K}) \geq \E{\bar \tau^\mathsf{1}}/\E{\bar \tau^\mathsf{2}}\}$. Then, again by the fact that $h(x) \slash x~\editTwo{>}~c$ for all $x$ such that $h(x) \geq \E{\bar \tau^\mathsf{1}} \slash \E{\bar \tau^\mathsf{2}}$, we can recognize that the value at the switch point must satisfy the same bound, 
\begin{align}
\frac{\tilde{\mathcal{K}}}{\E{\bar \tau^\mathsf{2}} h(\tilde{\mathcal{K}})}~\editTwo{<}~\frac{1}{c \E{\bar \tau^\mathsf{2}} }
,
\label{cBound}
\end{align}
and this now completes the proof \editTwo{of the throughput converging up to $1/(c\E{\bar \tau^\mathsf{2}})$ while being strictly less than $1/(c\E{\bar \tau^\mathsf{2}})$ for any finite $\mathcal{K}$}.}

\editTwo{Let us now observe how these convergence results for $\mathcal{G}(\mathcal{K})$, $\mathcal{I}(\mathcal{K})$, and $\mathcal{K}/\E{\tau(\mathcal{K})}$ imply that there always exists a larger concurrency with strictly faster throughput. By the fact that the guaranteed service rate converges up to $1/(c\E{\bar \tau^\mathsf{2}})$, we have that there exists a $\mathcal{K}'$ such that $\mathcal{G}(\mathcal{K}') \geq \mathcal{I}(\mathcal{K})$ for $\mathcal{K}$ as any arbitrary finite concurrency level. Moreover, because $\mathcal{G}(\mathcal{K}) < \mathcal{I}(\mathcal{K})$ for any finite, non-zero $\mathcal{K}$ by definition, we furthermore have that $\mathcal{K}' > \mathcal{K}$. Together with the service rate ordering via Corollary~\ref{rateBound}, we thus have
$$
\frac{\mathcal{K}}{\E{\tau(\mathcal{K})}}
\leq
\mathcal{I}(\mathcal{K})
<
\mathcal{G}(\mathcal{K}')
\leq 
\frac{\mathcal{K}'}{\E{\tau(\mathcal{K}')}}
,
$$
which achieves the stated result.}
\hfill\Halmos\endproof


\editTwo{Notice that if one relaxes the $h(x)/x > c$ condition in the statement of Corollary~\ref{noOptKcor} to $h(x)/x \geq c$, the inequality in Equation~\eqref{cBound} would hold in a less than or equal to fashion rather than being strictly less. Under these assumptions, the result of Corollary~\ref{noOptKcor} would instead be that there is always a strictly larger concurrency at which point the throughput is no worse than the current value, rather than strictly better. This weakened improvement guarantee follows from the fact that $\mathcal{I}(\mathcal{K})$ could be equal to $1/(c\E{\bar \tau^\mathsf{2}})$ for finite $\mathcal{K}$, rather than strictly less than it. Nevertheless, if one takes the $h(x)/x \geq c$ assumption and also knows that $\mathcal{K}/\E{\tau(\mathcal{K})} < 1/(c\E{\bar \tau^\mathsf{2}})$ for a given $\mathcal{K}$, then there will exist a $\mathcal{K}' > \mathcal{K}$ such that $\mathcal{K}/\E{\tau(\mathcal{K})} < \mathcal{K}'/\E{\tau(\mathcal{K}')}$, as $\mathcal{G}(\mathcal{K}) < 1/(c\E{\bar \tau^\mathsf{2}})$ for all finite $\mathcal{K}$.}

\subsection{Proof of Proposition~\ref{convexProp}}

\proof{Proof.}
To begin, let us prove that the $\mathcal{K}^*$ is the unique argument that \editTwo{maximizes} the guaranteed service rate, $\mathcal{G}(\mathcal{K})$.  The derivative of $\mathcal{G}(\mathcal{K})$ is
\begin{align*}
\frac{\mathrm{d}}{\mathrm{d}\mathcal{K}}\left( 
\frac{\mathcal{K}}{
\E{\bar \tau^\mathsf{1}} + \E{\bar \tau^\mathsf{2}} h(\mathcal{K})
}
\right)
&=
\frac{1}{
\E{\bar \tau^\mathsf{1}} + \E{\bar \tau^\mathsf{2}} h(\mathcal{K})
}
-
\frac{
\E{\bar\tau^\mathsf{2}}\mathcal{K}h'(\mathcal{K})
}{
\left(\E{\bar \tau^\mathsf{1}} + \E{\bar \tau^\mathsf{2}} h(\mathcal{K})\right)^2
}
,
\end{align*}
which, when set equal to 0, yields first order condition
\begin{align*}
\E{\bar \tau^\mathsf{1}} + \E{\bar \tau^\mathsf{2}} h(\mathcal{K})
&=
\E{\bar\tau^\mathsf{2}}\mathcal{K}h'(\mathcal{K})
,
\end{align*}
and this is equivalent to Equation~\eqref{convEq}. Now, we must further establish that Equation~\eqref{convEq} indeed has a unique solution. To see this, consider the derivative of its right-hand side:
\begin{align*}
\frac{\mathrm{d}}{\mathrm{d}\mathcal{K}}\left( 
\mathcal{K} h'(\mathcal{K}) - h(\mathcal{K})
\right)
&=
h'(\mathcal{K}) + \mathcal{K} h''(\mathcal{K}) - h'(\mathcal{K})
=
\mathcal{K} h''(\mathcal{K})
.
\end{align*}
Because we are given that $h(\cdot)$ is strictly convex, $\mathcal{K} h''(\mathcal{K}) > 0$ for all $\mathcal{K}$. Thus, $\mathcal{K} h'(\mathcal{K}) - h(\mathcal{K})$ can cross $\E{\bar \tau^\mathsf{1}}/\E{\bar \tau^\mathsf{2}}$ only once. Furthermore, because $\mathcal{K} h'(\mathcal{K}) - h(\mathcal{K}) = 0$ for $\mathcal{K} = 0$, the strict convexity of $h(\cdot)$ also assures that this crossing will happen on the positive reals.

To complete the proof, we will obtain $\underline{\mathcal{K}}$ and $\overline{\mathcal{K}}$ as concurrencies at which the idealized service rate obtains the maximal guaranteed service rate. Using Equation~\eqref{convEq} as an identity, we can simplify the optimal guaranteed service rate as
\begin{align*}
\mathcal{G}(\mathcal{K}^*)
&=
\frac{\mathcal{K}^*}{
\E{\bar \tau^\mathsf{1}} + \E{\bar \tau^\mathsf{2}} h(\mathcal{K}^*)
}
=
\frac{1}{\E{\bar \tau^\mathsf{2}} h'(\mathcal{K}^*)}
.
\end{align*}
For $\mathcal{K} < h^{-1}(\E{\bar \tau^\mathsf{1}}/\E{\bar \tau^\mathsf{2}})$, $\mathcal{I}(\mathcal{K}) = \mathcal{K}/\E{\bar \tau^\mathsf{1}}$, and this function will of course attain $\mathcal{G}(\mathcal{K}^*)$ before $\mathcal{K}^*$, as $\mathcal{G}(\mathcal{K}) \leq \mathcal{I}(\mathcal{K})$. Naturally, $\mathcal{K}/\E{\bar \tau^\mathsf{1}} = \mathcal{G}(\mathcal{K}^*)$ at $\mathcal{K} = \E{\bar \tau^\mathsf{1}}/(\E{\bar \tau^\mathsf{2}} h'(\mathcal{K}^*)) = \underline{\mathcal{K}}$. Then, for $\mathcal{K} \geq h^{-1}(\E{\bar \tau^\mathsf{1}}/\E{\bar \tau^\mathsf{2}})$, we have $\mathcal{I}(\mathcal{K}) = \mathcal{K}/(\E{\bar \tau^\mathsf{2}}h(\mathcal{K}))$, and this decreasing function is an upper bound for $\mathcal{G}(\mathcal{K})$. We can immediately see that this bound attains  $\mathcal{G}(\mathcal{K}^*)$ at the $\mathcal{K}$ such that $\mathcal{K}/h(\mathcal{K}) = 1/h'(\mathcal{K}^*)$, which is precisely $\overline{\mathcal{K}}$. Finally, because the true throughput is bounded between the guaranteed and idealized service rates, it must attain its maximum in the region spanned by where $\mathcal{I}(\mathcal{K}) \geq \mathcal{G}(\mathcal{K}^*)$, which is $[\underline{\mathcal{K}}, \overline{\mathcal{K}}]$.
\hfill\Halmos\endproof

\subsection{Proof of Proposition~\ref{polyOptProp}}

\proof{Proof.}
Suppose $\sigma > 1$. Connecting to Proposition~\ref{convexProp}, we have $h'(\mathcal{K}) = \sigma \mathcal{K}^{\sigma -1}$, and thus $\mathcal{K}^*$ is the solution to
\begin{align*}
\frac{
\E{\bar \tau^\mathsf{1}}
}{
\E{\bar \tau^\mathsf{2}}
}
=
(\sigma - 1)\mathcal{K}^{\,^\sigma}
,
\end{align*}
which is as stated in Equation~\eqref{polyOptKstarEq}. Because $h'(\mathcal{K}^*) = \sigma \left(
{\E{\bar \tau^\mathsf{1}}}/({(\sigma - 1) \E{\bar \tau^\mathsf{2}}})
\right)^{1 - {1}\slash{\sigma}}$, we can similarly find that $\underline{\mathcal{K}}$ is given by
\begin{align*}
\underline{\mathcal{K}} 
&= 
\frac{
\E{\bar \tau^\mathsf{1}}
}{
\sigma\E{\bar \tau^\mathsf{2}}
}
\left(
\frac{(\sigma - 1) \E{\bar \tau^\mathsf{2}}}{\E{\bar \tau^\mathsf{1}}}
\right)^{1 - {1}\slash{\sigma}}
=
\left(1 - \frac{1}\sigma\right)
\left(
\frac{\E{\bar \tau^\mathsf{1}}}{(\sigma - 1) \E{\bar \tau^\mathsf{2}}}
\right)^{{1}\slash{\sigma}}
=
\left(1 - \frac{1}\sigma\right)
\mathcal{K}^*
,
\end{align*}
and also that $\overline{\mathcal{K}}$ is the solution to
\begin{align*}
\sigma \left(
\frac{\E{\bar \tau^\mathsf{1}}}{(\sigma - 1) \E{\bar \tau^\mathsf{2}}}
\right)^{(\sigma - 1)\slash{\sigma}}
&=
\overline{\mathcal{K}}^{\sigma - 1}
,
\end{align*}
which yields $\overline{\mathcal{K}} = \sigma^{1/(\sigma - 1)} \mathcal{K}^*$. 

Now, suppose instead that $\sigma \leq 1$. In this case, the derivative of the guaranteed service rate is
\begin{align*}
\frac{\mathrm{d}}{\mathrm{d}\mathcal{K}}
\left(
\frac{\mathcal{K}}{
\E{\bar \tau^\mathsf{1}}
+
\E{\bar \tau^\mathsf{2}} \mathcal{K}^{\,^\sigma}
}
\right)
&=
\frac{1}{
\E{\bar \tau^\mathsf{1}}
+
\E{\bar \tau^\mathsf{2}} \mathcal{K}^{\,^\sigma}
}
-
\frac{\sigma \E{\bar \tau^\mathsf{2}} \mathcal{K}^{\,^\sigma}}{
\left(
\E{\bar \tau^\mathsf{1}}
+
\E{\bar \tau^\mathsf{2}} \mathcal{K}^{\,^\sigma}
\right)^2
}
\\
&=
\frac{1}{
\left(
\E{\bar \tau^\mathsf{1}}
+
\E{\bar \tau^\mathsf{2}} \mathcal{K}^{\,^\sigma}
\right)^2
}
\left(
\E{\bar \tau^\mathsf{1}}
+
(1-\sigma)
\E{\bar \tau^\mathsf{2}} \mathcal{K}^{\,^\sigma}
\right)
,
\end{align*}
and this is non-negative for all $\mathcal{K}$. Similarly, the derivatives of the two components of $\mathcal{I}(\mathcal{K})$ are $1/\E{\bar \tau^\mathsf{1}}$ and $(1-\sigma) / (\E{\bar \tau^\mathsf{1}} \mathcal{K}^{\,^\sigma})$, which are both also non-negative for all $\mathcal{K}$. Hence, both $\mathcal{G}(\mathcal{K})$ and $\mathcal{I}(\mathcal{K})$ are non-decreasing. Moreover, for $\sigma < 1$, both $\mathcal{G}(\mathcal{K}) \to \infty$ and $\mathcal{I}(\mathcal{K}) \to \infty$ as $\mathcal{K} \to \infty$, and if $\sigma = 1$, then both $\mathcal{G}(\mathcal{K})$ and $\mathcal{I}(\mathcal{K})$ converge to $1/\E{\bar \tau^\mathsf{2}}$ from below as $\mathcal{K} \to \infty$. Hence, for any $\sigma \leq 1$, there is no finite optimal concurrency.
\hfill\Halmos\endproof

\subsection{\edit{Proof of Proposition~\ref{symProp1}}}

\proof{Proof.}
\edit{This result is an almost immediate consequence of L'H\^opital's rule. By~\eqref{symRateDef}, the symmetric-slowdown throughput will be asymptotically monotonic if and only if $x/h(x)$ tends to 0 at one extreme of its domain and to infinity at the other. Because Assumption~\ref{conA} provides that $h(0) = 0$ and $\lim_{x \to \infty} h(x) = \infty$, we have by L'H\^opital that
$$
\lim_{x \to 0} \frac{x}{h(x)} = \frac{1}{h'(0)}
\quad
\text{ and }
\quad
\lim_{x \to \infty} \frac{x}{h(x)} = \frac{1}{ \lim_{x \to \infty} h'(x)}
.
$$
Thus, by Definition~\ref{nonuDef}, if $h'(\cdot) \in \mathcal{M}$, then either $\lim_{x \to 0} h(x) = \infty$ and $\lim_{x \to \infty} h(x) = 0$, or $\lim_{x \to 0} h(x) = 0$ and $\lim_{x \to \infty} h(x) = \infty$, yielding that $x/h(x) \in \mathcal{M}$ as well.
}\hfill\Halmos\endproof

\subsection{\edit{Proof of Proposition~\ref{symProp2}}}

\proof{Proof.}
\edit{By use of Equation~\eqref{symRateDef}, the derivative of the symmetric-slowdown {throughput} is}
\begin{align}
\edit{\frac{\mathrm{d}}{\mathrm{d}\mathcal{K}}
\left(
\frac{
\mathcal{K}
}{
\E{\hat \tau(1)} h(\mathcal{K})
}
\right)}
&\edit{=}
\edit{\frac{
1
}{
\E{\hat \tau(1)} 
}
\left(
 \frac{1}{h(\mathcal{K})}
-
 \frac{\mathcal{K} h'(\mathcal{K})}{h(\mathcal{K})^2}
\right)
.}
\label{symThruDeriv}
\end{align}
\edit{The stated non-decreasing result will now follow quickly by use of the assumed convexity or concavity. Let us start with the latter. Because we have $h(0) = 0$ by Assumption~\ref{conA}, concavity implies that 
$
h(x) = h(x) - h(0) \geq h'(x) (x - 0) = x h'(x)
$. 
Accordingly, the throughput of the derivative can be bounded}
\begin{align*}
\edit{\frac{\mathrm{d}}{\mathrm{d}\mathcal{K}}
\left(
\frac{
\mathcal{K}
}{
\E{\hat \tau(1)} h(\mathcal{K})
}
\right)}
&\edit{\geq}
\edit{\frac{
1
}{
\E{\hat \tau(1)} 
}
\left(
 \frac{1}{h(\mathcal{K})}
-
 \frac{1}{h(\mathcal{K})}
\right)
=
0
.}
\end{align*}
\edit{Hence, the throughput is non-decreasing if $h(\cdot)$ is concave. Using the same arguments with the analogous fact that $h(x) \leq x h'(x)$ for any convex $h(\cdot)$ with $h(0) = 0$, we similarly find that the throughput is non-increasing if $h(\cdot)$ is convex. Finally, under either concave or convex $h(\cdot)$, the lack of optimal concurrency follows immediately from the fact that the throughput can always be improved by either increasing or decreasing the concurrency level, respectively.}
\hfill\Halmos\endproof

\subsection{Proof of Theorem~\ref{uThm}}

\proof{Proof.}
Because $\rho_\mathsf{2} = \rho - \rho_\mathsf{1}$ by definition, we will simply phrase this proof entirely in terms of $\rho_\mathsf{1}$. From Proposition~\ref{convexProp}, we have that the guaranteed service rate is maximized by the unique $\mathcal{K}^*$ that solves $\E{\bar \tau^\mathsf{1}}/\E{\bar \tau^\mathsf{2}} = \mathcal{K}^* h'(\mathcal{K}^*) - h(\mathcal{K}^*)$. Hence, the concurrency-maximized guaranteed service rate can be written
\begin{align}
\mathcal{G}^* 
&=
\mathcal{G}(\mathcal{K}^*)
=
\frac{
\mathcal{K}^*
}{
\E{\bar \tau^\mathsf{1}}
+
\E{\bar \tau^\mathsf{2}}h(\mathcal{K}^*)
}
=
\frac{
1
}{
\E{\bar \tau^\mathsf{2}}h'(\mathcal{K}^*)
}
.
\label{gStarDef}
\end{align}
Similarly, let $\mathcal{I}^*$ be the concurrency-maximized idealized service rate (which need not occur at $\mathcal{K}^*$). Because $\mathcal{I}(\mathcal{K})$ can be seen to be the minimum between two functions of $\mathcal{K}$,
\begin{align*}
\mathcal{I}(\mathcal{K})
&=
\frac{1}{
\left(\E{\bar \tau^\mathsf{1}} \frac{1}{\mathcal{K}} \vee \E{\bar \tau^\mathsf{2}} \frac{h(\mathcal{K})}{\mathcal{K}}\right)
}
=
\left(\frac{\mathcal{K}}{\E{\bar \tau^\mathsf{1}}} \wedge \frac{\mathcal{K}}{\E{\bar \tau^\mathsf{2}} h(\mathcal{K})}\right) 
,
\end{align*}
let us consider the evolution of each component individually.

Clearly, $\mathcal{K}/\E{\bar \tau^\mathsf{1}}$ is increasing in $\mathcal{K}$. Then, because
\begin{align*}
\frac{\partial}{\partial \mathcal{K}} 
\frac{\mathcal{K}}{ h(\mathcal{K})}
&=
\frac{1}{ h(\mathcal{K})^2}
\left(
h(\mathcal{K})
-
\mathcal{K} h'(\mathcal{K})
\right)
\quad
\text{ and }
\quad
\frac{\partial}{\partial \mathcal{K}} 
\left(
h(\mathcal{K})
-
\mathcal{K} h'(\mathcal{K})
\right)
=
-\mathcal{K} h''(\mathcal{K})
,
\end{align*}
the strict convexity of $h(\cdot)$ implies that ${\mathcal{K}}/({\E{\bar \tau^\mathsf{2}} h(\mathcal{K})})$ is strictly decreasing on the positive reals. Hence, $\mathcal{I}^*$ must occur at the intersection of the two components: $h^{-1}(\E{\bar \tau^\mathsf{1}}/\E{\bar \tau^\mathsf{2}})$ (at which point, $\mathcal{K}/\E{\bar \tau^\mathsf{1}} = {\mathcal{K}}/({\E{\bar \tau^\mathsf{2}} h(\mathcal{K})})$). Thus, the concurrency-maximized idealized service rate can be expressed
\begin{align*}
\mathcal{I}^*
=
\frac{1}{\E{\bar \tau^\mathsf{1}}}
h^{-1}\left(\frac{\E{\bar \tau^\mathsf{1}}}{\E{\bar \tau^\mathsf{2}}}\right)
.
\end{align*}
Now, we can also recognize that because $\mathcal{G}^*$ is a unique maximum, 
\begin{align*}
\mathcal{G}^*
\geq
\mathcal{G}\left(h^{-1}\left(\frac{\E{\bar \tau^\mathsf{1}}}{\E{\bar \tau^\mathsf{2}}}\right)\right)
=
\frac{
1
}{
2\E{\bar \tau^\mathsf{1}}
}
h^{-1}\left(\frac{\E{\bar \tau^\mathsf{1}}}{\E{\bar \tau^\mathsf{2}}}\right)
,
\end{align*}
and, because by definition of the guaranteed and idealized service rates we have
\begin{align*}
\mathcal{G}^* \leq \max_{\mathcal{K} \in \mathbb{R}_+} \frac{\mathcal{K}}{\E{\tau(\mathcal{K})}} \leq \mathcal{I}^*
,
\end{align*}
we thus can together bound the concurrency-maximized throughput:
\begin{align}
\frac{
1
}{
2\E{\bar \tau^\mathsf{1}}
}
h^{-1}\left(\frac{\E{\bar \tau^\mathsf{1}}}{\E{\bar \tau^\mathsf{2}}}\right)
\leq 
\max_{\mathcal{K} \in \mathbb{R}_+} \frac{\mathcal{K}}{\E{\tau(\mathcal{K})}} 
\leq 
\frac{1}{\E{\bar \tau^\mathsf{1}}}
h^{-1}\left(\frac{\E{\bar \tau^\mathsf{1}}}{\E{\bar \tau^\mathsf{2}}}\right)
.
\end{align}

Let us now consider these bounds at the asymmetric extremes: agent self-production ($\rho_\mathsf{1} = 0$) and customer self-production ($\rho_\mathsf{1} = \rho$). Starting with the latter, we have by Lemma~\ref{rhoLemma} that $h^{-1}(\E{\bar \tau^\mathsf{1}}/\E{\bar \tau^\mathsf{2}})/\E{\bar \tau^\mathsf{1}} \to \infty$ as $\rho_\mathsf{1} \to \rho$, because $\lim_{x\to\infty} h(x) = \infty$ by definition. Then, turning to the former, by the inverse function rule and the definition of the derivative, let us now recognize that 
\begin{align*}
\lim_{r \to 0}
\frac{1}{r} h^{-1}(r)
&=
\lim_{r \to 0}
\frac{1}{r} \left(h^{-1}(r) - h^{-1}(0)\right)
=
\frac{1}{h'(h^{-1}(0))}
=
\frac{1}{h'(0)}
=
\infty
.
\end{align*}
Hence, again by Lemma~\ref{rhoLemma}, we have that $h^{-1}(\E{\bar \tau^\mathsf{1}}/\E{\bar \tau^\mathsf{2}})/\E{\bar \tau^\mathsf{1}} \to \infty$ as $\rho_\mathsf{1} \to 0$. Finally, because $\mathcal{I}^*$ is finite for all $\rho_\mathsf{1} \in (0, \rho)$, the maximal throughput must be as well, and thus the true concurrency-maximized throughput is $\U$-shaped in the interdependence. 
\hfill\Halmos\endproof

\section{Analysis and Decomposition of Univariate Marked Hawkes Clusters}\label{markedSec}


While not a focus in the big picture of this paper's analysis, marked Hawkes processes play a key role in the foundation of the parking function decomposition and the associated methodology we have employed throughout the main body. In this section of the appendix, we provide the supporting details for analyzing this generalization of the (one-sided) cluster model, in which each jump in the intensity may be of random size. By comparison to the two-sided, non-marked clusters studied in the main body of the paper, we will use $\bar{\cdot}$ notation to refer to the marked variants of the model. Following standard intensity-based notions of marked Hawkes point processes \citep[e.g.,][]{bremaud2002rate}, we begin by formalizing the marked Hawkes cluster through its intensity now in Definition~\ref{mhcDef}.


\begin{definition}[Marked Hawkes Cluster]\label{mhcDef}
Assuming $\bar \tau_0 = 0$, let $\bar N_t$ with $\bar N_0 = 1$ be the point process for  time $t \geq 0$ driven by a corresponding stochastic intensity $\bar \mu_t$, defined 
\begin{align}
\bar \mu_{t}
&=
\sum_{i=0}^{\bar N_{t} - 1} 
M_i \bar g(t - \bar \tau_i)
,
\label{intensityMarksDef}
\end{align}
where $\bar \tau_i$ is the $i$th epoch in the cluster, with $\{M_i \mid i \in \mathbb{N}\}$ as an independent and identically distributed sequence of positive random variables. That is,
\begin{align}
\PP{\bar N_{t+\delta} - \bar N_t = n \mid \bar{\mathcal{F}}_t} 
&
=
\begin{cases}
\bar \mu_t \delta + o(\delta) & n = 1,\\
1 - \bar \mu_t \delta + o(\delta) & n = 0,\\
o(\delta) & n > 1,
\end{cases}
\label{markHawkesDef1}
\end{align}
where $\bar{\mathcal{F}}_t$ is the natural filtration of the stochastic process.
Here, $\bar g: \mathbb{R}_+ \to \mathbb{R}_+$ is the response or decay function for each point and $M_i$ is the random mark of the $i$th point, so that $M_i \bar g(0)$ is the instantaneous impact of the $i$th point, and, generally, $M_i \bar g(u)$ is its impact $u \geq 0$ time after its occurrence. Finally, we will suppose that the cluster ends at the natural duration, $\bar \tau = \bar \tau_{N-1} = \inf\{t \geq 0\mid \bar N_t = \bar N\}$, with $\bar N = \lim_{t \to \infty} \bar N_t$ as the size of the cluster. Given this structure, let the points $\bar \tau_i$ for $i \in \{0, 1, \dots, N-1\}$ be the marked cluster model. \hfill \Halmos 
\end{definition}

Definition~\ref{mhcDef} is the marked analog of Definition~\ref{hcDef}. To build towards a marked analog of the parking-function-based Definition~\ref{pfcDef}, let us first define a generalized concept from probabilistic combinatorics that encapsulates parking functions, Dyck paths. Starting with deterministic Dyck paths, these combinatorial objects are simple yet central. While the underlying notion can appear in different contexts, they may be most naturally thought of as a sequence of steps on the lattice: let a $n$-step Dyck path be a non-decreasing sequence on the $(x,y)$ lattice from $(0,1)$ to $(n,n)$ that never exceeds $y \leq x+1$, where ``$n$-step'' refers to the fact that there are $n$ horizontal moves in total. Following the convention in \citet{daw2023conditional}, we record the Dyck path as the vector of the largest $y$-coordinate attained at each $x$-coordinate, for each $x \in \{0, 1, \dots, n-1\}$. As visualized in Figure 1 of \citet{daw2023conditional}, for example, the 3-step Dyck path $[1,1,3]$ refers to the sequence of steps $(0,1) \to (1,1) \to (2,1) \to (2,3) \to (3,3)$. We will let $\mathsf{D}_n \subset \mathbb{Z}_+^n$ be the set of all $n$-step Dyck paths. Hinting at the ubiquity of Dyck paths, it is well-known that they are enumerated by the Catalan numbers: $|\mathsf{D}_n| = C_n = (2n)!/((n+1)!n!)$ \citep[e.g.,][]{stanley1999enumerative}. Naturally, Dyck paths are also well-connected to (deterministic) parking functions, as there is a bijection between sorted parking functions of length $n$ and $n$-step Dyck paths; equivalently, there is a bijection between labeled or shuffled Dyck paths and parking functions  \citep{yan2015parking}.




In order to now connect these combinatorial objects to the stochastic cluster models, let us introduce a probabilistic version. For a given number of steps $n$, we will let a \emph{random Dyck path} refer to one path drawn from some distribution over $\mathsf{D}_n$. Notice that, from this perspective, uniformly random parking functions simply constitute one distribution over Dyck paths. Naturally, this allows us to introduce a generalization of Definition~\ref{pfcDef}. To do so, let us first establish a critical lemma that is essential \edit{to} \editTwo{achieving} an equivalence between definitions; in fact, readers may notice that this is already a key supporting piece in the proof of Theorem~\ref{equivThm}.  That is, in Lemma~\ref{dyckLemma}, we show that histograms of these random Dyck paths are equivalent in distribution to the number of direct descendants to each point in the marked Hawkes cluster.


\begin{lemma}\label{dyckLemma}
Let $X_0, X_1, \dots, X_n$ be the number of direct descendants for each epoch in a marked Hawkes process cluster with $\bar N = n + 1$ points and corresponding marks $M_0 = m_0, M_1 = m_1, \dots, M_n = m_n$. Additionally, suppose that $\pi$ is an $n$-step Dyck path randomly selected according to the weighted distribution,
\begin{align}
\PP{ \pi = d}
=
\frac{
\prod_{i=1}^n \frac{m_{i-1}^{\kappa_i(d)}}{\kappa_i(d)!}
}{
\sum_{\delta \in \mathsf{D}_n}
\prod_{j=1}^{n}
\frac{
m_{j-1}^{\kappa_{j}(\delta)}
}{
\kappa_{j}(\delta)!
}
}
,
\label{dyckDist}
\end{align}
for each $d \in \mathsf{D}_n$, where $\kappa_i(d) = \left|\{j \mid d_j = i\}\right|$ counts the occurrences of $i$ within $d$. Then, conditional on $\bar N = n + 1$ and $M_0 = m_0, M_1 = m_1, \dots, M_n = m_n$, the descendent pattern is such that $X_n = 0$ almost surely and 
\begin{align}
\left(X_0, X_1, \dots, X_{n-1}\right)
\stackrel{\mathsf{D}}{=}
\left(\kappa_1(\pi), \kappa_2(\pi), \dots, \kappa_n(\pi)\right)
,
\end{align}
for all $n \in \mathbb{Z}_+$ and $m \in \mathbb{R}_+^{n+1}$.
\end{lemma}
\proof{Proof.} Let us begin by showing that  $(X_0, X_1, \dots, X_{n-1})$ and $\left(\kappa_1(\pi), \kappa_2(\pi), \dots, \kappa_n(\pi)\right)$ share an equivalent sample space. From the \citet{hawkes1974cluster} representation of the process \citep[see, e.g.][for this perspective for a marked Hawkes process]{koops2017infinite}, we can view the time-agnostic descendant process as a discrete-time Poisson branching process. This can equivalently be seen as a random walk with jumps; the following arguments resemble that of the classic hitting time theorem \citep[e.g.,][]{dwass1969total,van2008elementary}, but here the jumps are not exchangeable because $m_i$ need not be equal to $m_j$ for $i \ne j$. 

Let $W_k = \sum_{i=0}^{k-1} X_i - k + 1$ for $k \geq 1$ track the number of active descendants in generation $k$, so that $W_k = W_{k-1} + X_{k-1} - 1$, with $W_0 = 1$ and $X_i \sim \mathrm{Pois}(\bar \rho m_i)$ mutually independent. Here, $\bar \rho m_i = m_i \int_0^\infty \bar g (u) \mathrm{d}u$ is the mean number of descendants across all time for mark $i$, and the Poisson distribution is a consequence of the time inhomogeneous offspring stream descending from each point. Then, given the sequence of $m_i$'s, the event that the cluster has size $n+1$ is equivalent to the event that the walk first reaches state 0 at time $n+1$:
\begin{align*}
\{\bar N = n + 1\} = \{W_{n+1} = 0\} \cap \bigcap_{k=1}^n \{W_k > 0\}
\end{align*}
Written in terms of $X_i$'s, this is
\begin{align*}
\{W_{n+1} = 0\} \cap \bigcap_{k=1}^n \{W_k > 0\}
=
\left\{\sum_{i=0}^{n} X_i = n\right\} \cap \bigcap_{k=1}^n \left\{\sum_{i=0}^{k-1} X_i > k-1\right\}
\end{align*}
For this event to hold, $X_n$ must be 0. On the other end, $X_0$ must be at least 1 and no more than $n$. Then, $X_1$ can take as much space as $X_0$ left it, meaning it can be as large as $n - X_0$ or as small as $2 - X_0$. Likewise, $X_2$ can take as much space as $X_0$ \textit{and} $X_1$ leave it, and this pattern continues successively. This structure exactly encodes a Dyck path: $X_0$ contains the number of steps at level 1, $X_1$ the number of steps at $2$, and so on. Immediately, we see that any walk must constitute a Dyck path, because the requirement that the walk does not hit 0 before time $n+1$ is equivalent to the requirement that the running height of Dyck path never exceeds its running index of its steps. Hence, likewise, any Dyck path must also encode a walk. Because this yields a bijection, we can sum across all $n$-step Dyck paths to enumerate all possibilities of $X_0, \dots, X_{n-1}$. This leads us now to the probabilities for the collection of events.

For a given Dyck path $d \in \mathsf{D}_n$, the (unconditional) probability of the branching process taking the walk of $X_0 = \kappa_1(d)$, $X_1 = \kappa_2(d)$, \dots, $X_{n-1} = \kappa_n(d)$, and $X_n = 0$ is
\begin{align*}
\PP{X_0 = \kappa_1(d), X_1 = \kappa_2(d), \dots, X_{n-1} = \kappa_n(d), X_n = 0}
&=
e^{-\bar \rho m_n}
\prod_{i=0}^{n-1}
\frac{e^{-\bar \rho m_i}}{\kappa_{i+1}(d)!}
\left( \bar \rho m_i \right)^{\kappa_{i+1}(d)}
,
\end{align*}
because the Poisson offspring streams are mutually independent, and thus so are the $X_i$ variables. Because $e^{-\bar \rho m_n} \prod_{i=1}^{n-1} e^{-\bar \rho m_i} = \exp(-\bar \rho \sum_{i=1}^n m_i)$ will appear as a coefficient in the probability for any given path $d$, the conditional probability simplifies to the distribution over Dyck paths given in Equation~\eqref{dyckDist}.
\hfill\Halmos\endproof


Building upon Lemma~\ref{dyckLemma}, let us now define a Dyck-path-based cluster model, where this distribution of descendants is directly invoked in the construction. In Definition~\ref{dpcDef}, we propose  a marked  cluster in which the size and marks are drawn according to a Poisson-type branching process and the number of descendants are drawn according to a random Dyck path as prescribed in Lemma~\ref{dyckLemma}, leaving the epochs in the cluster to be found akin to step (iv) of Definition~\ref{pfcDef}.

\begin{definition}[Dyck Path Cluster]\label{dpcDef}
{
For the kernel $\bar g: \mathbb{R}_+ \to \mathbb{R}_+$ and for $\{M_i \mid i \in \mathbb{N}\}$ as an independent and identically distributed sequence of positive random variables,
construct the marked cluster of points $0 = \bar \tau_0 < \bar \tau_1 < \dots < \bar \tau_{\bar N-1}$ as follows:
\begin{enumerate}[i)]
\item Let $\bar N$ and $M_0, \dots, M_{\bar N-1}$ be the size and marks of the cluster, which are jointly equivalent in distribution to the number of generations and mixture variables of a Poisson-mixture branching process.
\item Let $\pi \in \mathsf{D}_{\bar N-1}$ be  a random Dyck path encoding the number of descendants for each point, as given in the Dyck path distribution and associated structure in Lemma~\ref{dyckLemma}. 
\item For each $i \in \{1, \dots, \bar N-1\}$, let $\bar \Delta_i$ be independently generated according to the density $\bar g (\cdot)/\bar \rho$, and set $\bar T_i = \bar T_{\pi_{(i)}-1} + \bar \Delta_i$, with $\bar T_0 = 0$.
\end{enumerate}
Finally, let $\bar \tau_i = \bar T_{(i)}$ for each $i \in \{0, 1, \dots, \bar N-1\}$ be the Dyck path cluster. }\hfill \Halmos 
\end{definition}

As we saw in Theorem~\ref{equivThm} in the main body of the paper, these intensity- and combinatorially-based definitions can be seen to be equivalent.

\begin{theorem}\label{mEquivThm}
The marked Hawkes cluster (Definition~\ref{mhcDef}) and the Dyck path cluster (Definition~\ref{dpcDef}) are distributionally equivalent. 
\end{theorem}
\proof{Proof.}
In a manner similar to how the proof of Theorem~\ref{equivThm} followed the steps in Definition~\ref{pfcDef}, we will prove the present statement by walking through the steps of Definition~\ref{dpcDef}. However, steps (i) and (ii) are implied by Lemma~\ref{dyckLemma}, so we are simply left to show (iii). 

This final step follows similarly to the equivalence of the final step in Definition~\ref{pfcDef}. Given the \citet{hawkes1974cluster} definition, we know that the $\ell$th point generates direct offspring according to an inhomogeneous Poisson stream with rate $M_\ell \bar g(u)$ for \editTwo{$u$ time} since point $\ell$ occurred. Moreover, we have that this stream is independent of all others in the cluster. Hence, given that there are $\kappa_{\ell+1}(\pi)$ points in this stream, \citep[via, e.g., Section 2.1 of][]{daley2003introduction}, these can be viewed as $\kappa_{\ell+1}(\pi)$ order statistics of the density given by $\bar{g}(\cdot)/\bar \rho$. Ignoring the order, these are $\kappa_{\ell+1}(\pi)$ independent samples from $\bar{g}(\cdot)/\bar \rho$. 

Because the clusters of Definition~\ref{mhcDef} and Definition~\ref{dpcDef} are both fully characterized by the sequence of points, upon sorting we achieve the stated equivalence.
\hfill\Halmos\endproof


The equivalence of definitions shown in Theorem~\ref{mEquivThm} may be interesting and useful in its own right for the study of marked Hawkes clusters and point processes at large, but let us ground its insights back in the service asymmetry context of the main body. In particular, let us formalize what was hinted by the $\bar{\cdot}$ notation on the durations under $\mathsf{1}$- and $\mathsf{2}$-paced asynchrony: these random variables are equivalent to the durations of marked Hawkes clusters. As can be seen in their definitions in Section~\ref{limitSec}, these limiting objects obtained from the two-sided cluster duration as $\eta \to \infty$ and $\eta \to 0$ are such that all of response offsets on one side are equal to 0 almost surely. Because there may be a chain of responses on the instantaneous side, these points collapse into a mark driving further activty on the non-instantaneous side. We formalize this now in Corollary~\ref{barMarks}.

\begin{corollary}\label{barMarks}
The random variables $\bar \tau^\mathsf{1}$ and $\bar \tau^\mathsf{2}$ are each equivalent to the durations of marked Hawkes clusters. Specifically, $\bar \tau^\mathsf{1}$ is equivalent to the duration of a marked Hawkes cluster with response function $g_\mathsf{1}(\cdot)$ and mark distribution $\mathsf{Borel}(\rho_\mathsf{2})$, and $\bar \tau^\mathsf{2}$ is equivalent to the duration of a marked Hawkes cluster with response function $g_\mathsf{2}(\cdot)$ and mark distribution $\mathsf{Borel}(\rho_\mathsf{1})$.
\end{corollary}
\proof{Proof.}
This follows immediately from Theorems~\ref{equivThm},~\ref{convThm}, and~\ref{mEquivThm}, where we can recognize that the Poisson branching process descendant structure equivalent to parking function construction in Definition~\ref{pfcDef} will imply the Borel distributions of the marks, as the instantaneous side contributions can garner further immediate self-follow-ups at each non-instantaneous contribution point.
\hfill\Halmos\endproof

As briefly referred to in Section~\ref{limitSec}, one advantage of this understanding of marked Hawkes clusters is that, in light of the limits in Theorem~\ref{convThm}, there can be settings where the asymptotic, asynchronous durations are more readily analyzed than the two-sided duration is. For instance, when the response functions are each exponential but with different rates in the two-sided cluster, it is not clear if the cluster duration can be found in closed form. However, each of the asynchronous \emph{marked} cluster durations would only have one response function. In the case of exponential functions, this turns out to be a substantial simplification. In fact, in Theorem~\ref{expMarkedDist}, we show that the full chronology of the cluster, meaning all the epochs, can be seen to be sums of conditionally independent exponential random variables. 


\begin{theorem}\label{expMarkedDist}
Let $\bar g(x) = \alpha e^{-\beta x}$ for all $x \geq 0$ with $\bar \rho = \alpha \slash \beta < 1$. Then, the full marked cluster chronology is jointly equivalent to
\begin{align}
\bar \tau_{i}
\stackrel{\mathsf{D}}{=}
\frac{1}{\beta}
\sum_{\ell=0}^{i-1}
Z_{\pi,\ell}
~~
\forall~
i \in \{1, \dots, \bar N - 1\}
,
\end{align}
with $\bar \tau_{\editTwo{0}} = 0$, where given the marked cluster size $\bar N$ and the sequence of marks $M_0, \dots, M_{\bar N - 1}$, the Dyck path $\pi$ is distributed according to Equation~\eqref{dyckDist} and, given $\pi$, $Z_{\pi,\ell} \sim \mathsf{Exp}\left(\sum_{j=1}^{\ell +1} \kappa_j(\pi) - \ell\right)$ are mutually independent across $\ell \in \{0, \dots, \bar N -2\}$.
\end{theorem}
\proof{Proof.}
By Lemma~\ref{dyckLemma}, we have that the number of direct descendants of each point in the cluster is given by $\kappa_1(\pi)$, \dots, $\kappa_{\bar N-1}(\pi)$. Furthermore, by Theorem~\ref{mEquivThm}, we also know that the (unsorted) response offsets are distributed with density $\bar g(x) / \bar \rho = (\alpha e^{-\beta x})/(\alpha/\beta) = \beta e^{-\beta x}$ for $x \geq 0$; hence, the mutually dependent offsets are $\mathsf{Exp}(\beta)$ distributed. Leveraging the offset structure in step (iii) of Definition~\ref{dpcDef} and the properties of memorylessness and closure under minimums for exponential random variables, we have that the sorted cluster epochs will be governed by the following pattern. Starting at $\bar \tau_0 = 0$, the first of the $\kappa_1(\pi)$ responses will be $\mathsf{Exp}(\beta \kappa_1(\pi))$ distributed, and this time will be $\bar \tau_1$. Then, the time until the next point will be the minimum among the remaining $\kappa_1(\pi) - 1$ responses to the initial point and the $\kappa_2(\pi)$ responses to the first response, and thus $\bar \tau_2 - \bar \tau_1 \sim \mathsf{Exp}(\beta (\kappa_1(\pi) + \kappa_2(\pi) - 1))$. This pattern repeats, with $\bar \tau_{\ell+1} - \bar \tau_\ell \sim \mathsf{Exp}\left(\beta \left(\sum_{j=1}^{\ell+1} \kappa_j(\pi)- \ell\right)\right)$ up to $\bar \tau_{\bar N - 1} - \bar \tau_{\bar N-2} \sim \mathsf{Exp}\left(\beta \left(\sum_{j=1}^{\bar N - 1} \kappa_j(\pi)- \bar N + 2\right)\right)$, and, by memorylessness, these random variables are mutually independent. Finally, to reach the stated distributional equivalences to the sums of $Z_{\pi,\ell}$'s, we can recognize that any cluster epoch can naturally be seen to be the telescoping sum of these gaps, with the mutual $\beta$ coefficient moving to the front through the fact $\mathsf{Exp}(c) \stackrel{\mathsf{D}}{=} \frac{1}{c}\mathsf{Exp}(1)$ for any $c > 0$.
\hfill\Halmos\endproof


By providing the full chronology and extending to marked Hawkes clusters, Theorem~\ref{expMarkedDist} is a generalization of Theorem 2 in \citet{daw2023conditional}, although the rates of the conditionally independent exponential random variables may not seem the same on the surface. In Corollary~\ref{matchCor}, we verify that these expressions do indeed match.

\begin{corollary}\label{matchCor}
Let $\bar g(x) = \alpha e^{-\beta x}$ for all $x \geq 0$ with $\bar \rho = \alpha \slash \beta < 1$. Then,
\begin{align}
\bar \tau 
=
\bar \tau_{\bar N - 1}
\stackrel{\mathsf{D}}{=}
\frac{1}{\beta}
\sum_{i=1}^{\bar N-1}
Y_{\pi,i}
,
\end{align}
where given the marked cluster size $\bar N$ and the sequence of marks $M_0, \dots, M_{\bar N - 1}$, the Dyck path $\pi$ is distributed according to Equation~\eqref{dyckDist} and, given $\pi$, $Y_{\pi,i} \sim \mathsf{Exp}(i+1 - \sum_{j=\bar N-i}^{\bar N-1} \kappa_j(\pi))$ are independent.
\end{corollary}
\proof{Proof.}
Because the total count of the appearances of each integer in $\pi$ must sum to the length of the Dyck path ($\sum_{j=1}^{\bar N - 1}\kappa_j(\pi) = \bar N - 1$), we can recognize that each of the $Z_{\pi,\ell}$ rates can be re-expressed
$$
\sum_{j=1}^{\ell+1} \kappa_j(\pi)- \ell
=
\bar N - 1 - \ell - \sum_{j=\ell+2}^{\bar N - 1} \kappa_j(\pi)
.
$$
Then, by substituting $i+1$ for $\bar N - 1 - \ell$, or, equivalently, $\ell = \bar N - 2 - i$, we simplify this summation to
$$
\bar N - 1 - \ell - \sum_{j=\ell+2}^{\bar N - 1} \kappa_j(\pi)
=
i+1 - \sum_{j=\bar N - i}^{\bar N - 1} \kappa_j(\pi)
,
$$
which matches the stated distribution for $Y_{\pi,i}$.
\hfill\Halmos\endproof

Let us point out that Theorem~\ref{expMarkedDist} not only generalizes Theorem 2 from \citet{daw2023conditional}, it also provides a proof that is substantially simpler and more interpretable. That is, the aforementioned predecessor result is achieved by rather tedious and opaque manipulations of the conditional Laplace-Stieltjes transform (LST) of the duration. Fundamentally, the LST approach relies on the compensator transformation of the cluster epochs akin to the random time change theorem and the associated conditional uniformity of these transformed points, and this change from the time-space to the compensator-space is where the interpretability is lost. Instead, the approach in Theorem~\ref{expMarkedDist} never has to leave the time-space thanks to the discovery of parking functions (and, more generally, Dyck paths) in this chronological context through Theorems~\ref{equivThm} and~\ref{mEquivThm}.

As an aside, we can also recognize that this sum of exponential random variables seems somewhat reminiscent of an absorption time of some continuous time Markov chain. However, it is unclear if this can be properly recreated without the Dyck path upon which Definition~\ref{dpcDef} is built. While we are interested in further investigating whether there is any potential connection here in this exponential response functions setting, this ambiguity (at least at the surface level) demonstrates the benefits of using both the Dyck path construction and the end-state conditioning on $\bar N$ that enables it.

As a final result in this section of the appendix, in Lemma~\ref{rhoLemma}, we now provide the limits of the mean durations for clusters with Borel distributed marks and general response functions at the extremes of interdependence.  This technical result is used in the proof of Theorem~\ref{uThm}.

\begin{lemma}\label{rhoLemma}
Let $\rho \in (0,1)$. As $\rho_\mathsf{1} \to 0$ with $\rho_\mathsf{2} = \rho - \rho_\mathsf{1}$, $\E{\bar \tau^\mathsf{1}} \to 0$ and $\E{\bar \tau^\mathsf{2}} \to \bar{c}_\mathsf{2}$ for some $\bar{c}_\mathsf{2} > 0$. Likewise,  as $\rho_\mathsf{1} \to \rho$ with $\rho_\mathsf{2} = \rho - \rho_\mathsf{1}$, $\E{\bar \tau^\mathsf{2}} \to 0$ and $\E{\bar \tau^\mathsf{1}} \to \bar{c}_\mathsf{1}$ for some $\bar{c}_\mathsf{1} > 0$. 
\end{lemma}
\proof{Proof.}
From Corollary~\ref{barMarks}, we have that $\bar \tau^\mathsf{1}$ and $\bar \tau^\mathsf{2}$ are each equivalent to the duration of a marked Hawkes cluster. By the symmetry of their definitions, it suffices to  prove that $\E{\bar \tau^\mathsf{1}} \to 0$ and $\E{\bar \tau^\mathsf{2}} \to \bar c$ for some $\bar c > 0$ as $\rho_\mathsf{1} \to 0$ with $\rho_\mathsf{2} = \rho - \rho_\mathsf{1}$. Rather than prove these precise statements about the mean durations, we will instead show that $\lim_{\rho_\mathsf{1}\to 0}\PP{\bar \tau^\mathsf{1} = 0} = 1$ and $\lim_{\rho_\mathsf{1} \to 0} \PP{\bar \tau^\mathsf{2} = 0} < 1$. Because the durations are non-negative random variables, this immediately implies the stated results. Within the context of this proof, let us denote the cluster sizes associated with the durations $\bar \tau^\mathsf{1}$ and $\bar \tau^\mathsf{2}$ as $\bar N^\mathsf{1}$ and $\bar N^\mathsf{2}$, respectively.

Let us begin with $\E{\bar \tau^\mathsf{1}}$. By either of the (equivalent) definitions of Hawkes clusters, we can recognize that the event $\{\bar \tau^\mathsf{1} = 0\}$ is equivalent to $\{\bar N^\mathsf{1} = 1\}$: intuitively, the duration is zero when there are no responses. Conditioning on the initial mark $M_0 \sim \mathsf{Borel}(\rho_\mathsf{2})$ and using the conditionally non-stationary Poisson nature of the Hawkes process, the probability of zero duration can thus be expressed
$$
\PP{\bar \tau^\mathsf{1} = 0 }
=
\E{
\PP{\bar N^\mathsf{1} = 1 \mid M_0}
}
=
\E{
e^{-M_0 \int_0^\infty g_\mathsf{1}(u) \mathrm{d}u}
}
=
\E{
e^{- \rho_\mathsf{1} M_0 }
}
.
$$
Then, by the fact that $e^{-x} \geq 1 - x$ for $x \geq 0$, this probability can be lower bounded by
$$
\PP{\bar \tau^\mathsf{1} = 0 }
\geq
1
-
\rho_\mathsf{1} \E{M_0}
=
1
-
\frac{\rho_\mathsf{1}}{1 - \rho_\mathsf{2}}
=
1
-
\frac{\rho_\mathsf{1}}{1 - \rho + \rho_\mathsf{1}}
,
$$
and thus $\PP{\bar \tau^\mathsf{1} = 0} \to 1$ as $\rho_\mathsf{1} \to 0$.

Turning now to $\E{\bar \tau^\mathsf{2}}$, we can similarly argue that 
$$
\PP{\bar \tau^\mathsf{2} > 0}
=
\E{\PP{\bar N^\mathsf{2} > 1 \mid M_0}}
=
1 - \E{e^{-\rho_\mathsf{2} M_0}}
,
$$
where in this mirrored case, the mark distribution is now $\mathsf{Borel}(\rho_\mathsf{1})$. By the same $e^{-x} \geq 1 - x$ inequality, we have
$$
\PP{\bar \tau^\mathsf{2} > 0}
\leq
\rho_\mathsf{2}\E{M_0}
=
\frac{\rho_\mathsf{2}}{1 - \rho_\mathsf{1}}
=
\frac{\rho - \rho_\mathsf{1}}{1 - \rho_\mathsf{1}}
.
$$
Hence, $\lim_{\rho_\mathsf{1} \to 0}\PP{\bar \tau^\mathsf{2} > 0} \leq \rho < 1$.
\hfill\Halmos\endproof


While Lemma~\ref{rhoLemma} focuses only on  showing whether the mean durations converge to 0 or a positive constant, one can see that the proof shows intuition on convergence of the random variables themselves. That is, intuitively, as $\rho_\mathsf{1}$ tends to 0, then the $\mathsf{Borel}(\rho_\mathsf{1})$ marks in the $\bar \tau^\mathsf{2}$ cluster should simply \edit{each equal to one} deterministically. In this case, the limiting cluster should be that of a one-sided, non-marked Hawkes cluster, as was studied in \citet{daw2023conditional}.

\section{\edit{A Distinction Between Methodology: Two Different Parking Functions Per Cluster}}\label{agreeSec}


\edit{Though much of this paper has been primarily focused on the modeling and managerial insights that can be found at this micro-level perspective, the analysis of service interactions conducted here is indebted to Definition~\ref{pfcDef}, which provides a novel alternate construction of the Hawkes cluster model in terms of uniformly random parking functions. Furthermore, this decomposition of the cluster in Definition~\ref{pfcDef} may be of its own interest methodologically. (The same may be said for this notion's generalization to marked Hawkes clusters in Definition~\ref{dpcDef}.) Nevertheless, let us be clear that the present paper is not the first to find a connection between Hawkes clusters and parking functions: \citet{daw2023conditional} first identified these probabilistic combinatorial objects within Hawkes clusters through a conditional uniformity result akin to the random time change theorem. However, what may not be immediately clear is that the parking functions found through that compensator-based transformation are actually \emph{not} the same as the parking functions used in the temporal decomposition within the steps of Definition~\ref{pfcDef}. In this section of the appendix, we will prove their distinction at the sample-path level, and we will highlight and contrast the analytical advantages of each.}


\edit{To make this distinction, let us first introduce notation to clarify the two different ways in which parking functions can arise within any sample path of a Hawkes cluster model. For simplicity in this demonstration, we will focus on a univariate, non-marked cluster, meaning what Definition~\ref{pfcDef} reduces to if $g_1(\cdot) = g_2(\cdot)$ with $\eta = 1$ (or, equivalently, what Definition~\ref{dpcDef} reduces to if $\PP{M_1 = 1} = 1$). This is the stochastic model that is studied in \citet{daw2023conditional}. We will let this simplified univariate cluster have response function $g(\cdot)$, with $\rho = \int_0^\infty g(u) \mathrm{d}u$.}

\edit{When we say that both parking functions will be defined on the same sample path, we specifically mean that they are based around the same collection of points. Let $N \sim \mathsf{Borel}(\rho)$ be the size of this focal cluster, and likewise let $0 = \tau_0 < \tau_1 < \dots < \tau_{N-1}$ be the cluster's epochs. Let $N_t = |\{i \leq N - 1 | \tau_i \leq t\}|$ be the point process that counts these points, where $\lim_{t \to \infty}N_t = N$. Following the equivalence established in Theorem~\ref{equivThm}, we will now invoke Definition~\ref{pfcDef} to define $\pi^\mathsf{AS}$ as the parking function which encodes the response pattern of the points \emph{before} they are sorted (with $\mathsf{AS}$ being an abbreviation for this paper's title). Given the full context of this paper, we will not belabor $\pi^\mathsf{AS}$ with any further explanations. Based on its use in Definition~\ref{pfcDef}, we will refer to $\pi^\mathsf{AS}$ as the \emph{temporal} parking function.}

\edit{By contrast to the focus $\pi^\mathsf{AS}$ has received, we need to now introduce additional terminology and expressions to properly define the second parking function, $\pi^\mathsf{CU}$ \citep[where now $\mathsf{CU}$ abbreviates the title of][]{daw2023conditional}. First, let us define the continuous time \emph{compensator} $\Lambda(t)$ of the stochastic process at time $t$, which is the integral of the stochastic intensity $\mu_t$ from Definition~\ref{hcDef}:
\begin{align}
\Lambda(t)
&=
\int_0^t
\mu_u \mathrm{d}u
=
\sum_{i=0}^{N_t - 1}
G(t - \tau_i)
,
\end{align}
where $G(t) = \int_0^t g(u) \mathrm{d}u$. Then, let the \emph{compensator points} be defined as this continuous time stochastic process inspected specifically at the cluster epochs. That is, for each $k \in \mathbb{Z}_+$, let the $k$th compensator point be given by
\begin{align}
\Lambda_k 
&=
\Lambda(\tau_k)
=
\sum_{i=0}^{k - 1}
G(\tau_k - \tau_i)
.
\end{align}
Notice that now, because the fact that $g(\cdot) > 0$ implies that $G(\cdot)$ is strictly increasing and thus invertible, the cluster is  fully characterized by its collection of compensator points, $0 = \Lambda_0  < \Lambda_1 < \dots < \Lambda_{N-1}$. By additionally observing that $G(t) \leq \rho$ for all $t$ with $\lim_{t \to \infty} G(t) = \rho$, we also have that $\Lambda_k \leq \rho k$ for each $k \in \{1, \dots, N-1\}$. Through Theorem 1 of \citet{daw2023conditional}, we can now define the parking function $\pi^\mathsf{CU}$ via
\begin{align}
\pi^\mathsf{CU}_{(i)}
&=
\left\lceil \frac{\Lambda_i}{\rho} \right\rceil
,
\label{piCUdef}
\end{align}
where $\pi^\mathsf{CU}_{(i)}$ is \editTwo{the} $i$th value of $\pi^\mathsf{CU}$ when sorted non-decreasingly. Based on the manner of this construction, we will refer to $\pi^\mathsf{CU}$ as the \emph{compensator-based} parking function.}


\edit{One can immediately recognize that the two parking functions are equivalent in distribution. Given $N = n+1$ for any $\mathbb{Z}_+$, both $\pi^\mathsf{AS} \in \mathsf{PF}_n$ and $\pi^\mathsf{CU} \in \mathsf{PF}_n$ are uniformly random parking functions of length $n$. However, we can also now demonstrate that these two need not agree on any given sample path, as we have claimed.}

\edit{Consider the following example for a Hawkes cluster model with power law response function $g(x) = 1/(2+x)^2$, which yields $\rho = 1/2$. Suppose that $N  = 3$ (which occurs with probability $\PP{N=3} = e^{-3/2}(3/2)^2/3! \approx 8.4\%$), and additionally suppose that $\pi^\mathsf{AS} = [1, 1]$ (which occurs with probability $\PP{\pi^\mathsf{AS} = [1,1] \mid N=3} = 1/3$). Then, $\pi^\mathsf{CU}$ will instead be equal to $[1,1]$ if and only if
\begin{align}
\Lambda_2 
&= 
G(\tau_2 - \tau_1) + G(\tau_2)
=
\frac{\tau_2 - \tau_1}{2(\tau_2 - \tau_1) + 4}
+
\frac{\tau_2}{2\tau_2 + 4}
< 
\frac{1}{2}
=
\rho
,
\label{comp2exEq}
\end{align}
which will hold so long as $\tau_2 - 4/\tau_2 < \tau_1 < \tau_2$. Because $\pi^\mathsf{AS} = [1,1]$, we have by Definition~\ref{pfcDef} that $\tau_1$ and $\tau_2$ are ordered draws from the density $g(x)/\rho = 2/(2+x)^2$. Thus, \editTwo{one can explicitly compute the conditional probability that the event in \eqref{comp2exEq} occurs:}
\begin{align*}
\int_0^\infty \int_{(u_2-4/u_2)^+}^{u_2} \frac{8}{(2+u_2)^2(2+u_1)^2}\mathrm{d}u_1 \mathrm{d}u_2
&\approx
32.9\%
.
\end{align*}
Hence, this snippet alone is enough to justify the claim that the two parking functions need not agree. In fact, in this setting with power law response kernel, if $N=3$ and $\pi^\mathsf{AS} = [1,1]$, then this example shows that it is actually \emph{more likely \editTwo{than} not} that the compensator-based parking function will not match the temporal parking function.}

\edit{Throughout the remainder of this section, we will prove that this observation is not an isolated example. First, let us consider the case in which $g(\cdot)$ is an exponential function, in which case we can leverage Theorem~\ref{expMarkedDist} to demonstrate this claimed rarity. Specifically, in this exponential kernel setting, we will show that even \editTwo{when} the temporal parking function is conditioned to take on the likeliest value for a parking function (irrespective of ordering), there is a vanishingly small chance that the compensator-based parking function will match it (also irrespective of ordering). That is, for a uniformly random parking function $\pi \in \mathsf{PF}_n$, the most likely outcome after sorting non-decreasingly is
$$
\PP{\vec{\pi} =  [1, 2, \dots, n]}
=
\frac{n!}{(n+1)^{n-1}}
,
$$
where, as we have done in the proof of Theorem~\ref{equivThm}, we use the $\vec{\cdot}$  notation to indicate the non-decreasing sorting of a parking function (which, in effect, reduces to a Dyck path, as discussed in Appendix~\ref{markedSec}). As we now show in Proposition~\ref{expAgreeProp}, for large clusters, there is an exponentially small probability that the sorted compensator-based parking function is equal to $[1, 2, \dots, n]$ given that the sorted temporal parking function is equal to $[1,2,\dots, n]$. Moreover, as the cluster size grows large, this conditional probability shrinks to 0.}


\begin{proposition}\label{expAgreeProp}
\edit{Suppose that for all $x \geq 0$, $g(x) = \alpha e^{-\beta x}$ for some $\beta > \alpha > 0$. Then, given that $N = n + 1$,
\begin{align}\label{expAgreeEq}
\PP{
\vec{\pi}^\mathsf{CU} = [1, 2, \dots, n] 
\mid 
\vec{\pi}^\mathsf{AS} = [1, 2, \dots, n]}
&=
\log(2)^{n-1}
,
\end{align}
for all $n \in \mathbb{Z}_+$.}
\end{proposition}

\edit{Notice that though $N=n+1$ is not included among the conditional events within the probability function in Proposition~\ref{expAgreeProp} for the sake of brevity, it is implied by the parking functions being of length $n$. Before we prove this result, let us first introduce and obtain a supporting technical lemma.}

\begin{lemma}\label{uniLemma}
\edit{
Let $U_i \sim \mathsf{Uni}(0,1)$ for $i \in \{1, \dots, n\}$ be mutually independent. Then, 
\begin{align}
\PP{U_k \left( 1 + \sum_{i=1}^{k-1}\prod_{j=i}^{k-1}U_j\right) < 1 \,\forall\, k \in \{1, \dots, n\}}
&=
\log(2)^{n-1}
,
\label{uniLemmaMain}
\end{align}
where $n \in \mathbb{Z}_+$.}
\end{lemma}
\proof{Proof.}\edit{Let us start by instead proving that, for every $n \in \mathbb{Z}_+$, we have that the following expected value can be written
\begin{align}
\E{
\frac{1}{1 + \sum_{i=1}^n\prod_{j=i}^n U_j}
\prod_{k=1}^n
\mathbf{1}\left\{
U_k
\left(
1 + \sum_{i=1}^{k-1}\prod_{j=i}^{k-1} U_j
\right)
<
1
\right\}}
&=
\log(2)^n
.
\label{uniLemmaEq}
\end{align}
We will justify this claim through induction on $n$. Starting with the base case at $n = 1$, notice that this expression quickly simplifies to
$$
\E{
\frac{1}{1 + U_1}
\prod_{k=1}^n
\mathbf{1}\left\{
U_1
<
1
\right\}}
=
\int_0^1 \frac{1}{1+u_1}\mathrm{d}u_1
=
\log(2)
.
$$
Moving to the inductive step, we will assume that \eqref{uniLemmaEq} holds up to some value of $n$. Then, at $n+1$, we can condition on $\mathbf{U}_n = [U_1, \dots, U_n]$ to re-write this expected value as 
\begin{small}
\begin{align*}
&
\E{
\frac{1}{1 + \sum_{i=1}^{n+1}\prod_{j=i}^{n+1} U_j}
\prod_{k=1}^{n+1}
\mathbf{1}\left\{
U_k
\left(
1 + \sum_{i=1}^{k-1}\prod_{j=i}^{k-1} U_j
\right)
<
1
\right\}}
\\
&
=
\E{
\E{
\frac{1}{1 + U_{n+1}\left(1 + \sum_{i=1}^{n}\prod_{j=i}^{n} U_j\right)}
\mathbf{1}\left\{
U_{n+1}
<
\frac{1}{1 + \sum_{i=1}^{n}\prod_{j=i}^{n} U_j}
\right\} 
\,\Bigg|\, \mathbf{U}_n}
\prod_{k=1}^{n}
\mathbf{1}\left\{
U_k
\left(
1 + \sum_{i=1}^{k-1}\prod_{j=i}^{k-1} U_j
\right)
<
1
\right\} 
}
,
\end{align*}
\end{small}by way of the law of total expectation. Now, the inner conditional expectation (taken relative to $U_{n+1}$ only) can be simplified to
\begin{align*}
\E{
\frac{1}{1 + U_{n+1}\left(1 + \sum_{i=1}^{n}\prod_{j=i}^{n} U_j\right)}
\mathbf{1}\left\{
U_{n+1}
<
\frac{1}{1 + \sum_{i=1}^{n}\prod_{j=i}^{n} U_j}
\right\} 
\,\Bigg|\, \mathbf{U}_n}
&=
\frac{\log(2)}{1 + \sum_{i=1}^{n}\prod_{j=i}^{n} U_j}
,
\end{align*}
due to the fact that $\int_0^{1/c} 1/(1+c u) \mathrm{d}u = \log(2)/c$ for any $c > 0$. Thus, substituting into the original expectation we now have
\begin{align*}
&
\E{
\frac{1}{1 + \sum_{i=1}^{n+1}\prod_{j=i}^{n+1} U_j}
\prod_{k=1}^{n+1}
\mathbf{1}\left\{
U_k
\left(
1 + \sum_{i=1}^{k-1}\prod_{j=i}^{k-1} U_j
\right)
<
1
\right\}}
\\
&
\qquad
=
\log(2)
\E{
\frac{1}{1 + \sum_{i=1}^{n}\prod_{j=i}^{n} U_j}
\prod_{k=1}^{n}
\mathbf{1}\left\{
U_k
\left(
1 + \sum_{i=1}^{k-1}\prod_{j=i}^{k-1} U_j
\right)
<
1
\right\} 
}
,
\end{align*}
and by the inductive hypothesis, we complete the proof of \eqref{uniLemmaEq}.}

\edit{It now remains to prove \eqref{uniLemmaMain} via use of \eqref{uniLemmaEq}. Notice that \editTwo{for} $n=1$ and $n=2$, \eqref{uniLemmaMain} reduces to $\PP{U_1 < 1} = 1$ and $\PP{U_1(1+U_2) < 1} = \E{1/(1+U_1)} = \log(2)$, respectively, so let us consider $n \geq 3$. For any such $n$, the probability that all $n$ events hold can be written as an expectation of the product of corresponding indicator random variables:
\begin{align*}
\PP{U_k \left( 1 + \sum_{i=1}^{k-1}\prod_{j=i}^{k-1}U_j\right) < 1 \,\forall\, k \in \{1, \dots, n\}}
&=
\E{
\prod_{k=1}^n
\mathbf{1}\left\{
U_k \left( 1 + \sum_{i=1}^{k-1}\prod_{j=i}^{k-1}U_j\right) < 1
\right\}
}
.
\end{align*}
By a similar total expectation approach conditioning on $\mathbf{U}_{n-1} = [U_1, \dots, U_{n-1}]$, we can observe that
\begin{align*}
&
\E{
\prod_{k=1}^n
\mathbf{1}\left\{
U_k \left( 1 + \sum_{i=1}^{k-1}\prod_{j=i}^{k-1}U_j\right) < 1
\right\}
}
\\
&\qquad=
\E{
\E{\mathbf{1}\left\{
U_n \left( 1 + \sum_{i=1}^{n-1}\prod_{j=i}^{n-1}U_j\right) < 1
\right\}
\, \bigg| \,
\mathbf{U}_{n-1}
}
\prod_{k=1}^{n-1}
\mathbf{1}\left\{
U_k \left( 1 + \sum_{i=1}^{k-1}\prod_{j=i}^{k-1}U_j\right) < 1
\right\}
}
\\
&\qquad=
\E{
\frac{1}{ 1 + \sum_{i=1}^{n-1}\prod_{j=i}^{n-1}U_j}
\prod_{k=1}^{n-1}
\mathbf{1}\left\{
U_k \left( 1 + \sum_{i=1}^{k-1}\prod_{j=i}^{k-1}U_j\right) < 1
\right\}
}
,
\end{align*}
and by \eqref{uniLemmaEq}, we have that this simplifies to $\log(2)^{n-1}$, which yields the stated result.}
\hfill\Halmos\endproof

\edit{We can now proceed with proving Proposition~\ref{expAgreeProp}, which follows relatively quickly from this technical lemma used in conjunction with the distributional identities for the case of the exponential response kernel, as shown in Theorem~\ref{expMarkedDist}.}

\proof{Proof of Proposition~\ref{expAgreeProp}.}\edit{To begin, let us notice that $\vec{\pi}^\mathsf{CU} = [1, \dots, n]$ if and only if $\lceil \Lambda_k /\rho \rceil > k - 1$ for every $k \in \{1, \dots, n\}$. Equivalently, for the sorted compensator-based parking function to be equal to this likeliest Dyck path, it must be that  $\rho k - \Lambda_k < \rho$ for every index $k$. Then, under this exponential kernel, the Hawkes compensator points are known to satisfy the recurrence relation
\begin{align}
\rho k - \Lambda_k
=
\left(
\rho k - \Lambda_{k-1}
\right)
e^{-\beta (\tau_k - \tau_{k-1})}
,
\label{lamRec}
\end{align}
for each $k \geq 2$, where $\rho - \Lambda_1 = \rho e^{-\beta \tau_1}$ \citep[see, e.g., equation (14) of][]{daw2023conditional}.}

\edit{Now, by Theorem~\ref{expMarkedDist}, given that $\vec{\pi}^\mathsf{AS} = [1, \dots, n]$, the inter-epoch times are independently and identically distributed such that $\tau_i - \tau_{k-1} \sim \mathsf{Exp}(\beta)$ for each $k \in \{1, \dots, n\}$, with $\tau_0 = 0$. Thus, conditional on $\vec{\pi}^\mathsf{AS} = [1, \dots, n]$, we have that 
$$
e^{-\beta (\tau_k - \tau_{k-1})}
\stackrel{\mathsf{D}}{=}
U_k
\stackrel{\mathsf{iid}}{\sim} 
\mathsf{Uni}(0,1)
,
$$
for each $k \in \{1, \dots, n\}$. By consequence of this fact, we can re-express the  $\rho k - \Lambda_k$ in \eqref{lamRec} through these standard uniform random variables as $\rho k - \Lambda_k = (\rho  + \rho(k-1) - \Lambda_{k-1}) U_k$, with $\Lambda_1 = \rho U_1$. One can quickly verify that the unique equation satisfying this recursion is
\begin{align*}
\rho k - \Lambda_k
&=
\rho
\sum_{i=1}^k \prod_{j=i}^k U_j
=
\rho
\left(
1
+
\sum_{i=1}^{k-1} \prod_{j=i}^{k-1} U_j
\right)
U_k
.
\end{align*}
Accordingly, given $\vec{\pi}^\mathsf{AS} = [1, \dots, n]$, the event that $\rho k - \Lambda_k < \rho$ is equivalent to
\begin{align*}
\left\{\rho k - \Lambda_k < \rho \right\}
=
\left\{
U_k
\left(
1
+
\sum_{i=1}^{k-1} \prod_{j=i}^{k-1} U_j
\right)
< 
1
\right\}
,
\end{align*}
for each $k \in \{1, \dots, n\}$. Moreover, the conditional agreement probability is equal to the probability of the intersection of these events, and thus by Lemma~\ref{uniLemma} we achieve~ \eqref{expAgreeEq}.}
\hfill\Halmos\endproof

\edit{Now, to demonstrate the distinction between $\pi^\mathsf{AS}$ and $\pi^\mathsf{CU}$ broadly and generally, let us invoke another family of combinatorial objects. Specifically, let us now introduce \emph{generalized parking functions}, which add an extra degree of freedom to the classic parking function concept \citep[see, e.g.,][]{stanley2002polytope,yan2015parking,kenyon2023parking}. Relative to the parking functions given in Definition~\ref{pfDef} and in a re-purposing of the original car metaphor, the essential difference is that the number of parking spaces now need to simply be at least as large as the number of vehicles, rather than exactly the same. Following the literature, we formalize this notion now in Definition~\ref{gpfDef}.}


\begin{definition}[Generalized Parking Function]\label{gpfDef}
\edit{For $m, n \in \mathbb{Z}_+$ with $m \leq n$, $\pi \in \mathbb{Z}_+^m$ is a \emph{generalized parking function} of length $m$ over $n$ preferences if and only if it is such that, when sorted $\pi_{(1)} \leq \dots \leq \pi_{(m)}$, $\pi_{(i)} \leq n - m + i$ for each $i \in \{1, \dots, m\}$.
} \hfill \Halmos 
\end{definition}

\edit{In the cars and spaces analogy, $m$ is the number of cars, $n$ is the number of spaces, and $\pi_i \in \{1, \dots, n\}$ is the preferred parking space for car $i \in \{1, \dots, m\}$. If $\pi$ satisfies the conditions of Definition~\ref{gpfDef}, then all cars will be able to park. Much like how we let $\mathsf{PF}_k$ denote the set of all parking functions of length $k$, we will let $\mathsf{PF}_{m,n}$ denote the set of all generalized parking functions of length $m$ over $n$ preferences. A key fact we will use from the literature is that the cardinality of this set is  $|\mathsf{PF}_{m,n}| = (n-m+1)(n+1)^{m-1}$, as formally stated in Proposition~\ref{gpfCard}.}

\begin{proposition}\label{gpfCard}
\edit{There are $(n-m+1)(n+1)^{m-1}$ generalized parking functions of length $m \in \mathbb{Z}_+$ over a space of $n  \in \mathbb{Z}_+$ preferences, where $m \leq n$.}
\end{proposition}
\proof{Proof.}\edit{This can be shown through a natural generalization of Pollak's circle argument; see, e.g., Theorem 2.3 of \citet{kenyon2023parking}.}
\hfill\Halmos\endproof










\edit{Though generalized parking functions can likely be used to construct interesting cluster models on their own, our present uses for them will actually be tools for analysis of the (non-generalized)  parking functions we have already used throughout the paper. We begin this now with the following technical lemma. For a uniformly random parking function of length $n \in \mathbb{Z}_+$, in Lemma~\ref{k1Lemma} we explicitly obtain the distribution for the number of appearances of the value $1$. Fittingly for these objects, its proof will be combinatorial in nature.}

\begin{lemma}\label{k1Lemma}
\edit{Let $\pi \in \mathsf{PF}_n$ be a uniformly random parking function of length $n$. Then, the number of appearances of the value $1$ in $\pi$, $\kappa_1(\pi) = \left|\{j ~|~ \pi_j = 1\}\right|$, is distributed such that 
\begin{align}
\left(\kappa_1(\pi) - 1\right) \sim \mathsf{Bin}\left(n-1, \frac{1}{n+1}\right)
,
\end{align}
with the probability mass function of $\kappa_1(\pi)$ given by
\begin{align}
\PP{\kappa_1(\pi) = m}
&=
{n-1 \choose m-1}
\frac{n^{n-m}}{(n+1)^{n-1}}
,
\label{k1PMF}
\end{align}
where $m \in \{1, \dots, n\}$.}
\end{lemma}
\proof{Proof.}
\edit{If the value $1$ appears exactly $m \in \{1, \dots, n\}$ times in a parking function of length $n$, then the remaining $n-m$ coordinates must take on values from 2 to $n$. Moreover, if the remaining values are to be sorted, the $i$th largest value cannot exceed $i + m$; otherwise, the original vector would not be a parking function. Subtracting $1$ from all of the remaining values, this remnant collection becomes $n-m$ entries on $\{1, \dots, n-1\}$ such that, when sorted non-decreasingly, the $i$th value is no more than $m - 1 + i = (n-1) - (n-m) + i$ for each $i \in \{1, \dots, n-m\}$. Hence, the remaining non-$1$ values from the length $n$ parking function form a generalized parking function of length $n-m$ over $n-1$ preferences.}

\edit{By Proposition~\ref{gpfCard}, there are $(n-1 - (n-m) + 1)n^{n-m-1} = m n^{n-m-1}$ generalized parking functions of such length for so many preferences. Furthermore, given a generalized parking function of length $n-m$ on values $2$ to $n$, there are ${n \choose m}$ ways to intersperse $m$ copies of the value $1$ to form a vector of length $n$. Hence, there are 
$$
{n \choose m}
m n^{n-m-1}
=
\frac{n!}{m!(n-m)!}
m n^{n-m-1}
=
\frac{(n-1)!}{(m-1)!(n-m)!}
n^{n-m}
=
{n-1 \choose m-1}
n^{n-m}
$$
parking functions of length $n$ with exactly $m$ values equal to $1$. Because there are $(n+1)^{n-1}$ parking functions of length $n$, we achieve \eqref{k1PMF} as the probability that a uniformly random parking function of this length contains exactly $m$ copies of $1$. By further recognizing that
$$
{n-1 \choose m-1}
\frac{n^{n-m}}{(n+1)^{n-1}}
=
{n-1 \choose m-1}
\left(
\frac{1}{n+1}
\right)^{m-1}
\left(
\frac{n}{n+1}
\right)^{n-m}
,
$$
for each $m \in \{1, \dots, n\}$, we complete the proof.}
\hfill\Halmos\endproof




\edit{Thanks to this characterization of the distribution of the number of values equal to $1$ in a uniformly random parking function, we can now provide a simple yet general demonstration that disagreement of $\pi^\mathsf{AS}$ and $\pi^\mathsf{CU}$ cannot be entirely avoided, even at the first opportunity for these objects to be mismatched. We will now show that agreement between the number of values equal to $1$ in the temporal and compensator-based parking functions cannot be guaranteed for any cluster with size 3 or larger. Specifically, in Proposition~\ref{k1ConProp}, given that the cluster is not trivially small (so that the parking functions are well-defined and not simply equal to $[1]$), \editTwo{we show that there is a strictly positive probability that $\pi^\mathsf{CU}$ will have more than one entry equal to $1$ when given that there is only entry equal to $1$ in $\pi^\mathsf{AS}$}. As we will show, general unavoidability of disagreement between the parking functions will follow as an immediate consequence of this result.}

\begin{proposition}\label{k1ConProp}
\edit{Let $n \in \mathbb{Z}_+$. Given that the Hawkes cluster contains $N = n + 1$ points, the probability that the compensator-based parking function, $\pi^\mathsf{CU}$, contains only one occurrence of the value 1 given that the temporal parking function, $\pi^\mathsf{AS}$, contains only one 1 is
\begin{align}
\PP{\kappa_1(\pi^\mathsf{CU}) = 1 \mid \kappa_1(\pi^\mathsf{AS}) = 1, N = n+1}
&=
1
-
\mathcal{E}(n)
,
\end{align}
where
\begin{align}
\mathcal{E}(n)
&=
\frac{1}{\rho}
\left(
1
-
\frac{1}{n}
\right)
\int_0^{\frac{1}{2}}
G\left(
G^{-1}\left(\rho ( 1 - u ) \right)
-
G^{-1}\left(\rho u\right)
\right)
\left(
2 - u
\right)
\left(
1 - \frac{u}{n}
\right)^{n-3}
\mathrm{d}u
>
0
,
\label{calEdef}
\end{align}
for all $n \geq 1$.}
\end{proposition}
\proof{Proof.}
\edit{Much like the discussion throughout this section, this proof will start with $\pi^\mathsf{AS}$ as obtained via Definition~\ref{pfcDef}, and we will then build towards identifying $\pi^\mathsf{CU}$ on the same cluster sample path. Specifically, for this result, we will focus on the number of appearances of the value $1$ in each parking function, $\kappa_1\left(\pi^\mathsf{CU}\right)$ and $\kappa_1\left(\pi^\mathsf{AS}\right)$. From Definition~\ref{pfcDef}, it is clear that the event $\{\kappa_1\left(\pi^\mathsf{AS}\right) = 1\}$ means that there is precisely one response directly to the initial point. For $\kappa_1\left(\pi^\mathsf{CU}\right)$, there is naturally less immediate temporal interpretation, but we can quickly observe that this implies that the second compensator point must be at least as large as $\rho$:
\begin{align}
\left\{\kappa_1(\pi^\mathsf{CU}) = 1\right\}
=
\left\{\pi_{(2)}^\mathsf{CU} = 2\right\}
&=
\left\{
\left\lceil
\frac{1}{\rho}
\left(
G(\tau_2)
+ 
G(\tau_2 - \tau_1)
\right)
\right\rceil
=
2
\right\}
=
\left\{
G(\tau_2)
+ 
G(\tau_2 - \tau_1)
\geq
\rho
\right\}
.
\label{cuEventEq}
\end{align}
Let us now analyze this event through the conditional distributions of $\tau_1$ and $\tau_2$.}

\edit{Given that $\kappa_1\left(\pi^\mathsf{AS}\right) = 1$, following the structure from Definition~\ref{pfcDef} as discussed above, there exist a pair of independent random variables which we will denote as $T_1$ and $T_{(1,\kappa_2)}$ such that $\tau_1 =  T_1$ and $\tau_2 = T_1 + T_{(1,\kappa_2)}$, where $T_1$ is drawn according to the distribution given by $G(\cdot)/\rho$, and where $T_{(1,\kappa_2)}$ is the minimum of $\kappa_2\left(\pi^\mathsf{AS}\right)$ i.i.d.~draws from $G(\cdot)/\rho$. Because the fact that $g(\cdot)>0$ implies that $G(\cdot)$ is invertible, $T_1$ and $T_{(1,\kappa_2)}$ can expressed in terms of inversely transformed standard uniform random variables (or order statistics thereof). That is, for $U_1 \sim \mathsf{Uni}(0,1)$ and $U_{(1,\kappa_2)} \sim \mathsf{Beta}\left(1, \kappa_2\left(\pi^\mathsf{AS}\right)\right)$ independent, $T_1 \stackrel{\mathsf{D}}{=} G^{-1}\left(\rho U_1\right)$ and $T_{(1,\kappa_2)} \stackrel{\mathsf{D}}{=} G^{-1}\left(\rho U_{(1,\kappa_2)}\right)$.}


\edit{Let us now additionally condition on $\kappa_2\left(\pi^\mathsf{AS}\right)$, where, for the sake of brevity, we will frequently suppress the dependence on the temporal parking function and simply let $\kappa_2 := \kappa_2\left(\pi^\mathsf{AS}\right)$ when the context is clear. (So as to assuage any concerns of ambiguity in this reduced notation, let us note that this proof will not be analyzing $\kappa_2\left(\pi^\mathsf{CU}\right)$.) Using this value along with the fact that $\kappa_1\left(\pi^\mathsf{AS}\right) = 1$, we can express the conditional probability of the agreement event from Equation~\eqref{cuEventEq} in terms of the underlying uniform random variables as
\begin{align*}
\PP{ 
\kappa_1\left(\pi^\mathsf{CU}\right) = 1
\mid 
\kappa_2\left(\pi^\mathsf{AS}\right), \kappa_1\left(\pi^\mathsf{AS}\right) = 1}
&=
\PP{ 
G\left(
G^{-1}\left(\rho U_1\right)
+ 
G^{-1}\left(\rho U_{(1,\kappa_2)}\right)
\right)
+ 
\rho U_{(1,\kappa_2)}
> 
\rho
\,\big|\,
\kappa_2}
\\
&=
\PP{ 
U_1
> 
\frac{1}{\rho}
G\left(
G^{-1}\left(\rho ( 1 - U_{(1,\kappa_2)} ) \right)
-
G^{-1}\left(\rho U_{(1,\kappa_2)}\right)
\right)
\,\big|\,
\kappa_2}
.
\end{align*}
Then, by applying the density of $U_{(1,\kappa_2)}$, we can further re-express this probability as an integral over probability statements relative to $U_1$. That is,
\begin{align*}
&
\PP{ 
U_1
> 
\frac{1}{\rho}
G\left(
G^{-1}\left(\rho ( 1 - U_{(1,\kappa_2)} ) \right)
-
G^{-1}\left(\rho U_{(1,\kappa_2)}\right)
\right)
\,\big|\,
\kappa_2}
\\
&\qquad=
\int_0^1
\PP{ 
U_1
> 
\frac{1}{\rho}
G\left(
G^{-1}\left(\rho ( 1 - u ) \right)
-
G^{-1}\left(\rho u\right)
\right)
}
\kappa_2(1-u)^{\kappa_2 - 1}
\mathrm{d}u
.
\end{align*}
Now, because $G^{-1}(\cdot)$ is an increasing function by consequence of the fact that $G(\cdot)$ is increasing, we can split the interval of this integral over $u$ into two parts. Specifically, when $u > 1/2$ then $G^{-1}(\rho(1-u)) < G^{-1}(\rho u)$ and thus the event within the probability function is vacuously true, whereas when $u \leq 1/2$ the probability will be non-trivial. Applying this logic and simplifying, we find
\begin{align*}
&
\int_0^1
\PP{ 
U_1
> 
\frac{1}{\rho}
G\left(
G^{-1}\left(\rho ( 1 - u ) \right)
-
G^{-1}\left(\rho u\right)
\right)
}
\kappa_2(1-u)^{\kappa_2 - 1}
\mathrm{d}u
\\
&\qquad
=
\left(\frac{1}{2}\right)^{\kappa_2}
+
\int_0^{\frac{1}{2}}
\left(
1
-
\frac{1}{\rho}
G\left(
G^{-1}\left(\rho ( 1 - u ) \right)
-
G^{-1}\left(\rho u\right)
\right)
\right)
\kappa_2(1-u)^{\kappa_2 - 1}
\mathrm{d}u
\\
&\qquad
=
1
-
\frac{1}{\rho}
\int_0^{\frac{1}{2}}
G\left(
G^{-1}\left(\rho ( 1 - u ) \right)
-
G^{-1}\left(\rho u\right)
\right)
\kappa_2(1-u)^{\kappa_2 - 1}
\mathrm{d}u
,
\end{align*}
which now leaves us to carefully consider $\kappa_2$.}

\edit{As can be surmised from Proposition~\ref{gpfCard} (\editTwo{and} rigorously attained from Definition~\ref{gpfDef} in the proof of Lemma~\ref{k1Lemma}), if a parking function $\pi \in \mathsf{PF}_k$ for any $k \geq 2$ is such that $\kappa_1(\pi) = 1$, then the remaining elements $\pi_{(2)}, \dots, \pi_{(k)}$ will themselves form a parking function of length $k-1$ on the values 2 through $k$. Thus, given $\kappa_1(\pi) = 1$, by Lemma~\ref{k1Lemma}, $(\kappa_2 - 1) \sim \mathsf{Bin}\left(k-2, 1/k\right)$. Hence, in this context, given that $\kappa_1\left(\pi^\mathsf{AS}\right) = 1$, we have that $\left(\kappa_2\left(\pi^\mathsf{AS}\right) - 1\right) \sim \mathsf{Bin}\left(n-2, 1/n\right)$.} 

\edit{Now, if $X$ is a random variable such $X \sim \mathsf{Bin}(k,p)$ with $k \in \mathbb{N}$ and $p \in (0,1)$, then one has that
\begin{align*}
\E{(X + 1)z^X}
=
k p z (1 - p + pz)^{k-1}
+
(1-p+pz)^k
=
\left(
1 - p + (k+1)pz
\right)
\left(
1-p + pz
\right)^{k-1}
,
\end{align*}
which follows from the fact that $\E{z^X} = (1-p + pz)^k$ for any $z \in \mathbb{R}$. Applying this fact to the expectation relative to $\kappa_2\left(\pi^\mathsf{AS}\right)$ given that $\kappa_1\left(\pi^\mathsf{AS}\right) = 1$, we find now that
\begin{align*}
&
\E{
1
-
\frac{1}{\rho}
\int_0^{\frac{1}{2}}
G\left(
G^{-1}\left(\rho ( 1 - u ) \right)
-
G^{-1}\left(\rho u\right)
\right)
\kappa_2(1-u)^{\kappa_2 - 1}
\mathrm{d}u
\,\,\Bigg|\,\,
\kappa_1\left(\pi^\mathsf{AS}\right) = 1
}
\\
&\qquad=
1
-
\frac{1}{\rho}
\left(
1
-
\frac{1}{n}
\right)
\int_0^{\frac{1}{2}}
G\left(
G^{-1}\left(\rho ( 1 - u ) \right)
-
G^{-1}\left(\rho u\right)
\right)
\left(
2 - u
\right)
\left(
1 - \frac{u}{n}
\right)^{n-3}
\mathrm{d}u
,
\end{align*}
and through simplification we thus complete the proof.}
\hfill\Halmos\endproof

\edit{Though Proposition~\ref{k1ConProp} focuses on a conditional probability of a specific type of disagreement, it now prepares us to show that, universally, agreement cannot be guaranteed. In Theorem~\ref{disagreeThm}, the generality of this disagreement result spans across all Hawkes clusters, and the probability statement is not conditioned on a particular parking function nor on the overall cluster size. In the original, time-forward cluster language of Definition~\ref{hcDef}, this can be thought of as a probability statement made before the cluster is initialized with the time 0 point. Hence, we now show that for a realization of an arbitrary Hawkes cluster model, there is a strictly positive probability that the temporal parking function and the compensator-based parking function are not the same at the sample-path level (even when ignoring the order of the values).}


\begin{theorem}\label{disagreeThm}
\edit{Let $\pi^\mathsf{CU}$ and $\pi^\mathsf{AS}$ be the compensator-based and temporal parking functions, respectively, for a Hawkes cluster model with $\rho = \int_0^\infty g(u) \mathrm{d}u$. The probability that the two parking functions agree, irrespective of order, is no more than
\begin{align}
\PP{\vec{\pi}^\mathsf{CU} = \vec{\pi}^\mathsf{AS}}
\leq
1
-
\rho e^{-\rho}
\int_0^{\frac{1}{2}}
G\left(
G^{-1}\left(\rho ( 1 - u ) \right)
-
G^{-1}\left(\rho u\right)
\right)
e^{-\rho u}
\mathrm{d}u
,
\end{align}
and thus $\PP{\vec{\pi}^\mathsf{CU} = \vec{\pi}^\mathsf{AS}} < 1$ for any $\rho \in (0, 1)$.}
\end{theorem}
\proof{Proof.}
\edit{
We will prove this by successively conditioning on the layers of cluster model as ordered in Definition~\ref{pfcDef} and then invoking the results established in this section of the appendix. First, let us decompose the event according to the size of the cluster,
\begin{align*}
\PP{\vec{\pi}^\mathsf{CU} = \vec{\pi}^\mathsf{AS}}
&=
\sum_{n=0}^\infty
\PP{\vec{\pi}^\mathsf{CU} = \vec{\pi}^\mathsf{AS} \mid N = n+1}
\PP{N=n+1}
, 
\end{align*}
and then additionally condition the on whether the value $1$ appears in $\pi^\mathsf{AS}$ once or more,
\begin{align*}
\PP{\vec{\pi}^\mathsf{CU} = \vec{\pi}^\mathsf{AS} \mid N = n+1}
&=
\PP{\vec{\pi}^\mathsf{CU} = \vec{\pi}^\mathsf{AS} \mid \kappa_1\left(\pi^\mathsf{AS}\right) = 1,  N = n+1}
\left(\frac{n}{n+1}\right)^{n-1}
\\
&\qquad
+
\PP{\vec{\pi}^\mathsf{CU} = \vec{\pi}^\mathsf{AS} \mid \kappa_1\left(\pi^\mathsf{AS}\right) > 1, N = n+1}
\left(
1
-
\left(\frac{n}{n+1}\right)^{n-1}
\right)
, 
\end{align*}
where we have applied Lemma~\ref{k1Lemma} in conditioning. Now, we can notice that the event that the value $1$ appears in the two parking functions the same number of times, $\{\kappa_1(\pi^\mathsf{CU}) = \kappa_1(\pi^\mathsf{AS})\}$, is a subset of the full agreement event, $\{\vec{\pi}^\mathsf{CU} = \vec{\pi}^\mathsf{AS}\}$, and thus we upper bound $\PP{\vec{\pi}^\mathsf{CU} = \vec{\pi}^\mathsf{AS} \mid \kappa_1\left(\pi^\mathsf{AS}\right) = 1,  N = n+1} \leq \PP{\kappa_1\left(\pi^\mathsf{CU}\right) = 1 \mid \kappa_1\left(\pi^\mathsf{AS}\right) = 1,  N = n+1} = 1 - \mathcal{E}(n)$ via Proposition~\ref{k1ConProp}. Rather than characterize the more complicated case when $\kappa_1(\pi^\mathsf{AS}) > 1$, we will simply take a crude upper bound of this probability: $\PP{\vec{\pi}^\mathsf{CU} = \vec{\pi}^\mathsf{AS} \mid \kappa_1\left(\pi^\mathsf{AS}\right) > 1, N = n+1} \leq 1$.}

\edit{Together, after simplification, this makes the probability of the agreement event upper bounded by
\begin{align*}
\PP{\vec{\pi}^\mathsf{CU} = \vec{\pi}^\mathsf{AS}}
&\leq
1
-
\sum_{n=1}^\infty
\mathcal{E}(n)
\left(\frac{n}{n+1}\right)^{n-1}
\PP{N=n+1}
, 
\end{align*}
where we have dropped the $n=0$ summand to account for $\mathcal{E}(0) = 0$, letting us otherwise take $\mathcal{E}(n)$ on $n \in \mathbb{Z}_+$ as given in Proposition~\ref{k1ConProp}. Now, by substituting the probability mass function for the Borel distribution, let us further recognize that
$$
\left(\frac{n}{n+1}\right)^{n-1}
\PP{N=n+1}
=
\left(\frac{n}{n+1}\right)^{n-1}
\frac{\left(\rho(n+1)\right)^{n}e^{-\rho(n+1)}}{(n+1)!}
=
n^{n-1}
\rho^n \frac{e^{-\rho(n+1)}}{n!}
=
\rho e^{-\rho}
\PP{N=n}
,
$$ 
for all $n \in \mathbb{Z}_+$. Hence we can simplify the bound on the agreement probability to
\begin{align}
\PP{\vec{\pi}^\mathsf{CU} = \vec{\pi}^\mathsf{AS}}
&\leq
1
-
\rho e^{-\rho}
\E{\mathcal{E}(N)}
.
\label{probBoundExp}
\end{align}
Next, we will evaluate the expectation, $\E{\mathcal{E}(N)}$. From \eqref{calEdef}, we can see that, in terms of $N$, this expectation is proportional to
$$
\E{\left(1 - \frac{1}{N}\right)\left(1 - \frac{u}{N}\right)^{N-3}}
=
\sum_{n=2}^\infty
\left(1 - \frac{1}{n}\right)\left(1 - \frac{u}{n}\right)^{n-3}
\left(\rho n\right)^{n-1}
\frac{e^{-\rho n}}{n!}
=
\sum_{n=2}^\infty
(n - u)^{n-3}
\rho^{n-1}
\frac{e^{-\rho n}}{(n-2)!}
,
$$
where $u \in (0,1/2)$.} 

\edit{Now, if $X$ is a random variable defined such that $X = \sum_{i=1}^Y B_i$ where $Y \sim \mathsf{Pois}(\theta \mu)$ and $B_i \stackrel{\mathsf{iid}}\sim \mathsf{Borel}(\mu)$ are mutually independent with $\mu \in (0,1)$ and $\theta > 0$, then $X$ has probability mass function
\begin{align}
\PP{X = x}
&= 
\sum_{y=1}^x
\PP{X = x \mid Y = y}
\frac{\left(\theta \mu\right)^y e^{-\theta \mu}}{y!}
\nonumber
\\
&=
\sum_{y=1}^x
\frac{y}{x}
\frac{e^{-\mu x}(\mu x)^{x-y}}{(x-y)!}
\frac{\left(\theta \mu\right)^y e^{-\theta \mu}}{y!}
\nonumber
\\
&=
\theta e^{-\mu (x + \theta)}
\frac{\mu^x }{x!}
\sum_{y=1}^x
{x - 1 \choose y - 1}
x^{x-y}
\theta^{y-1}
\nonumber
\\
&=
\theta
(x+\theta)^{x-1}
e^{-\mu (x + \theta)}
\frac{ 
\mu^x }{x!}
,
\label{PBTeq}
\end{align}
for each $x \in \mathbb{Z}_+$, with $\PP{X = 0} = \PP{Y = 0} = e^{-\theta \mu}$, where $X \mid Y$ is Borel-Tanner distributed with rate $\mu$ and size $y$ \citep[c.f.][and one may observe that the Borel-Tanner reduces to the Borel when $y=1$]{haight1960borel}. The final equality follows from the third through the binomial theorem. We will refer to the distribution in \eqref{PBTeq} as the Poisson-Borel-Tanner with endogenous rate $\mu$ and exogenous rate $\theta$ ($\mathsf{PBT}(\mu, \theta)$), which by construction is a valid distribution on the non-negative integers for all $\mu \in (0,1)$ and all $\theta > 0$. Notice that, though \eqref{PBTeq} was developed for $x \geq 1$, this formula for the $\mathsf{PBT}(\mu,\theta)$ probability mass function also holds at $x = 0$.}

\edit{With this $\mathsf{PBT}(\mu,\theta)$ distribution in hand, we can see that $\E{\mathcal{E}(N)}$ is proportional to
$$
\E{\left(1 - \frac{1}{N}\right)\left(1 - \frac{u}{N}\right)^{N-3}}
=
\rho e^{-\rho u}
\sum_{j=0}^\infty
(j + 2 - u)^{j-1}
\rho^{j}
\frac{e^{-\rho (j+2 - u)}}{j!}
=
\frac{\rho e^{-\rho u}}{2- u}
.
$$
Substituting back into \eqref{calEdef}, we are left with
\begin{align*}
\E{\mathcal{E}(N)}
&=
\frac{1}{\rho}
\int_0^{\frac{1}{2}}
G\left(
G^{-1}\left(\rho ( 1 - u ) \right)
-
G^{-1}\left(\rho u\right)
\right)
\left(
2 - u
\right)
\E{
\left(
1
-
\frac{1}{N}
\right)
\left(
1 - \frac{u}{N}
\right)^{N-3}
}
\mathrm{d}u
\\
&=
\int_0^{\frac{1}{2}}
G\left(
G^{-1}\left(\rho ( 1 - u ) \right)
-
G^{-1}\left(\rho u\right)
\right)
e^{-\rho u}
\mathrm{d}u
,
\end{align*}
and by \eqref{probBoundExp}, this completes the proof.}
\hfill\Halmos\endproof



\edit{One can easily refine this result when given even slightly more specific assumptions. For instance, in a settings that covers many typical choices for the response function $g(\cdot)$, we can provide a bound on the integral and further simplify the overall expression.}

\begin{corollary}\label{nonIncCor}
\edit{Suppose that $g(\cdot)$ is non-increasing. Then,
\begin{align}
\PP{\vec{\pi}^\mathsf{CU} = \vec{\pi}^\mathsf{AS}}
\leq
1
- 
e^{-\rho}
\left(
\rho
+
2 e^{-\rho/2}
-
2
\right)
,
\end{align}
for all $\rho \in (0,1)$.}
\end{corollary}
\proof{Proof.}
\edit{Given that $g(\cdot)$ is non-increasing, we have that $G(\cdot)$ is concave. Because we also know that $G(\cdot) \geq 0$, we furthermore have that $G(\cdot)$ is sub-additive, which implies that
$$
G\left(
G^{-1}\left(\rho ( 1 - u ) \right)
-
G^{-1}\left(\rho u\right)
\right)
\geq
G\left(
G^{-1}\left(\rho ( 1 - u ) \right)
\right)
-
G\left(
G^{-1}\left(\rho u\right)
\right)
=
\rho(1-2u)
.
$$
By substituting this term into the integral in Theorem~\ref{disagreeThm} and simplifying, we complete the proof.}
\hfill\Halmos\endproof


\edit{On some level, it is actually somewhat remarkable that the same combinatorial objects appear with the same marginal distributions in two separate places for the Hawkes cluster model, and it is even more interesting now that Theorem~\ref{disagreeThm} (and, more simply, Corollary~\ref{nonIncCor}) has shown that these temporal and compensator-based parking functions are truly two separate and non-equal objects. In total, though the two parking function notions are indeed isomorphic as combinatorial objects, this series of results has shown that such equality cannot be guaranteed on any given sample path.} 

\edit{Of course, one would expect the bounds we have obtained here to be loose, as they were primarily built off of the probability that both parking functions would each contain only one $1$. To that end, let us now close this demonstration with Table~\ref{simTable}, which simulates many different clusters across several different parameter settings and shows just how rare parking function agreement may be.\endnote{\edit{Table~\ref{simTable} reported for simulation experiments with $\beta = 1/(1-\rho)$, but these values were also reproduced (up to statistically insignificant simulation noise) for the same number of replications with the $\rho$ values held fixed and $\beta$ twice as large.}} (Methodologically, the simulation procedure is precisely provided by the steps of Definition~\ref{pfcDef}.) Specifically, in the following simulation experiments, we numerically evaluate the probability of agreement (as is bounded in Theorem~\ref{disagreeThm}), the probability of \emph{non-trivial agreement} (because at $N = 1$, the parking functions are empty by definition, and, at $N = 2$, they are each always equal to $[1]$), and the probability of agreement given the size of the cluster. Like Table~\ref{simTable} shows, as the clusters become large, whether that be precisely or on average, it becomes exceedingly unlikely that $\pi^\mathsf{CU}$ and $\pi^\mathsf{AS}$ will be the same. By convention, a $1$ or $0$ without decimal values means that the event was observed in all or none of the replications, respectively, whereas these values presented with decimal places means that the fraction of observations round to such extremes, yet there were both observations and non-observations across the full experiment.}

\begin{table}[h]
\caption{\edit{Simulated probability of sample path agreement between the compensator-based parking function definition ($\pi^\mathsf{CU}$) and temporal parking function definition ($\pi^\mathsf{AS}$) for increasing average cluster size ($g(x) = \alpha e^{-\beta x}$, $2^{24}$ replications).}}\label{simTable}
\renewcommand{\arraystretch}{1.4}
\centering
\edit{
\begin{tabular}{c | c c c c c c c c}
& $\rho = \frac{1}{2}$ & $\rho = \frac{3}{4}$ & $\rho = \frac{7}{8}$ & $\rho = \frac{15}{16}$ & $\rho = \frac{31}{32}$ & $\rho = \frac{63}{64}$ & $\rho = \frac{127}{128}$ & $\rho = \frac{255}{256}$
\\
\hline
$\PP{\vec{\pi}^\mathsf{CU} = \vec{\pi}^\mathsf{AS}}$ & 0.8590 & 0.7176 & 0.6427 & 0.6059 & 0.5877 & 0.5786 & 0.5742 & 0.5722 
\\
$\PP{\vec{\pi}^\mathsf{CU} = \vec{\pi}^\mathsf{AS}, N > 2}$ & 0.0685 & 0.0779 & 0.0738 & 0.0705 & 0.0686 & 0.0675 & 0.0670 & 0.0670 
\\
$\PP{\vec{\pi}^\mathsf{CU} = \vec{\pi}^\mathsf{AS} \mid N = {1}/({1-\rho})}$ & 1 & 0.3027 & 0.0118 & 0.0000 & 0 & 0 & 0 & 0 \\
\end{tabular}
}
\end{table}


\begin{remark}[\editTwo{Different Parking Functions, Different Uses}]\label{pfUseRemark}
\edit{As a postlude to this section's demonstration of parking function disagreement, let us remark that not only are the two parking functions meaningfully different at the sample path level in terms of interpretation, they are also each of different potential use. For instance, the proofs throughout this paper can serve as a demonstration of the benefit of the temporal parking function for interpretable analysis of the cluster model, particularly so when the scope of that analysis is rooted in the cluster's chronology. To this end, this methodological value underscores \editTwo{our claim} that this temporal decomposition not only reinforces the notion \editTwo{that} ``parking functions are hidden spines within Hawkes clusters,'' \editTwo{as first remarked by \citet{daw2023conditional}  for $\pi^\mathsf{CU}$}, it makes that connection more direct and more interpretable.
Specifically, the compensator-based definition cannot produce this paper's foundational temporal analysis of the cluster as conducted in Appendix~\ref{smProofs}, and, moreover, the effort in the proofs within this present section is a testament to how much $\pi^\mathsf{AS}$ and $\pi^\mathsf{CU}$ can differ in analytical use.
However, let us also be clear that the temporal parking function identified in Definition~\ref{pfcDef} does not make \editTwo{the} compensator-based definition from \citet{daw2023conditional} redundant. For instance, \citet[][Theorem 2]{daw2021co} used the compensator-based definition and its associated conditional uniformity property to establish a novel extension of the ``residual analysis'' goodness-of-fit technique, and, because we have now shown that these notions are distinct, we also now know that the temporal parking function cannot provide that method of evaluating the model's fit to data.}
\end{remark}






\section{\edit{Performance of Bounds and Non-Asymptotic Non-Monotonicity: N-Shape Case Study}}\label{hSec}

\edit{At various points in the main body of the paper, we have claimed that more specific shapes of performance could be obtained through similarly specific assumptions on the agent-side slowdown function. This section of the appendix will be devoted to further exploration and discussion around $h(\cdot)$. First, as both a demonstration of the performance of the bounds in Proposition~\ref{boundProp} and Corollary~\ref{rateBound} and an investigation of shapes that are more general than the $\IU$- and $\U$-shapes, we will consider a particular agent-side slowdown function. By scrutinizing the bounds under a specific $h(\cdot)$, we  demonstrate how this interaction model can also produce (and, thus, be used to analyze) non-monotonic shapes that are more detailed than the definitions based on asymptotic extremes that we have analyzed in the main body of the paper. Then, drawing further inspiration from the literature, we will discuss how the conditions of Assumption~\ref{conA} could actually be relaxed while still producing some key results of the paper. Specifically, by dropping the requirement of Assumption~\ref{conA} that $h(\cdot)$ is non-decreasing, we will show that insights like Theorem~\ref{iuThm} immediately generalize, thus revealing additional connections to observations in the empirical literature.}

\edit{
For the sake of the first demonstration, let us now assume a specific form of $h(\cdot)$. That is, let us take 
\begin{align}
h(x)
&=
\begin{cases}
x^2 & 
\text{ if }
x \leq 12 \\
144 + 24\left(1 - e^{-(x-12)}\right) + \sqrt{x - 12} - \frac{\sqrt{\pi}}{2}  \mathsf{Erf}\left(\sqrt{x - 12}\right)
&
\text{ if }
x > 12
,
\end{cases}
\label{nhDef}
\end{align}
which may actually be more interpretable as the solution to differential equation
\begin{align}
h'(x)
&=
\begin{cases}
2 x
&
\text{ if }
x \leq 12
\\
24 e^{-(x-12)}
+
\frac{1}{2 \sqrt{x-12}} \left(1 - e^{-(x-12)}\right)
&
\text{ if }
x > 12
.
\end{cases}
\label{nhDefDE}
\end{align}
This $h(\cdot)$ is visualized in Figure~\ref{nFig-h}.}


\begin{figure}[ht]
\centering
\includegraphics[width=0.75\textwidth]{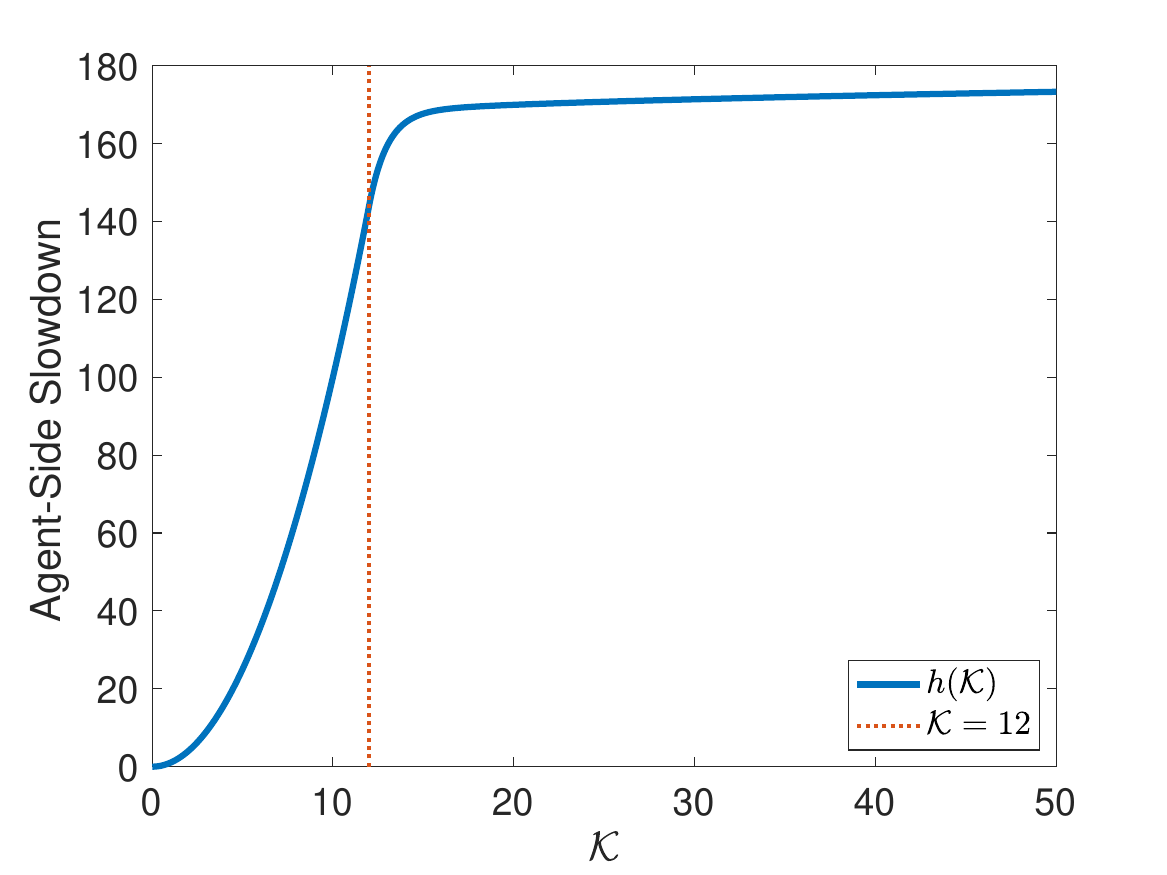}
\caption{\edit{Agent-side slowdown function as given in~\eqref{nhDef}.}}
\label{nFig-h}
\end{figure}

\edit{In the differential equation form of~\eqref{nhDefDE}, we can see the idea behind this choice of $h(\cdot)$. Up to the breakpoint at $\mathcal{K} = 12$, the slowdown function grows super-linearly according to a quadratic function. Beyond $\mathcal{K} = 12$, though, the growth of the slowdown function gradually switches to that of a square root function, which is, of course, strictly sub-linear. Borrowing the language of \citet{berry2016past}, one could say that this $h(\cdot)$ has a ``tipping point,'' after which the agent-side slowdown switches from an effect that is stronger than time to one that is weaker than time. Hence, as in accordance with Corollary~\ref{monoCor}, because $h(\cdot)$ is sub-linear in its tail, we know that, in the asymptotic characterization, the throughput tends toward infinity as the concurrency grows large: $\mathcal{K}/\E{\tau(\mathcal{K})} \in \mathcal{M}$. Both through simulation of the duration and by the bounds of the guaranteed and idealized service rates, let us now show that there is ample non-asymptotic detail to be seen when given this level of specificity.}



\begin{figure}[ht]
\centering
\includegraphics[width=0.495\textwidth]{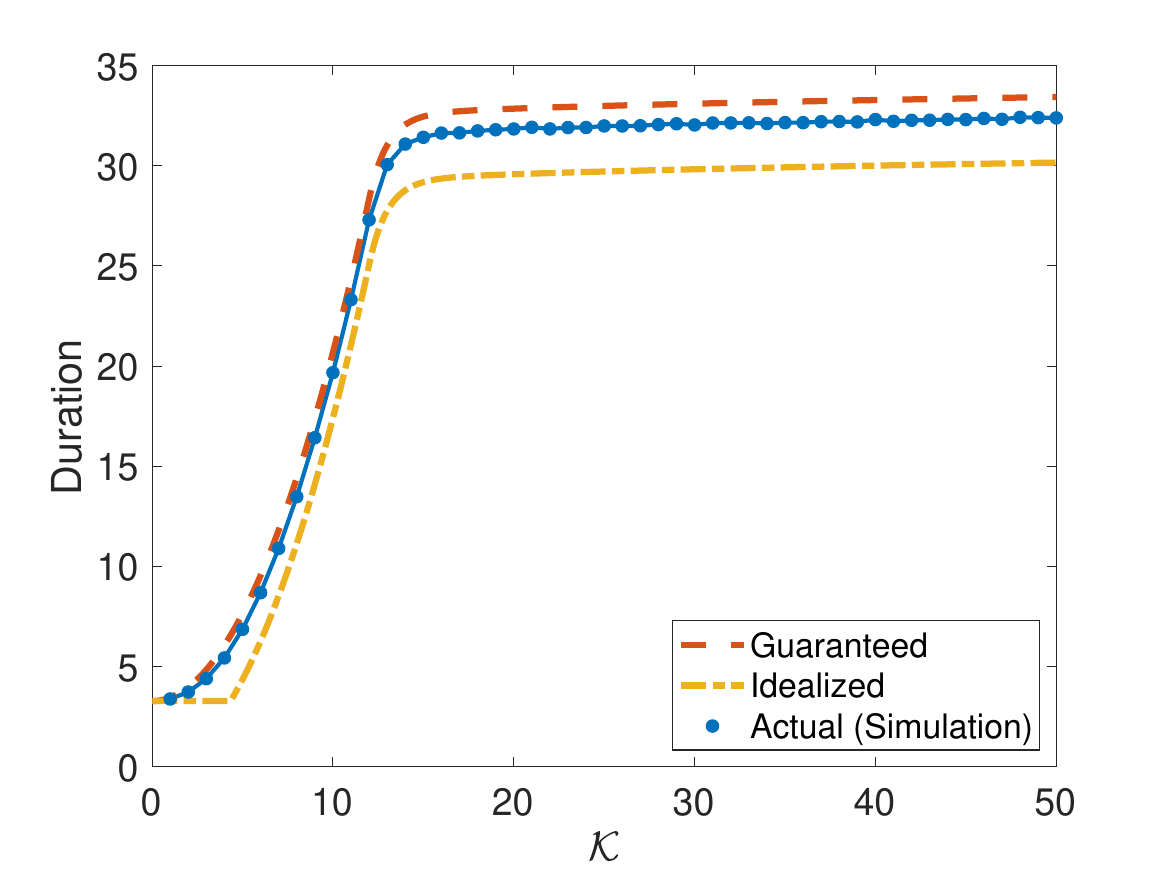}
\includegraphics[width=0.495\textwidth]{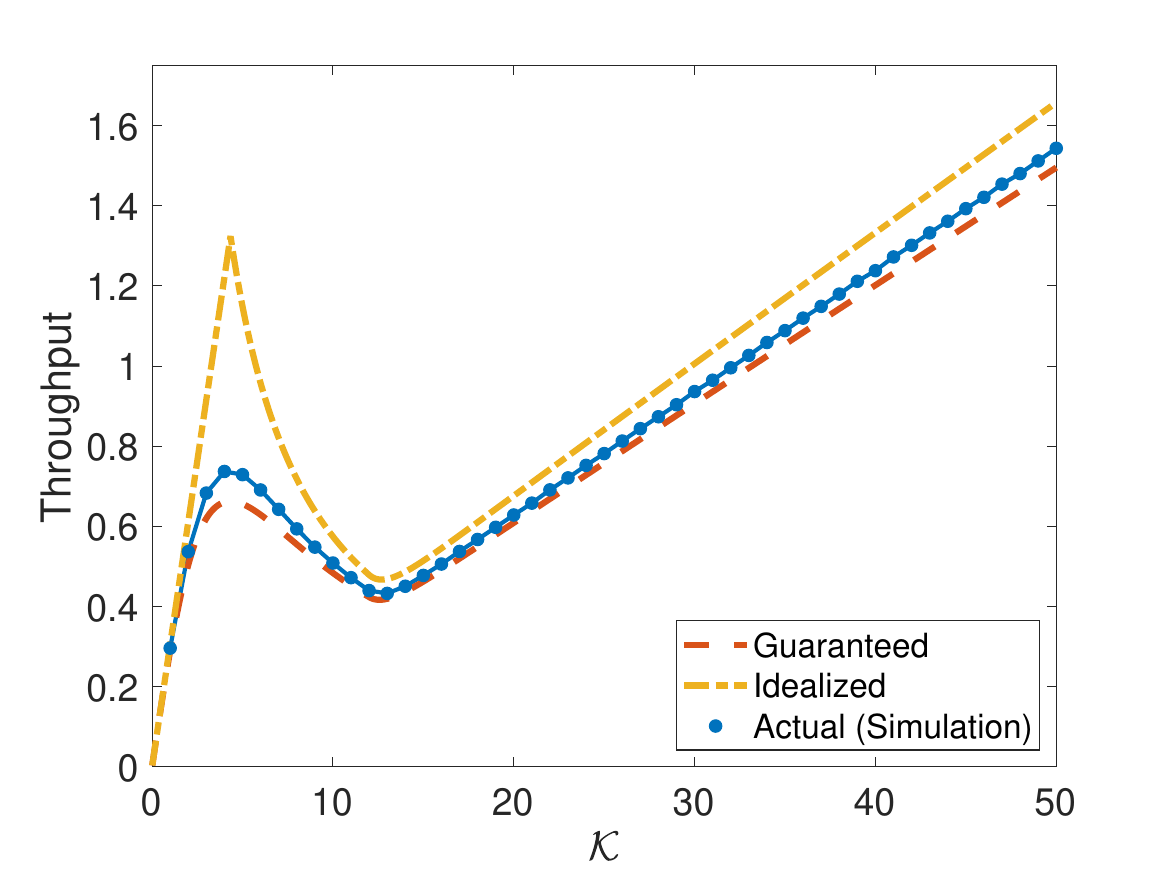}
\caption{\edit{Simulated duration (left) and throughput (right) under the agent-side slowdown function given in~\eqref{nhDef}, sandwiched by the upper and lower bounds provided through the guaranteed and idealized service rates.}}\label{nFig}
\end{figure}


\edit{In Figure~\ref{nFig}, for both the duration and the throughput, we plot the simulated actual performance, guaranteed service rate approximation, and idealized service rate approximation. The actual values are simulated precisely through the steps of Definition~\ref{pfcDef}, where the simulation generates $2^{22}$ interactions for each positive integer concurrency level. The guaranteed and idealized approximations each invoke $\E{\bar \tau^\mathsf{1}}$ and $\E{\bar \tau^\mathsf{2}}$ obtained via $2^{24}$ replications of the pair of marked Hawkes clusters, generated according to the steps of Definition~\ref{dpcDef} (with steps (i) and (ii) performed jointly by simply simulating a Poisson-mixture branching process with Borel distributed marks). Specifically, for $\bar \tau^\mathsf{1}$, the simulation conducts the steps of Definition~\ref{dpcDef} with $\bar g(\cdot) = g_\mathsf{1}(\cdot)$ and $M_i \sim \mathsf{Borel}(\rho_\mathsf{2})$; for $\bar \tau^\mathsf{2}$, the simulation uses $\bar g(\cdot) = g_\mathsf{2}(\cdot)$ and $M_i \sim \mathsf{Borel}(\rho_\mathsf{1})$. With $\E{\bar \tau^\mathsf{1}}$ and $\E{\bar \tau^\mathsf{2}}$ in hand, the guaranteed and idealized approximations follow immediately and deterministically for any $\mathcal{K}$.}


\edit{As one can see in Figure~\ref{nFig}, the throughput follows an \emph{N-shape}: the throughput rises, then falls, and then rises again as a function of the agent's concurrency. To the left of the tipping point at $\mathcal{K} = 12$, the throughput appears to be $\IU$-shaped, and indeed it would be $\mathcal{N}_\IU$ if $h(x) = x^2$ for all $x$. However, to the right of $\mathcal{K} = 12$, the throughput is monotonically increasing, and, likewise, the throughput would indeed belong to $\mathcal{M}$ if $h(x) = \sqrt{x}$ for all $x$.
In the blue line with circle marks, we see that this shape is what was observed numerically in the simulation, and by the combination of the dashed yellow and orange lines, we see that the simulated shape is not just by luck, as the lower right-hand valley point of the ``N'' for the idealized service rate is lower than the upper left-hand peak for the guaranteed service rate.} 

\edit{Following that intuition, we can readily formalize conditions on $h(\cdot)$ that produce this non-monotonic N-shape. We do so now in Corollary~\ref{nCor}, which follows immediately from the combination of Proposition~\ref{convexProp} and Corollary~\ref{monoCor}.}


\begin{corollary}\label{nCor}
\edit{Suppose that there exists some $\overline{x} \geq \overline{\mathcal{K}}$ such that $h(x)$ is strictly convex for all $x \in [0, \overline{x}]$, where $\overline{\mathcal{K}}$ is as defined in Proposition~\ref{convexProp}. If it is also the case that $h(x)/x \to 0$ as $x \to \infty$, then both of the following statements are true:
\begin{enumerate}[i)]
\item there exists a unique optimal concurrency on the open interval $(0,\overline{x})$ which maximizes the throughput on this interval,
\item the throughput is asymptotically monotonic: $\mathcal{K} / \E{\tau(\mathcal{K})} \in \mathcal{M}$.
\end{enumerate}}
\end{corollary}






\edit{We now close this section with a series of tangential, but perhaps notable, observations found in Figure~\ref{nFig}.}

\begin{remark}[\edit{Notable Service Interaction Model Properties Visualized in Figure~\ref{nFig}}]\label{propertiesRemark}
\edit{\editTwo{Consider the following findings from} Figure~\ref{nFig}.
\begin{enumerate}[i)]
\item First, let us notice that as the agent's concurrency shrinks infinitesimally small, the average duration of the service does not. Instead, we can see that the simulated actual value, the guaranteed upper bound, and the idealized lower bound all converge to the same strictly positive value. This is not a quirk of this particular parameter setting; one can see by Proposition~\ref{boundProp} that $\E{\tau} \to \E{\bar{\tau}^\mathsf{1}} > 0$ as $\eta \to \infty$ (which occurs as $\mathcal{K} \to 0$ because $h(0) = 0$ by Assumption~\ref{conA}).
\item Similarly, we can also observe from Proposition~\ref{boundProp} that the gap in the upper and lower bounds on the mean duration increases up to $\E{\bar{\tau}^\mathsf{1}}$ as $\mathcal{K} \to \infty$. This gap can likewise be seen in the left-hand side plot of Figure~\ref{nFig}.
\item  In the right-hand side plot, one can observe much of the intuition \editTwo{for} Theorem~\ref{iuThm} and Corollary~\ref{monoCor}. That is, if $h(\cdot)$ is strictly super-linear (like the example in Figure~\ref{nFig}  would appear to be before the tipping point), then the gap in the throughput bounds is asymptotically subsumed by the impacts of the agent-side slowdown, and both the guaranteed service rate and idealized service rate will go to 0. On the other hand, if $h(\cdot)$ is strictly sub-linear (as in Figure~\ref{nFig} beyond the tipping point), then the gap in the throughput bounds persists, but both the guaranteed and idealized service rates tend towards infinity, and thus so does the throughput. In each case, the gap in the throughput bounds is precisely the gap in the individual-level mean duration bounds converted into the scale of the agent-level service rate. 
\item One may notice from the right-hand side plot that the bounds on the throughput are specifically at their worst at the top-left peak of the N, meaning what would be the peak of the $\IU$-shape in the case that the agent-side slowdown is actually super-linear. Certainly, this optimized throughput may merit better bounds, and thus tightening this gap is likely worthy of future research attention.
\item In either plot within Figure~\ref{nFig}, one can notice that the guaranteed service rate (or corresponding bound on the duration) appears to be a better approximation for the true throughput (or mean duration, respectively), relative to the idealized service rate. Anecdotally, we have also observed this in further numerical experiments, and accordingly we view $\mathcal{G}(\cdot)$ as the more promising proxy for the throughput at present (and the same applies for $\E{\bar{\tau}^\mathsf{1}} + \E{\bar{\tau}^\mathsf{2}} / \eta$ as a proxy for $\E{\tau}$).
\end{enumerate}}
\end{remark}

\section{Numerical Demonstrations and Extension of Insights\edit{: Directionally Dependent Interactions}}\label{numerSec}


To close our investigation of customer-agent asymmetry and its implications for operating concurrent services, we conduct numerical demonstrations which serve a two-fold purpose. First and perhaps also foremost, these experiments will show that the shapes and insights we have identified transcend to a greater level of generality. We will simulate a system-level model, composed of interactions that have higher degrees of freedom in relationship modeling and that are closed systematically, rather than naturally. Hence, these demonstrations show a generality of our analytical results. Then, as the second purpose, these experiments will  use parameter estimates obtained directly from data in \citet{daw2021co}, in which it was shown that the Hawkes service model fits well to contact center timestamp data. Hence, these demonstrations will also ground our insights in true practice and show how the model can be used to evaluate alternate service designs that are obtained from perturbations from the real-world parameter estimates.



As aligned with the first aim, let us now generalize the contribution rate intensity from the two-sided expression in Definition~\ref{hcDef} to a construction that is both two-sided and quad-directional. Relative to the definition of $\mu_t$ in Equation~\eqref{intensityDef}, let us now define contribution rate\edit{s} $\mu_t^{\mathsf{1}}$ and $\mu_t^{\mathsf{2}}$ to be
\begin{align}
\mu_t^{\mathsf{1}}
&=
\int_0^t g_\mathsf{1,1}(t-s)\mathrm{d}N_s^\mathsf{1}
+
\int_0^t g_\mathsf{1,2}(t-s)\mathrm{d}N_s^\mathsf{2}
,
\label{mu1Def}
\end{align}
and
\begin{align}
\mu_t^{\mathsf{2}}
&=
\eta \cdot \int_0^t g_\mathsf{2,1}\left(\eta \cdot (t-s) \right)\mathrm{d}N_s^\mathsf{1}
+
\eta \cdot \int_0^t g_\mathsf{2,2}\left(\eta \cdot (t-s) \right)\mathrm{d}N_s^\mathsf{2}
,
\label{mu2Def}
\end{align}
where $N_t^\mathsf{1}$ and $N_t^\mathsf{2}$ count the number of contributions on Sides $\mathsf{1}$ and $\mathsf{2}$, respectively. By comparison to Equation~\eqref{intensityDef}, the model defined by Equations~\eqref{mu1Def} and~\eqref{mu2Def} captures the relationships within the service in greater detail, where the response functions capture four directions rather than merely two sides. Here, $g_{x,y}(\cdot)$ governs the excitement in the Side $x$  rate generated from prior contributions by Side $y$, where $x,y \in \{\mathsf{1}, \mathsf{2}\}$. While these added dimensions elevate the modeling degrees of freedom, they also significantly diminish its analytical tractability. Indeed, the parking function decomposition and its associated analytical techniques no longer apply, and thus we will demonstrate the robustness of our insights through simulation of this generalized model.\endnote{Note that we should \emph{not} expect Definition~\ref{pfcDef} to be equivalent to the cluster defined by Equations~\eqref{mu1Def} and~\eqref{mu2Def} because of the added dimensions.} To contrast with the two-sided model,  we will refer to the model given by Equations~\eqref{mu1Def} and~\eqref{mu2Def} as the \emph{quad-directional model}.

We conduct simulation experiments from two perspectives: agent-level and system-level. The former considers the quad-\editTwo{directional} model in a setting equivalent to that in which we have analyzed the two-sided model: the duration of a customer-agent service interaction given that the agent is operating at a fixed concurrency level. Then, the latter embeds this interaction model within a queueing model of the agent's total assignment, which may evolve throughout time. Instead of a fixed concurrency level, we will suppose that the agent can have a concurrency of at most $\kappa \in \mathbb{Z}_+$. Then, the queueing model can then be thought of as a $M/G/\kappa+M$ system in which the service duration is the duration of the Hawkes service model. Following Section~\ref{nonMonSec}, we will let the synchronicity be determined by the concurrency through $\eta(\mathcal{K}_t) = 1/h(\mathcal{K}_t)$, where now $\mathcal{K}_t$ is the agent's concurrency at time $t$. Hence, the service duration is system-state-dependent in the system-level model. To deepen the contrast (and thus further demonstrate the robustness of our analytical insights), we will maintain the assumption of naturally closed services (meaning the service ends at the last contribution)  in the agent-level model, but in the system-level model we will suppose that services are closed systematically. That is, in the system-level case, we will suppose that a given service will be ended when the probability that there will be no more contributions (which is a function of the contribution rates) first hits a certain target level, say $p > 0$. From \citet{daw2021co}, this systematic closure policy corresponds to a stopping time defined on the natural filtration of the quad-directional model, and thus we can use the methodology from \citet{daw2023markov} to simulate this system. 

For both experiments, we will use four parameter settings to capture a range of interdependence.\endnote{These four parameter settings were also considered in a prior, publicly available draft of \citet{daw2021co} in an entirely different numerical experiment, but they do not appear in the final, published version of the paper.} Drawing inspiration from \citet{roels2014optimal}, we perturb the estimated parameters in \citet{daw2021co} to consider hypothetical interdependence arrangements as alternate  realities of the co-produced service captured in that contact center data. Specifically, the modeled response functions in \citet{daw2021co} were all of the form $g_{x,y}(t) = \alpha_{x,y} e^{-\beta_{x,y} t}$ for some jump size $\alpha_{x,y}$ and decay rate $\beta_{x,y}$ in each of the four $(x,y) \in \{\mathsf{1},\mathsf{2}\}^2$ directions of the service. In the context of the contact center data, these were estimated to be $\alpha_{\mathsf{1,1}} = 0.843$, $\alpha_\mathsf{1,2} = 14.083$, $\alpha_{2,1} = 17.102$, and $\alpha_{2,2} = 113.719$ and $\beta_\mathsf{1,1} = 3.64$, $\beta_\mathsf{1,2} = 38.388$, $\beta_\mathsf{2,1} = 20.374$, and $\beta_\mathsf{2,2} = 260.1$ (see the \textsf{SysBHP} parameters from Table 2 therein, where we are taking $\alpha_{x,y}$ as the sum of the pair of corresponding coefficients since we are not considering the sentiment or word count random variables in this experiment). Let us now describe how we update these values to create the four focal parameter settings.


Using Corollary 1 of \citet{daw2021co}, we have that the expected size, or mean number of contributions in the service interaction, is given by 
\begin{align}
\E{N}
&=
\frac{
1 - \frac{\alpha_\mathsf{2,2}}{\beta_\mathsf{2,2}} + \frac{\alpha_{\mathsf{2,1}}}{\beta_{\mathsf{2,1}}}
}{
\left( 
1 - \frac{\alpha_{\mathsf{1,1}}}{\beta_\mathsf{1,1}}
\right)
\left( 
1 - \frac{\alpha_{\mathsf{2,2}}}{\beta_\mathsf{2,2}}
\right)
-
\frac{\alpha_{\mathsf{1,2}}}{\beta_\mathsf{1,2}}
\frac{\alpha_{\mathsf{2,1}}}{\beta_\mathsf{2,1}}
}
.
\end{align}
Similarly, as informed by the \citet{hawkes1974cluster} cluster perspective of the model, we can observe that $\alpha_\mathsf{1,1}/\beta_\mathsf{1,1} + \alpha_\mathsf{2,1}/\beta_\mathsf{2,1}$ is the expected number of total direct responses to a given Side $\mathsf{1}$ contribution, and, likewise, $\alpha_\mathsf{1,2}/\beta_\mathsf{1,2} + \alpha_\mathsf{2,2}/\beta_\mathsf{2,2}$ is the expected number of direct responses to a given Side $\mathsf{2}$ point. 

These three average quantities give us a way of designing experimental service settings that preserve the cumulative task load in the interaction, but change how it is divided among the customer and agent. Letting $\epsilon_\mathsf{1}, \epsilon_\mathsf{2} \in \mathbb{R}$, notice that if we perturb the estimated instantaneous impacts to become $\alpha_{x,y} + (\epsilon_y \mathbf{1}\{x = y\} - \epsilon_y \mathbf{1}\{x \ne y\}) \beta_{x,y}$ for each $x,y \in \{\mathsf{1}, \mathsf{2}\}$ (meaning, updated to be $\alpha_\mathsf{1,1} = 0.843 + 3.64 \epsilon_\mathsf{1} $, $\alpha_\mathsf{1,2} = 14.083 - 38.388 \epsilon_\mathsf{2}$, $\alpha_\mathsf{2,1} = 17.012 -  20.374 \epsilon_\mathsf{1}$, and $\alpha_\mathsf{2,2} = 113.719 +  260.1 \epsilon_\mathsf{2}$), then the mean number of direct responses will remain the same for a given contribution from each side, yet as $\epsilon_\mathsf{1}$ and $\epsilon_\mathsf{2}$ change, the replies become more skewed towards tending from one side over the other. Specifically, for $\epsilon > 0$, there would be a larger expected number of self-replies, and for $\epsilon < 0$, there would be a larger expected number of cross-replies. Let us notice that if, additionally, $\epsilon_\mathsf{2} = \beta_\mathsf{2,1}/\alpha_\mathsf{2,1} (1 - \alpha_\mathsf{1,1}/\beta_\mathsf{1,1}) \epsilon_\mathsf{1}$, then the total expected number of messages, $\E{N}$, will furthermore be unchanged. By consequence of this relationship, we can further recognize that the stability condition in this quad-directional model \citep[i.e., Theorem 1 of][]{daw2021co} implies that $\epsilon_\mathsf{1}$ and $\epsilon_\mathsf{2}$ will match in sign.

Together, this $\epsilon_\mathsf{1}$ and $\epsilon_\mathsf{2}$ construction allows us to consider a quad-directional generalization of the interdependence spectrum: the average total taskload and average number of direct replies all stay the same, but which parties contribute more (on average) shifts with the $\epsilon$'s. Positive perturbations will increase the rates at which both parties follow-up on their own prior points and decrease the rates at which they respond to the other side; negative perturbations will conversely create more cross-responses and less self-replies. Hence, as the $\epsilon$'s increase, the interaction becomes increasingly \emph{mutually self-productive}, in which the customer and agent largely work independently but may each contribute significantly; as the $\epsilon$'s decrease, the interaction becomes increasingly \emph{mutually co-productive}, in which the service is characterized by a significant degree of customer-agent collaboration.

With this spectrum in mind, let us interpret the parameters as originally estimated in the data of \citet{daw2021co} by way of the four \emph{responsiveness ratios} in the quad-directional model, meaning $\alpha_{x,y} / \beta_{x,y}$ for each $(x,y) \in \{\mathsf{1},\mathsf{2}\}^2$, which are the average number of direct responses in each direction of the service interaction. In this unperturbed setting with $\epsilon_\mathsf{1} = \epsilon_\mathsf{2} = 0$, we see that both the customer and agent respond to the other party more than they follow-up to their own messages, $0.232 = \alpha_\mathsf{1,1}/\beta_\mathsf{1,1} < \alpha_\mathsf{1,2}/\beta_\mathsf{1,2} = 0.367$ and $0.437 = \alpha_\mathsf{2,2}/\beta_\mathsf{2,2} < \alpha_\mathsf{2,1}/\beta_\mathsf{2,1} = 0.839$, but it is not the case that the two cross-response directions of the service interaction, $(\mathsf{1},\mathsf{2})$ and $(\mathsf{2},\mathsf{1})$, have the two largest responsiveness ratios overall. Hence, we will deem this setting as \emph{moderately co-productive}. To form three other parameter settings across the range of interdependence, we will set $\epsilon_\mathsf{1}$ (and, by direct consequence, $\epsilon_\mathsf{2}$) values that shift the weight of the interaction among these responsiveness ratios. First, for a \emph{highly co-productive} setting, we will take $\epsilon_\mathsf{1} = -0.2$, so that responses to the other party strictly dominate self-responses: $0.032 = \alpha_\mathsf{1,1}/\beta_\mathsf{1,1} < \alpha_\mathsf{1,2}/\beta_\mathsf{1,2} = 0.501$ and $0.303 = \alpha_\mathsf{2,2}/\beta_\mathsf{2,2} < \alpha_\mathsf{2,1}/\beta_\mathsf{2,1} = 1.039$ with $\min\{\alpha_\mathsf{1,2}/\beta_\mathsf{1,2}, \alpha_\mathsf{2,1}/\beta_\mathsf{2,1} \} > \max\{\alpha_\mathsf{1,1}/\beta_\mathsf{1,1}, \alpha_\mathsf{2,2}/\beta_\mathsf{2,2} \}$. Then, turning to mutual self-production, for a \emph{moderately self-productive} setting, we will take $\epsilon_\mathsf{1} = 0.25$, so that each party follows-up to their own messages more often than they \editTwo{respond} to the other side, but follow-ups do not globally dominate cross-responses: $0.482 = \alpha_\mathsf{1,1}/\beta_\mathsf{1,1} > \alpha_\mathsf{1,2}/\beta_\mathsf{1,2} = 0.199$ and $0.605 = \alpha_\mathsf{2,2}/\beta_\mathsf{2,2} > \alpha_\mathsf{2,1}/\beta_\mathsf{2,1} = 0.589$, yet $\max\{\alpha_\mathsf{1,2}/\beta_\mathsf{1,2}, \alpha_\mathsf{2,1}/\beta_\mathsf{2,1} \} > \min\{\alpha_\mathsf{1,1}/\beta_\mathsf{1,1}, \alpha_\mathsf{2,2}/\beta_\mathsf{2,2} \}$. Finally, for a \emph{highly self-productive} setting in which both the customer and agent contribute but mostly do so in their own silos, we will take $\epsilon_\mathsf{2} = 0.5$: $0.732 = \alpha_\mathsf{1,1}/\beta_\mathsf{1,1} > \alpha_\mathsf{1,2}/\beta_\mathsf{1,2} = 0.032$ and $0.772 = \alpha_\mathsf{2,2}/\beta_\mathsf{2,2} > \alpha_\mathsf{2,1}/\beta_\mathsf{2,1} = 0.339$, so now $\max\{\alpha_\mathsf{1,2}/\beta_\mathsf{1,2}, \alpha_\mathsf{2,1}/\beta_\mathsf{2,1} \} < \min\{\alpha_\mathsf{1,1}/\beta_\mathsf{1,1}, \alpha_\mathsf{2,2}/\beta_\mathsf{2,2} \}$. 


Let us emphasize that these four parameter settings  constitute just four points along a  slice of the space of what service interactions could be modeled. However, we believe that these are natural alterations of the service system captured in the original contact center data. These perturbations constitute service conversations with different instantaneous impacts upon each message ($\alpha_{x,y}$'s), the same rates of processing the impacts ($\beta_{x,y}$'s), different responsiveness ratios (each individual $\alpha_{x,y}/\beta_{x,y}$), the same mean number of responses per message from each party (each $\alpha_{\mathsf{1},x}/\beta_{\mathsf{1},x} + \alpha_{\mathsf{2},x}/\beta_{\mathsf{2},x}$), and the same mean number of messages overall ($\E{N}$). In this way, these constructions preserve the overall average work within the service interaction, but the particular responsibilities shift across the relationship between customer and agent.


\begin{figure}[htb]
\centering
\subfigure[Highly co-productive service setting]{
\includegraphics[width=0.33\textwidth]{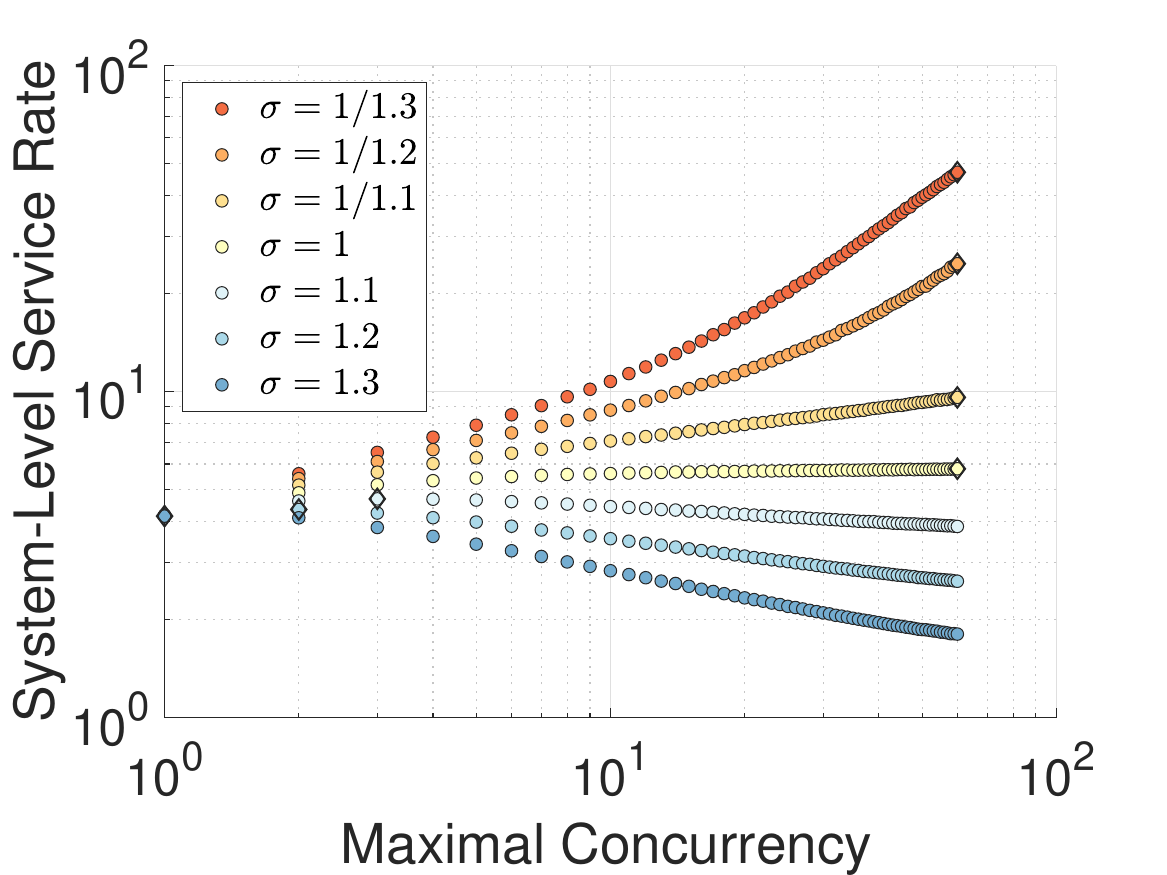}
\includegraphics[width=0.33\textwidth]{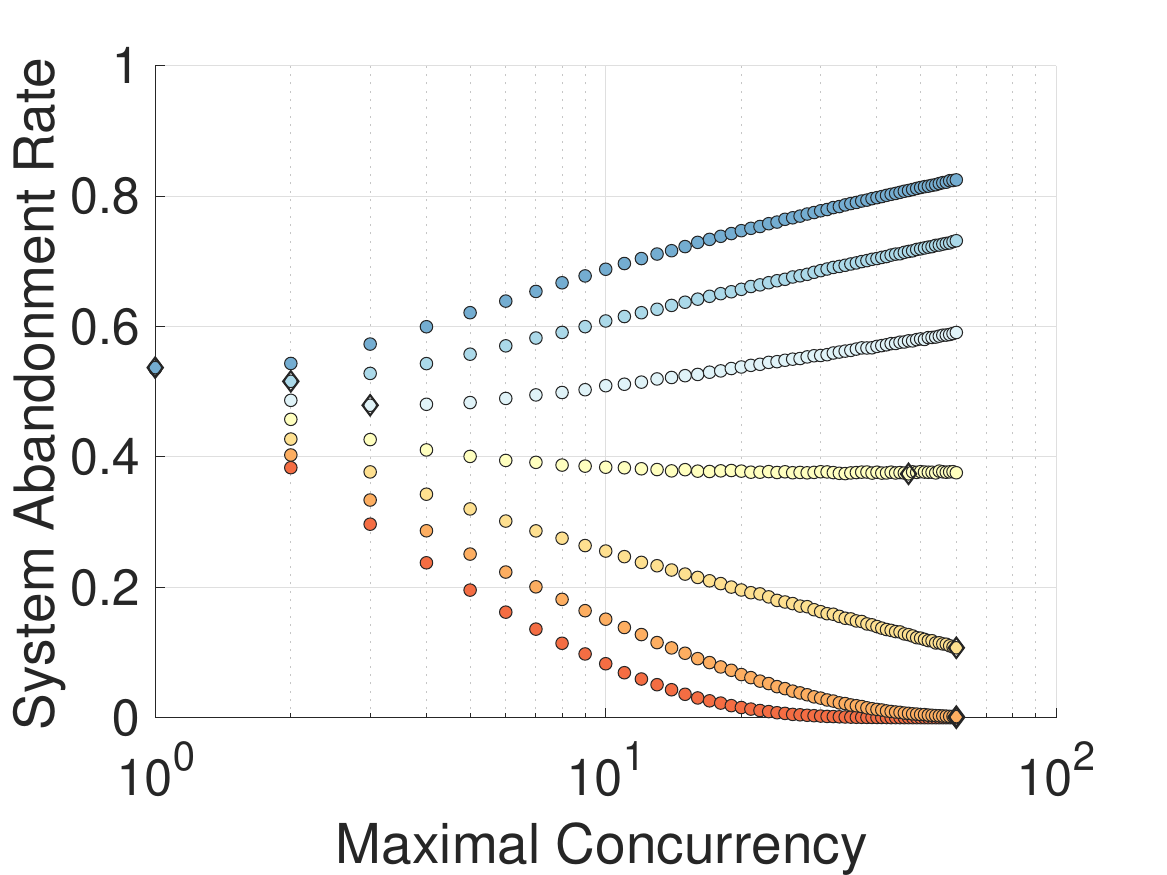}
\includegraphics[width=0.33\textwidth]{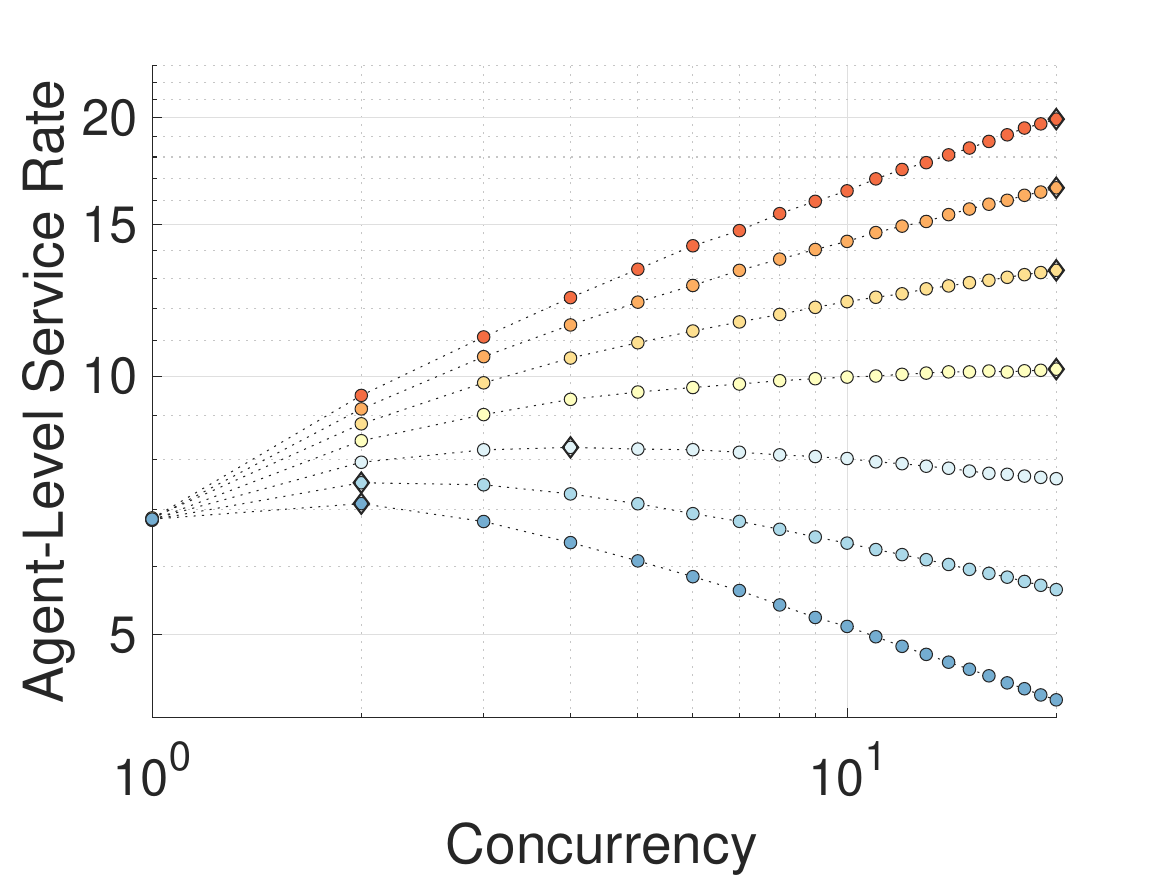}
} 
\subfigure[Moderately co-productive service setting]{
\includegraphics[width=0.33\textwidth]{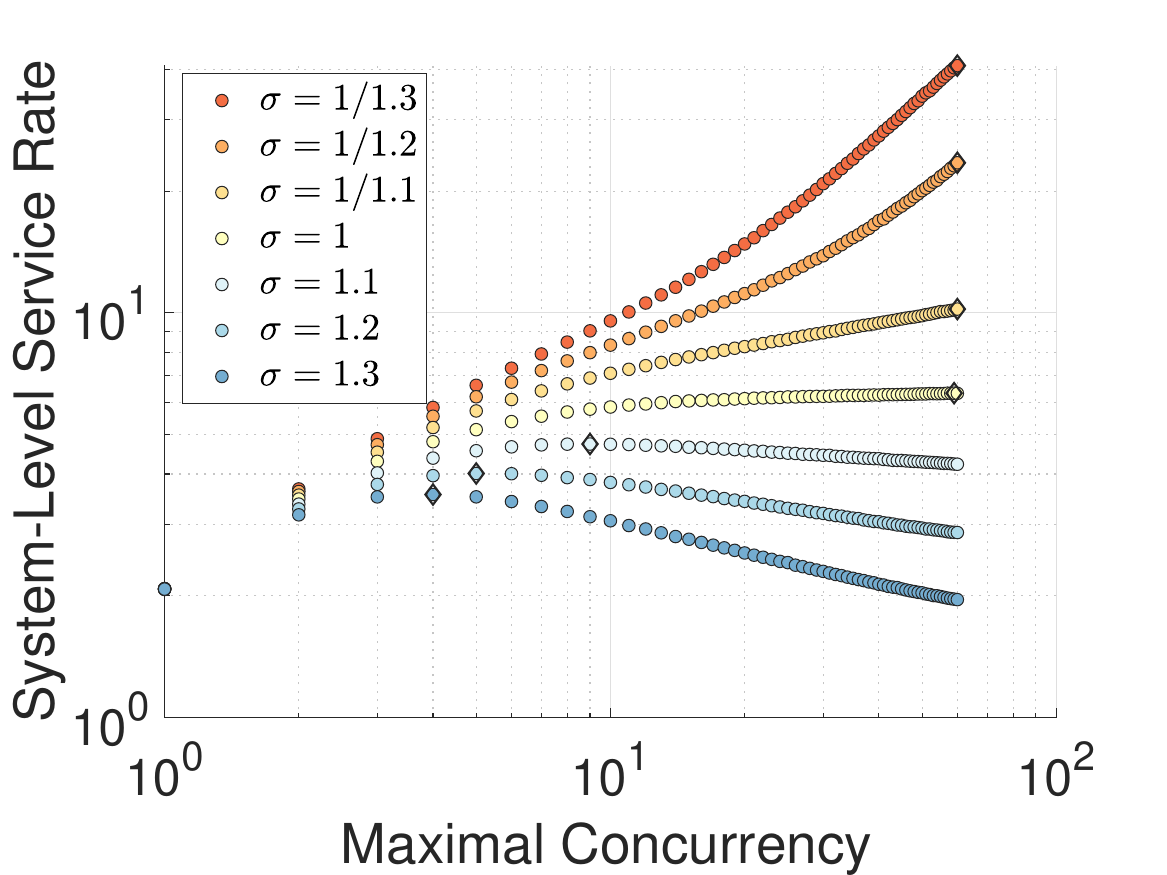}
\includegraphics[width=0.33\textwidth]{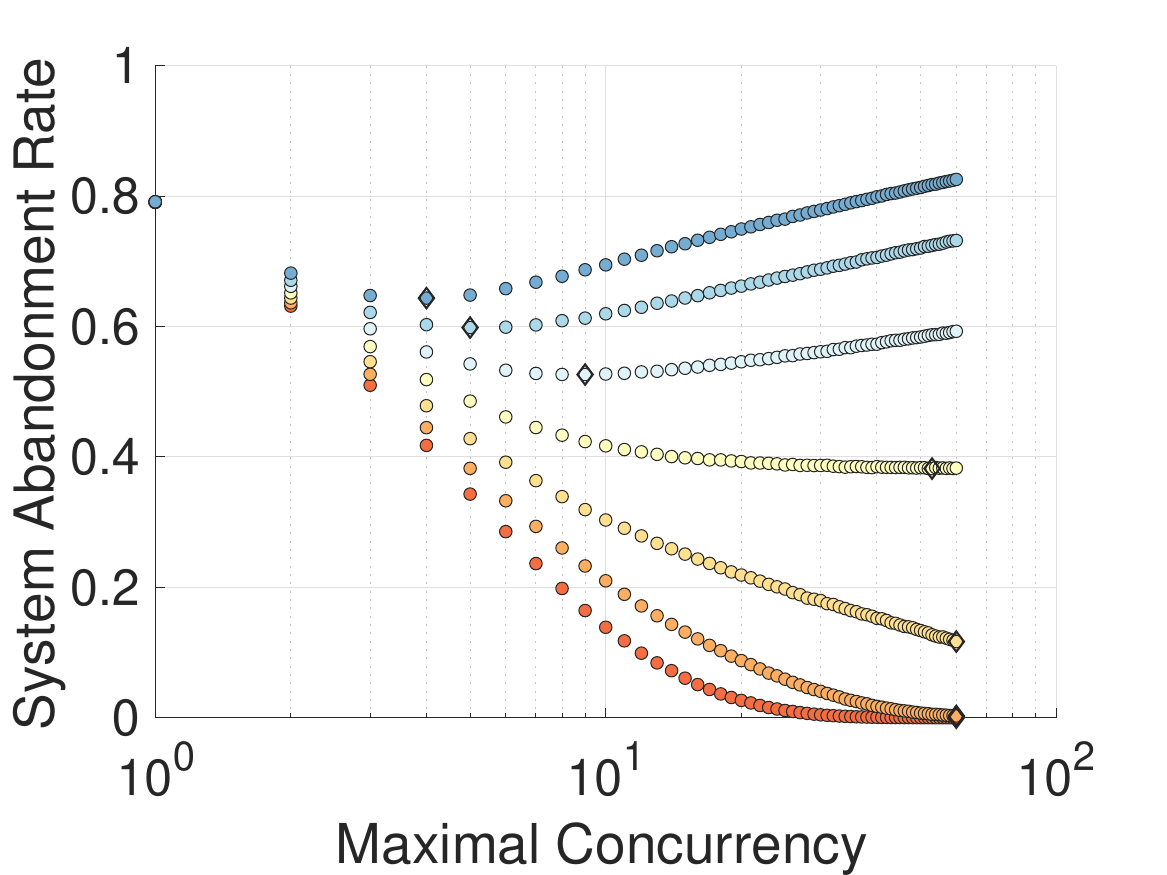}
\includegraphics[width=0.33\textwidth]{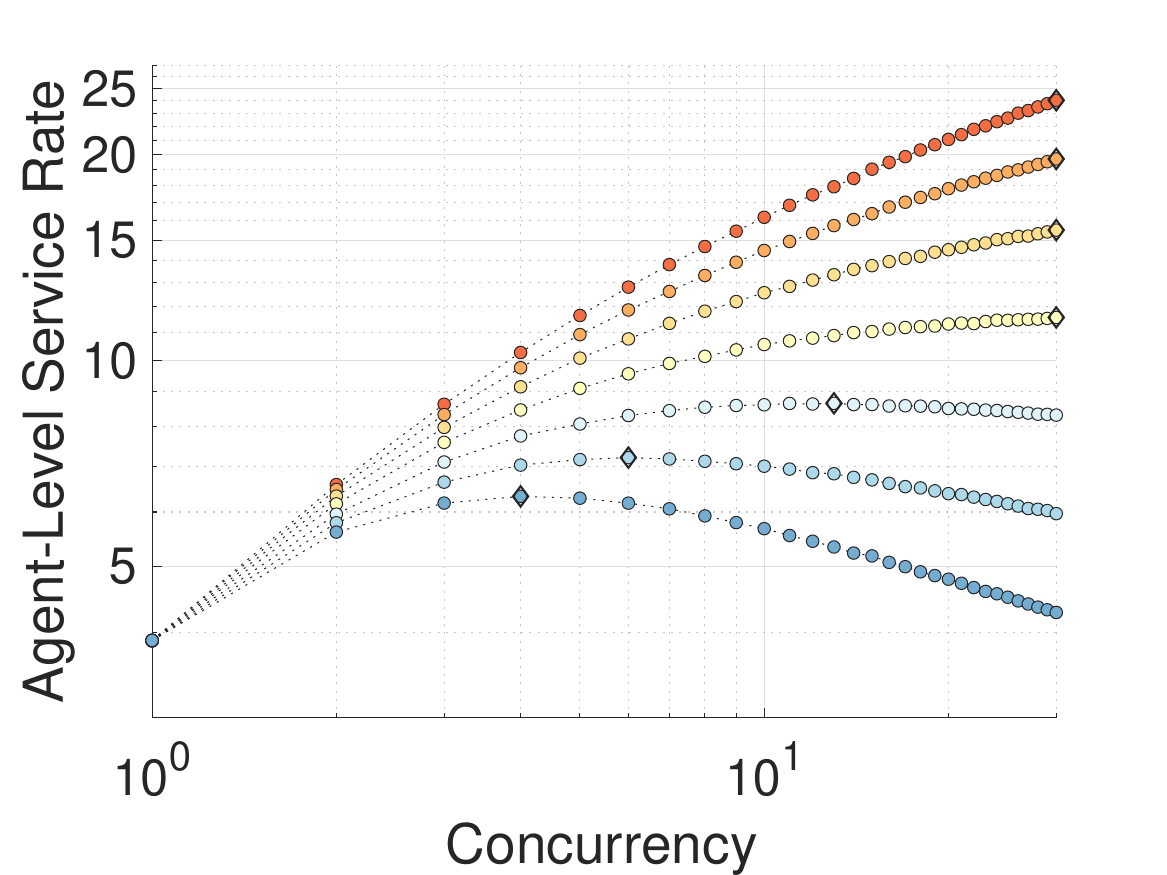}
} 
\subfigure[Moderately (mutually) self-productive service setting]{
\includegraphics[width=0.33\textwidth]{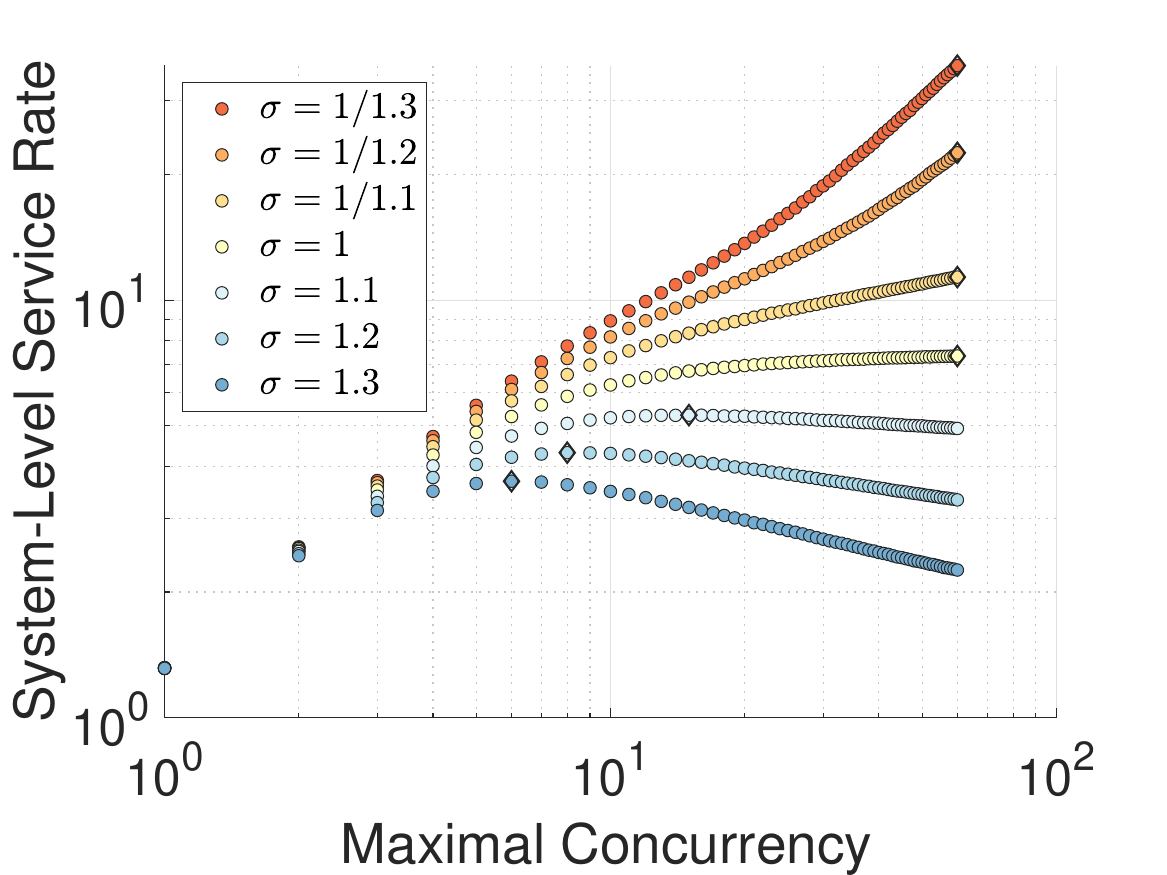}
\includegraphics[width=0.33\textwidth]{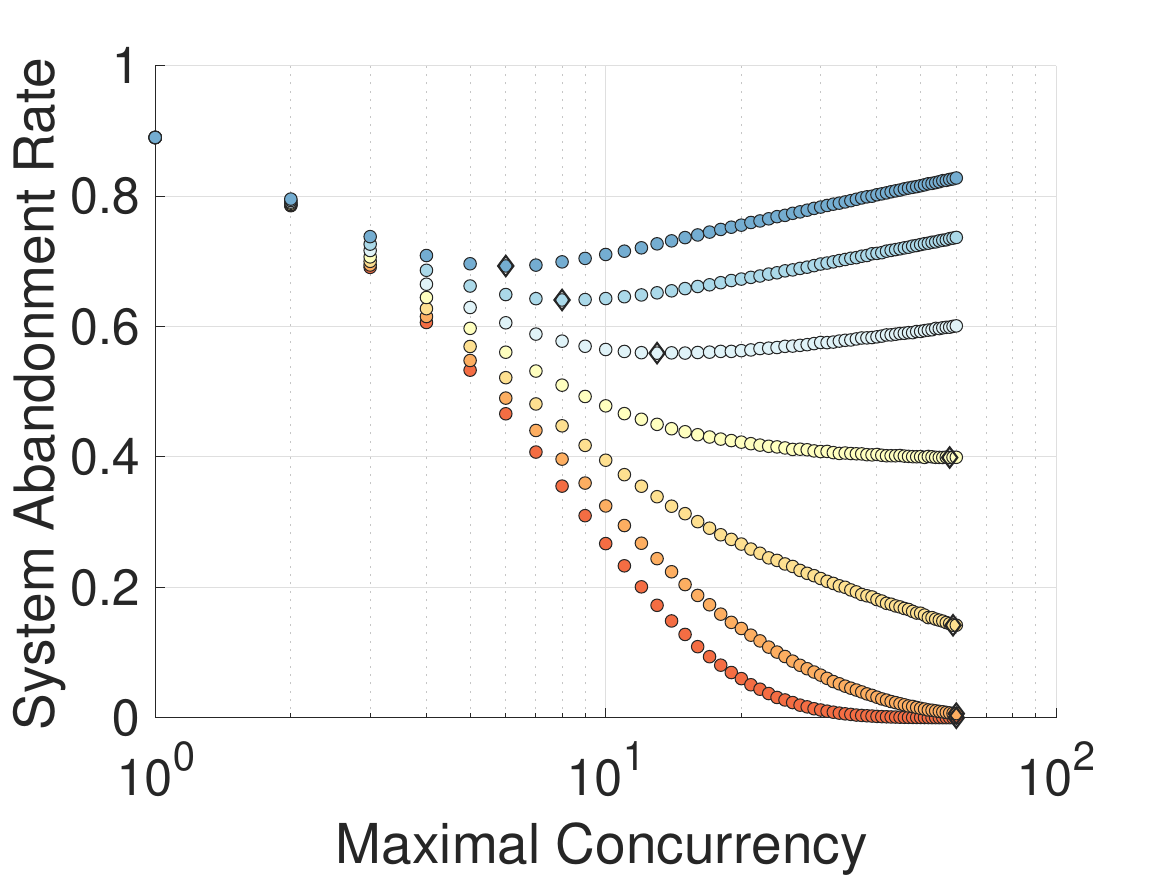}
\includegraphics[width=0.33\textwidth]{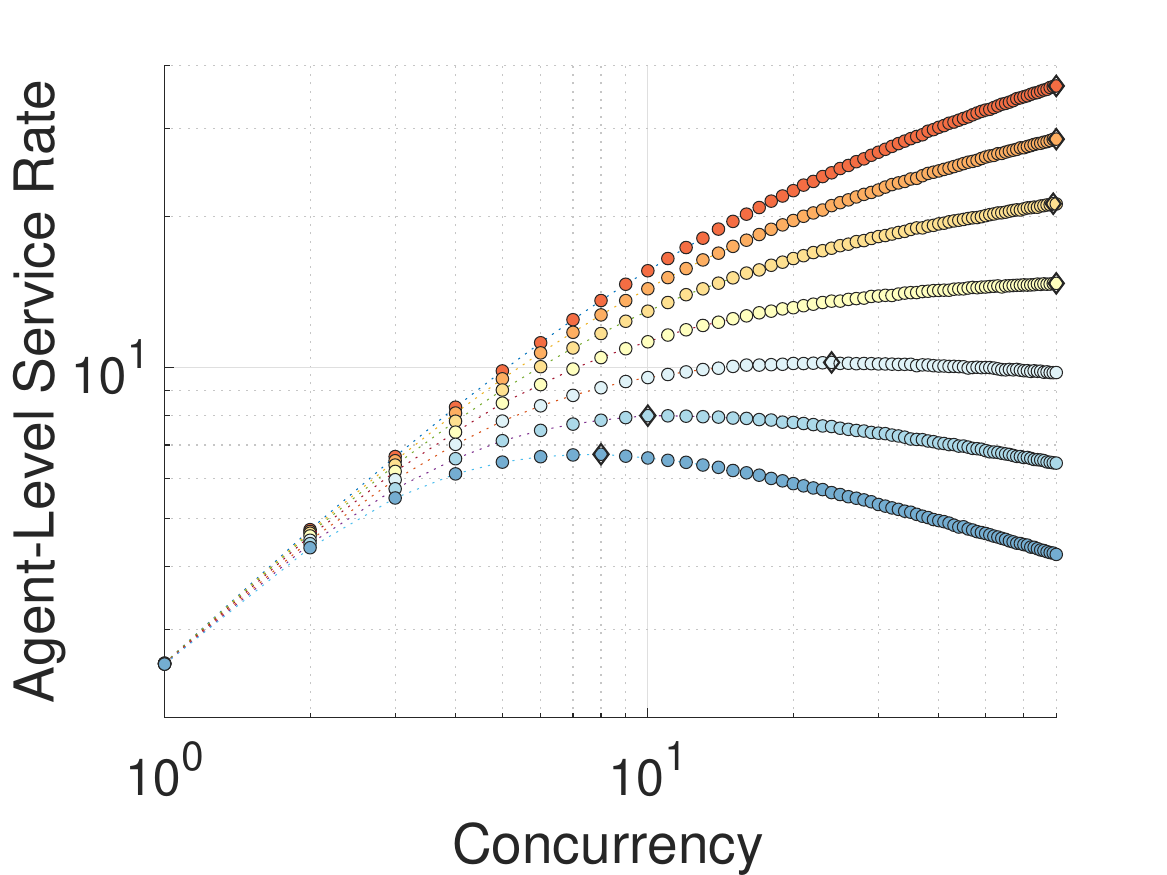}
} 
\subfigure[Highly (mutually) self-productive service setting]{
\includegraphics[width=0.33\textwidth]{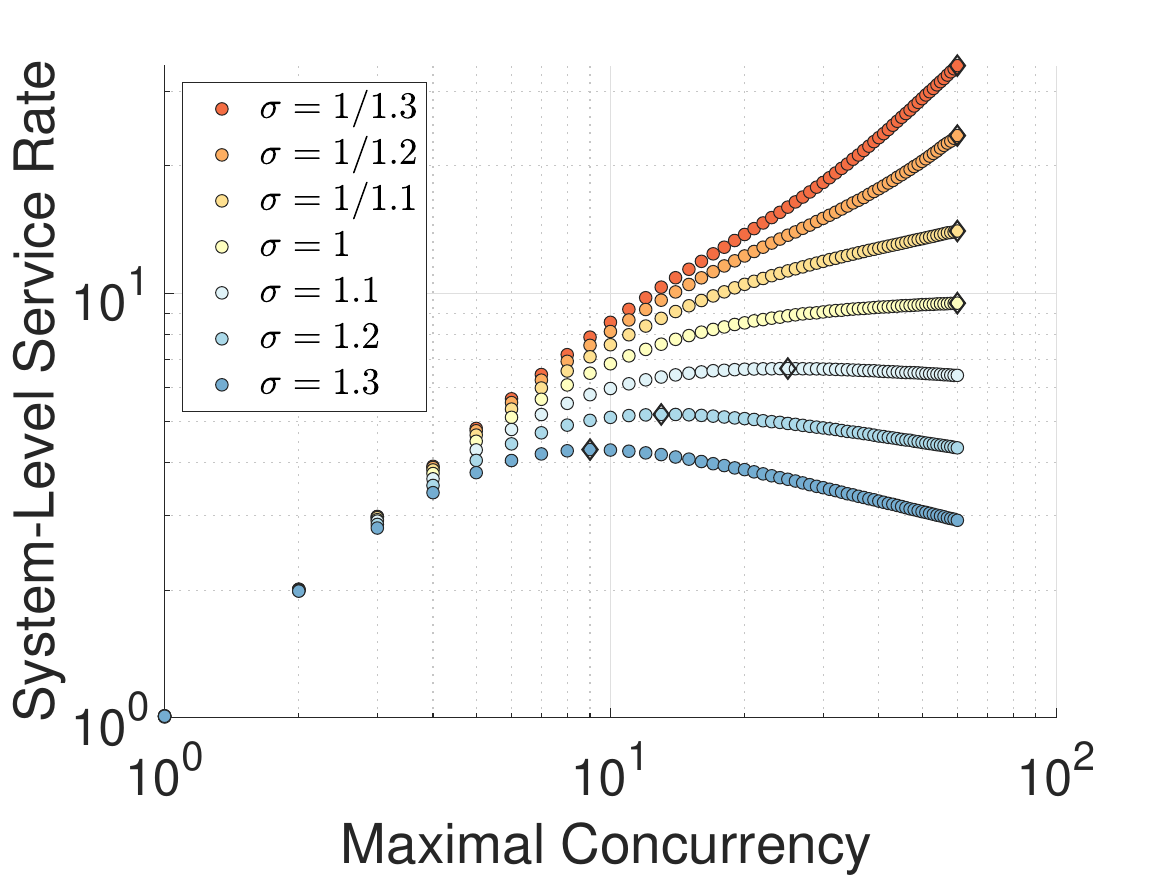}
\includegraphics[width=0.33\textwidth]{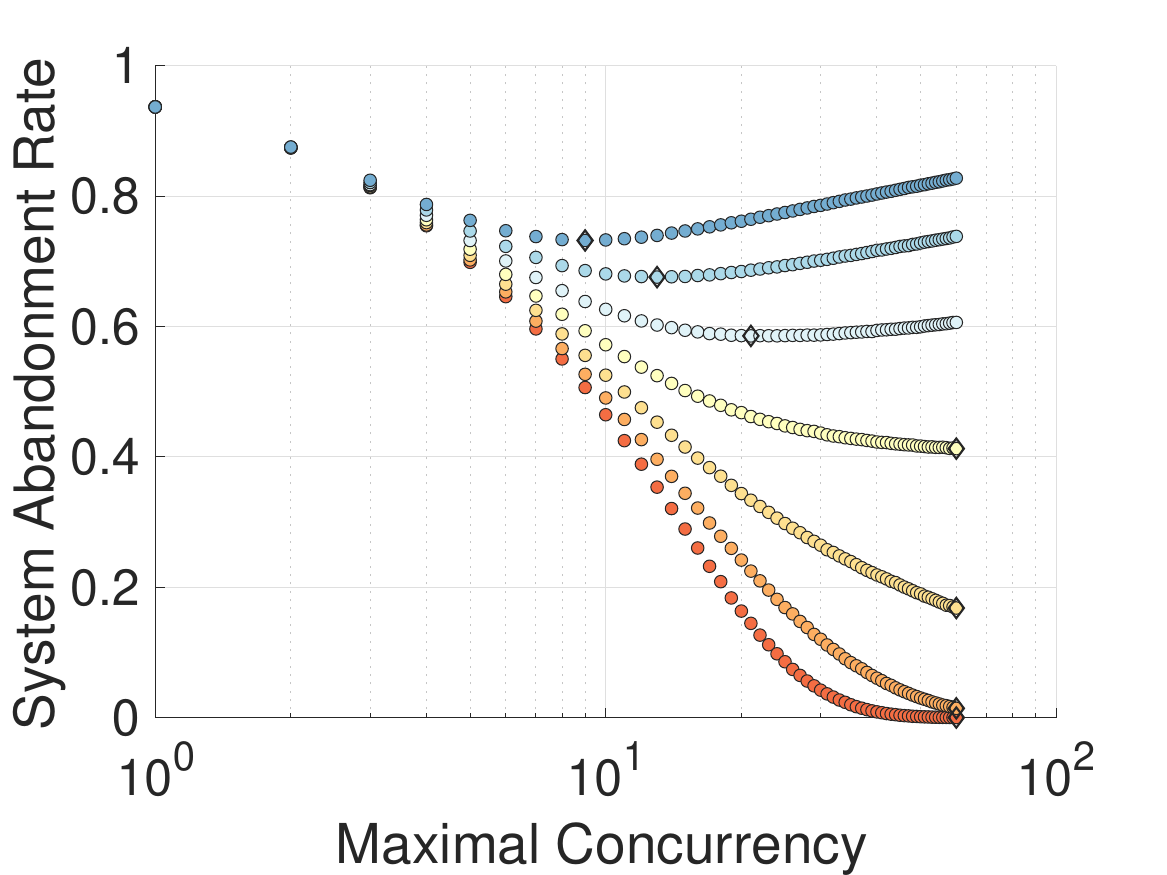}
\includegraphics[width=0.33\textwidth]{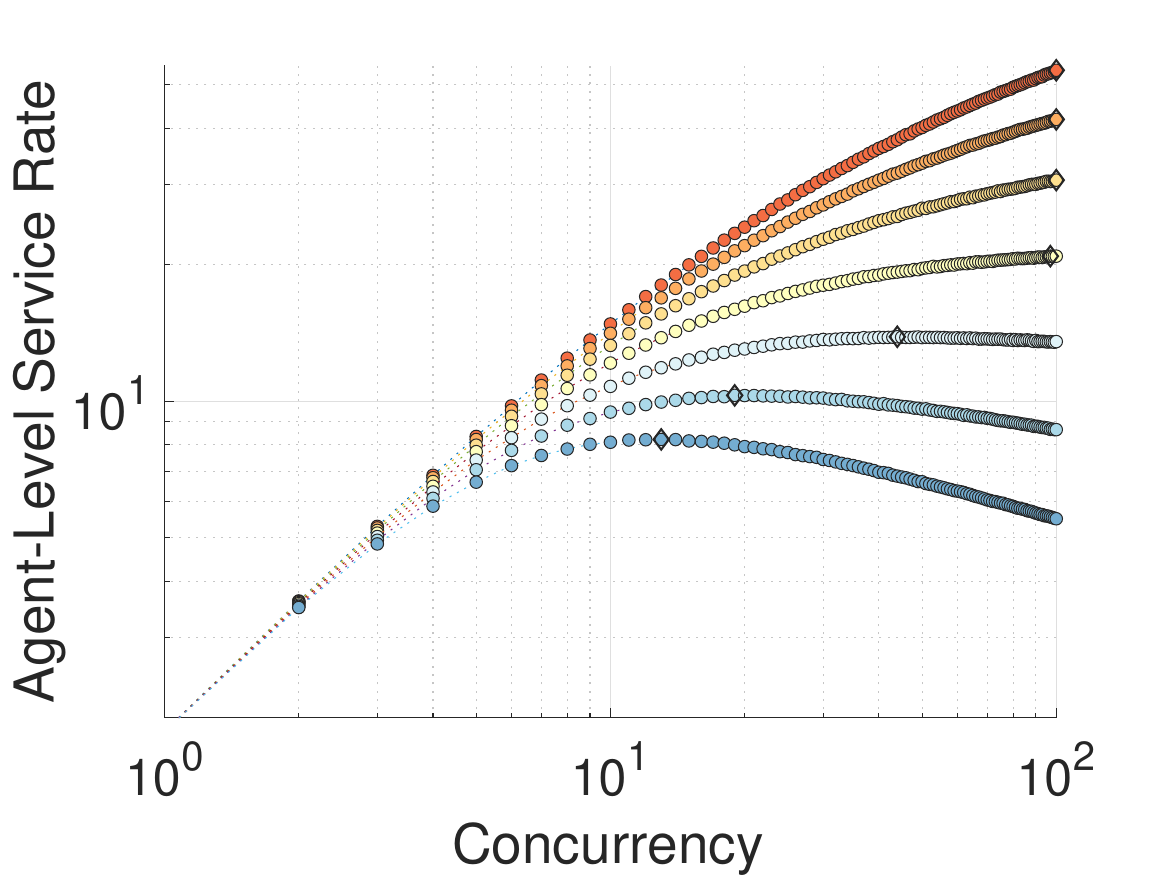}
}
\caption{Demonstration of conditions for non-monotonic performance in agent-level and system-level simulations. Diamonds mark the empirical maximum (minimum) of the throughput (abandonment) rates.}\label{iuCbhpFig}
\end{figure}


With these four settings in hand, let us now precisely describe the two experiments we conduct for each combination of parameters. First, we simulate at the agent-level, meaning we simulate customer-agent interactions under a fixed, constant concurrency level and repeat the experiment for different concurrency levels. In essence, this is simulating the model analyzed throughout the paper, with the addition of the quad-directional representation of the interaction.  Services are closed at their natural duration (\edit{occurring at} the final contribution in the cluster), and hence the focal performance metric is agent-level service rate, meaning the fixed concurrency level divided by the mean service duration at that level.\endnote{In the main body, we simply use the term ``throughput'' to refer to agent-level service rate, but here we take a more precise terminology, given the two different levels of the simulation experiments.} As we studied in Section~\ref{polySec}, we take the agent-side slowdown function to be $h(\mathcal{K}) = \mathcal{K}^\sigma$, where $\mathcal{K}$ is the concurrency level and where $\sigma > 0$. Then, the system-level experiments take this agent-level interaction model and embed it within a queueing theoretic model of the agent's evolving concurrent workload. Specifically, we simulate a \edit{$M/HC/\kappa+M$} model for the customers either waiting for or receiving service from one agent\edit{, where we use $HC$ to denote that the service duration is obtained via the Hawkes cluster model of the service interaction, which may depend on the state of the system. Specifically,}  the argument of the slowdown function is the agent's concurrency at the current time, $\mathcal{K}_t$. We vary the maximal concurrency level $\kappa$ across experiments. To extend beyond the agent-level experiments, we also now suppose that services are closed systematically, given by the stopping time when the probability of no more activity first reaches $p = 0.9$. For these synthetic experiments, the external arrival rate is $\lambda = 16$, and the abandonment rate is $\theta = 0.5$. All agent-level simulations contain $2^{20}$ replications, and all system-level simulations are comprised of $2^{10}$ replications, each with a time horizon of 200.

From right-to-left, the columns of Figure~\ref{iuCbhpFig} show the agent-level service rate (or fixed-level throughput) at each fixed concurrency, the system-level abandonment rate at each maximal concurrency, and system-level service rate (or average throughput) at each maximal concurrency, with each row labeled for one of the four aforementioned service interaction parameter settings. In each plot combining one interdependence setting and one performance metric, the experiment results are shown for slowdown function degree ranging over $\sigma \in \{1/1.3, 1/1.2, 1/1.1, 1, 1.1, 1.2, 1.3\}$. As seen in the analytical results of Proposition~\ref{polyOptProp}, Figure~\ref{iuCbhpFig} shows that performance is a non-monotonic function of the agent's concurrency when the slowdown effect is stronger-than-time ($\sigma > 1$, shaded in blue), and monotonicity appears when the slowdown effect is equal-to or weaker-than-time ($\sigma \leq 1$, shaded in orange). 

This \editTwo{condition on $\sigma$} is consistent in both the agent-level and system-level experiments, suggesting that the insights around, and mere existence of, an optimal concurrency generalize beyond the modeling framework we have analyzed. For example, in the right-most column of plots, the peaks of the throughput all occur within the interior of the concurrency values when the slowdown effect on the agent is super-linear. We then see this \editTwo{phenomenon} extend to the system-level experiments in the middle and left-most columns, where the same conditions produce $\IU$-shaped average throughput and $\U$-shaped abandonment rates. This should not be taken for granted, and one can quickly reason that the occurrence of this monotonicity must also depend on other system features. For example, if the arrival rate was trivially low, then there could not be sufficient customer volume to produce $\IU$-shaped average throughput, even if the agent-level analysis and experiments show that the agent's underlying production would yield it for larger customer volumes. Nevertheless, the fact that the agent-level monotonicity conditions \emph{can} translate to the system-level performance is valuable. Through this extension of the results, the insights also become increasingly practical: in many services, it may be more feasible to set a target maximal concurrency level, rather than to prescribe a target fixed concurrency level, as the latter relies on the existence of something like a consistent customer stream. \editTwo{These findings raise} interesting questions around determining maximal concurrency, which may further connect to classic staffing decisions and the recent notion of metastability \citep[e.g.][]{dong2015service,dong2022metastability}, and we look forward to pursuing these in future research.



\theendnotes


\end{APPENDIX}

\end{document}